\documentclass[11pt,a4paper]{article}
\usepackage[hmargin=1.1in,vmargin={1.1in,1.1in}]{geometry}
\PassOptionsToPackage{hyphens}{url}\usepackage[colorlinks]{hyperref}
\usepackage{amsmath}
\usepackage{multirow}
\usepackage{mathtools}
\usepackage{amssymb}
\usepackage{bbm}
\usepackage{url}
\usepackage{float}
\usepackage{tabularx}
\usepackage{natbib}
\usepackage[font=footnotesize]{caption,subcaption} 
\usepackage{graphicx}
\usepackage[affil-it]{authblk}
\usepackage{array} 
\usepackage{pdflscape}
\usepackage{verbatim}
\usepackage{placeins} 
\usepackage[dvipsnames,table,xcdraw,svgnames]{xcolor}
\usepackage{pdfpages}
\usepackage{lineno} 
\usepackage[nottoc,numbib]{tocbibind} 
\usepackage[percent]{overpic} 
\usepackage{chngcntr} 
\usepackage{fullpage}

\newcolumntype{L}[1]{>{\let\newline\\\arraybackslash\hspace{0pt}}m{#1}}

\AtBeginDocument{%
 \hypersetup{
  citecolor=Bittersweet,
    linkcolor=Blue,
  urlcolor=Blue}
}

\title{Production networks and epidemic spreading: \\
How to restart the UK economy?}

\author{Anton Pichler$^{1,2,3}$, Marco Pangallo$^{4}$, R. Maria del Rio-Chanona$^{1,2}$,\hspace{5cm}
Fran\c{c}ois Lafond$^{1,2}$, J. Doyne Farmer$^{1,2,5}$ \\

\vspace{0.25cm}

\footnotesize{$^{1}$ Institute for New Economic Thinking at the Oxford Martin School, University of Oxford, UK} \\
\footnotesize{$^{2}$ Mathematical Institute, University of Oxford, UK}\\
\footnotesize{$^3$ Complexity Science Hub Vienna, Austria} \\
\footnotesize{$^{4}$Sant’Anna School of Advanced Studies, Pisa, Italy} \\
\footnotesize{$^{5}$ Santa Fe Institute, US}\\
}

\date{\today}

\begin{document}

\maketitle

\begin{abstract}
We analyse the economics and epidemiology of different scenarios for a phased restart of the UK economy.  Our economic model is designed to address the unique features of the COVID-19 pandemic.   Social distancing measures affect both supply and demand, and input-output constraints play a key role in restricting economic output. Standard models for production functions are not adequate to model the short-term effects of lockdown. A survey of industry analysts conducted by IHS Markit allows us to evaluate which inputs for each industry are absolutely necessary for production over a two month period. Our model also includes inventory dynamics and feedback between unemployment and consumption. We demonstrate that economic outcomes are very sensitive to the choice of production function, show how supply constraints cause strong network effects, and find some counter-intuitive effects, such as that reopening only a few industries can actually lower aggregate output.  Occupation-specific data and contact surveys allow us to estimate how different industries affect the transmission rate of the disease. We investigate six different re-opening scenarios, presenting our best estimates for the increase in $R_0$ and the increase in GDP. Our results suggest that there is a reasonable compromise that yields a relatively small increase in $R_0$ and delivers a substantial boost in economic output.  This corresponds to a situation in which all non-consumer facing industries reopen, schools are open only for workers who need childcare, and everyone who can work from home continues to work from home. \\

\end{abstract}

\vfill

\footnotesize{
\noindent
\emph{Acknowledgements:} We would like to thank David Van Dijcke, David Vines, Eric Beinhocker, Spencer Fox and John Muellbauer for many useful comments and discussions. We thank Baillie Gifford, IARPA, the Oxford Martin School and JSMF for the funding that made this possible. We appreciate that IHS Markit provided us with a survey on critical vs. non-critical inputs.  (Note that JDF is on their advisory board).\\
\\
\noindent
\emph{Contacts:}
anton.pichler@maths.ox.ac.uk, 
marco.pangallo@santannapisa.it,
rita.delriochanona@maths.ox.ac.uk,
francois.lafond@inet.ox.ac.uk,
doyne.farmer@inet.ox.ac.uk
}
\normalsize

\newpage

\section{Introduction} \label{sec:intro}

The social distancing measures imposed to combat the COVID-19 pandemic have created severe disruptions to economic output.  Governments throughout the world are contemplating or implementing measures to ease social distancing and reopen the economy, which may involve a tradeoff between increasing economic output vs. increasing the expected number of deaths due to the pandemic.  Here we investigate several scenarios for the phased reopening of the economy. At one extreme, we find that reopening only a very limited number of industries can create supply chain mis-coordination problems that in some cases might actually decrease aggregate output. In contrast, reopening all industries would most likely increase $R_0$ above 1. We find a good scenario in-between these extremes: reopening a large part of the upstream industries, while consumer-facing industries stay closed, limits supply chain mis-coordination while providing a large boost to output and a relatively small increase in infection rates.

The shocks to the economy caused by social distancing are highly industry specific.  Some industries are nearly entirely shut down by lack of demand, others are restricted by lack of labor, and many are largely unaffected.  Feedback effects amplify the initial shocks.  The lack of demand for final goods such as restaurants or transportation propagates upstream, reducing demand for the intermediate goods that supply these industries.   Supply constraints due to lack of labor under social distancing propagate downstream, by creating input scarcity that can limit production even in cases where the availability of labor and demand would not have been an issue. The resulting supply and demand constraints interact to create bottlenecks in production. The resulting decreases in production may lead to unemployment, decreasing consumption and causing additional amplification of shocks that further decrease final demand.

Understanding these effects requires a model at the level of individual industries. 
Most of the economic analysis of the COVID-19 pandemic uses relatively aggregate macro models \citep{eichenbaum2020macroeconomics,bodenstein2020social}, with only a few studies predicting the economic effects using input-output (IO) models. IO models are particularly relevant to evaluate the consequences of crises such as COVID-19, where different sectors are affected differently, and the propagation of shocks through supply chains is likely to amplify the initial effects. Table \ref{tab:lit_table} summarizes the main features of several IO models that have been put forward recently to evaluate the macroeconomic effects of the COVID-19 crisis. Our paper differs in a number of important ways from the literature. On one hand, we provide comprehensive scenarios, an estimation of the epidemic spreading, non-equilibrium dynamics, and explicit demand shocks together with a sophisticated consumption response. On the other hand, we do not model prices, as we argue that price changes during the lockdown are relatively small.

The most important conceptual difference that distinguishes our model is our treatment of the production function, which dictates most of the behavior of the models listed in Table \ref{tab:lit_table}.
Essentially, the literature can be ordered by the degree to which the production function allows substitutions between inputs. At one extreme, the Leontief production function assumes a fixed recipe for production, allowing no substitutions and restricting production based on the limiting input \citep{inoue2020propagation}.  Under the Leontief production function, if a single input is severely reduced, overall production will be reduced proportionately, even if that input is ordinarily relatively small. This can lead to unrealistic behaviours. For example, the steel industry has restaurants as an input, presumably because steel companies have a workplace canteen and sometimes entertain their clients and employees.  A literal application of the Leontief production function would predict that a sharp drop in the output of the restaurant industry will dramatically reduce steel output. This is unrealistic, particularly in the short run.

\begin{table}[H]
\tiny
\begin{center}
    \begin{tabular}{ | p{16mm} | p{16mm}  | p{16mm} | p{16mm} | p{16mm} | p{16mm} | p{16mm} |p{17mm}|}
    \hline
    Feature                 & \citet{inoue2020propagation}   & \citet{barrot2020sectoral}   & \citet{mandel2020economic} & \citet{fadinger2020effects} & \citet{bonadio2020global} &\citet{baqaee2020nonlinear}  & \textbf{This paper} \\ 
    \hline

 Date & 31 Mar & 2 April&  29 Apr & April & April & 7 May& \today \\
    \hline
Geographical scope   & Japan & France + some EU countries&  World             & Germany                        &  64 countries     & US & UK \\
\hline
Data/ aggreg. level & 1.67M firms & INSEE 38 + WIOD 54 sectors                  & WIOD 55 sectors                    & Eurostat 62 sectors  & OECD 33 sectors & BEA 66 sectors & WIOD 55 sectors \\
    \hline
Production function  &Leontief & CES              & CES    & CD                                      & CD nesting CES  & nested CES & Leontief for ``critical inputs'' + Lack of non-critical inputs does not harm output \\
    \hline
Utility and/ or consumption function        &  Fixed demand &  CES           & CES                                          & CD & 2 nests CES        & CRS non parametric & Linear \\
    \hline
Supply shocks         & Essential activities & WFH + Essential industries + School closures  & Plausibility considerations from \citet{dorn2020economic}   & WFH & WFH + lockdown stringency + Essential for Health only & WFH + some share ``not removed'' & WFH + Essential industries \\
    \hline
Final demand shocks         & No &  No                                  &  No              &  No                         & No  & Theory only & Yes \\
    \hline
Equilibrium              & No   & Yes                                     & Transient          & Yes                    & Yes & Yes & No\\
    \hline
Prices                       & No   & Yes                                     &  Yes, sticky           & Yes              & Yes & Yes & No\\
    \hline
Inventories                   &Inputs inventories  & No & Output inventories                                                                          & No & No & No& Inputs inventories \\
    \hline
Fiscal/ monetary policy &No  & No                                     &  No                       &  No                &     No & No & Yes\\
    \hline
Reopening policy        &Yes & Individual industries          &  Yes (lockdown lifted)    & Individual industries      & Yes, symmetric to lockdown & No & Yes, scenarios \\
    \hline
Epidemic spreading                    &No     & No             & No                             & No                                     & No & No & Yes\\
     \hline
Other                   &      &      &  Frictional input shares adjustment; hysteresis in final demand                                                                      & & Counter factual for renationalizing supply chains &  Productivity shocks (theory) & V-shape or instantaneous recovery consumption function\\
    \hline
    \end{tabular}
\end{center}
\caption{Summary of features in IO papers modelling the effect of COVID-19 on the economy. WFH: Work From Home. CD: Cobb Douglas. CES: Constant Elasticity of Substitution. }
\label{tab:lit_table}
\normalsize
\end{table}

The alternatives used in the literature are the Cobb-Douglas production function \citep{fadinger2020effects}, which has an elasticity of substitution of 1, and the CES production function, where typically calibration for short term analysis uses an elasticity of substitution less than 1 \citep{barrot2020sectoral,mandel2020economic,bonadio2020global}. Some papers \citep{baqaee2020nonlinear} consider a nested CES production function, which can accommodate a wide range of technologies. In principle, nests could allow for substitution between some inputs and forbid it between others, in different ways for different industries. However, it is hard to calibrate all these elasticities, so that in practice many models end up using very limited nesting structure or assuming uniform substitutability. Consider again our example of the steel industry. With common calibrations of the (nested) CES production function, firms could substitute iron for energy, while still producing the same output. To the extent that certain production processes are encoded in fixed technological ``recipes'', this is clearly unrealistic\footnote{
Papers using CES production functions also assume equilibrium mechanisms for substitution and prices that are not completely realistic for the short run dynamics in the current context. Consider again our restaurant example. In an equilibrium model with a CES production function, if the output of restaurants is constrained, the relative price of restaurants will immediately increase, and firms will choose a relatively lower level of restaurant inputs, and a relatively higher level of other inputs. 
In the current context the closure of work canteens in steel factories is not driven by increased prices but policy and it is not clear that it is affecting steel output, or that it is substituted by other inputs.
}.

We argue that modeling production during the COVID-19 crisis requires a new approach to production functions, that is different from both standard Leontief and CES production functions. In this paper, we mostly keep the basic Leontief assumption that firms cannot substitute one input for another. However, we depart from the Leontief assumption in that we allow firms to keep producing as long as they have the inputs that are absolutely necessary, which we call ``critical inputs''. The steel industry cannot produce steel without iron and energy, but it can operate for a considerable period of time without restaurants or logistics consultants. 
Specifically, we make the assumption that if restaurants cannot supply the steel industry, the steel industry simply keeps producing at the same rate. This is of course only an approximation. To keep the same example, by not using restaurants, the costs of the steel industry are reduced and, ceteris paribus, its profits increase.   
In reality, non-critical inputs may have an impact on steel output that could be modeled as a shock to productivity. However, we think that during the short time-scales of the pandemic, these problems are second-order effects, and our production function provides a better assumption than Leontief or CES production functions.

In order to determine which inputs are critical and which are not, we use a survey that IHS Markit performed at our request. In this survey they asked ``Can production continue in industry X if input Y is not available for two months?''. The list of possible industries X and Y was drawn from the 55 industries in the  World  Input-Output  Database.  This question was  presented  to  30  different  industry analysts who were experts in industry X.  Each of them was asked to rate the importance of each of its inputs Y. They assigned a score of 1 if they believed input Y is critical, 0 if it is not critical, and 0.5 if it is in-between, with the possibility of a rating of NA if they could not make a judgement.  We then apply the Leontief function to the list of critical inputs, ignoring non-critical inputs.  We experimented with several possible treatments for industries with ratings of 0.5 and found that we get somewhat better empirical results by treating them as non-critical (though at present we do not have sufficient evidence to resolve this question unambiguously).

Besides the bespoke production function discussed above, we also introduce a COVID-19-specific treatment of consumption. Most models do not incorporate the demand shocks that are caused by changes in consumer preferences in order to minimize risk of infection. The vast majority of the literature has focused on the ability to work from home, and some studies incorporate lists of essential vs. inessential industries, but almost no papers have also explicitly added shocks to consumer preferences.  (\cite{baqaee2020nonlinear} is an exception, but the treatment is only theoretical). Here we use the estimates from \citet{del2020supply}, which are taken from a prospective study by the \citet{CBO2006}. These estimates are crude, but we are not aware of estimates that are any better. As we write, data on actual consumption starts to become available; what we have seen so far is qualitatively consistent with the shocks predicted by the CBO, with massive shocks to the hospitality industry, travel and recreation, milder (but large) shocks elsewhere, and increased expenses in groceries/food retail \citep{andersen2020consumer,carvalho2020tracking,chen2020impact}. Besides the initial shock, we also attempt to introduce realistic dynamics for recovery and for savings.  The shocks to on-site consumption industries are more long lasting, and savings from the lack of consumption of specific goods and services during lockdown are only partially reallocated to other expenses.

The first step in the analysis of our model is empirical validation: We compare model predictions to the economic data that statistical agencies have started to disseminate. To compare to UK data, we start the lockdown in our model on March 23$^{rd}$, and keep it for two months. For 2020Q1, we find a 1.7\% reduction in GDP compared to 2019Q4, which is close to the 2\% early estimate recently released by the Office for National Statistics. For 2020Q2, we forecast that GDP would be 21.5\% lower than in 2019Q4, which is in the range of forecasts provided by economic institutions and consulting firms. We also compare model predictions to sectoral unemployment data, finding good agreement.

As a second step, we investigate some theoretical properties of the model. Our analysis makes it clear that bottlenecks in supply chains can strongly suppress aggregate economic output.  The extent to which this is true depends on the production function.  These effects are extremely strong with the Leontief production function, are much weaker with a linear production function (which allows unrealistically strong substitutions) and have an intermediate effect with our modified Leontief function.  Network effects can strongly inhibit recovery, and can cause counter-intuitive results, such as situations in which reopening a few industries can actually depress economic output.

Our third step, which is the key aim of this paper, is to find a good compromise between the economic benefit of reopening industries and negative health consequences of doing so. It is worth keeping in mind that many health outcomes depend on the state of the economy, so that keeping the economy closed also has negative health consequences. 

The fundamental principles of epidemic spreading are relatively well understood, and it is clear that social distancing measures reduce the spreading of COVID-19 \citep{jarvis2020quantifying,maier2020effective,arenas2020mathematical}. The difficulty comes with obtaining good estimates of the key parameters that govern the fate of an epidemic, and in particular, the reproduction number $R_0$, which gives the number of secondary cases for each primary case in a largely unaffected population. If $R_0$ is above one, the disease spreads to a given percentage of the population, otherwise the epidemic dies out. In this paper, we side-track the problem of developing a full-fledged epidemic spreading model, and focus on estimating $R_0$. We decompose the reproduction number into the infections caused by contacts during work, during consumption, during public transport, and in other contexts, i.e. home and other social interactions. We use recent contact survey data from Sweden to estimate the share of infection due to each type of contact.  For each industry we estimate its relative contribution to overall work and consumption infections. For instance, the Health sector is responsible for more work-related infections than the Forestry sector.  This is because workers have more contacts, contacts are more risky, and there are more Health workers than Forestry workers. As another example, the Retail and Restaurant sectors contribute much more to consumption infections than the Mining sector, because there are virtually no direct consumption activities in the Mining industry.

We estimate the epidemiological consequences of scenarios for coming out of lockdown. Lifting the lockdown for a specific industry has several effects: workers of this industry contribute to increased work-related infection; consumers of this industry (if any) contribute to increased consumption-related infections; workers of this industry contribute to increased public transport infections; and finally, children of these workers go back to school if the workers cannot work from home, contributing to increased school-related infections. We assume in all the scenarios that workers who can work from home continue working remotely.

We present a summary of our re-opening scenarios results in the next section. We then present  in detail our economic model and its calibration in Section \ref{sec:model}. We show our model predictions for the UK economy in Section \ref{sec:econimpact} and discuss production network effects and re-opening single industries in Section \ref{sec:supply_demand_network}.
We introduce the epidemic model and present effects of re-starting the economy on infectious contagion in Section \ref{sec:epidemic_model}. 
We conclude in Section \ref{sec:discuss}.

\section{A sweet spot for partially reopening the economy with only a minimal boost to the epidemic} \label{sec:result_highlights}

\begin{figure}[H]
    \centering
\includegraphics[width = 1.0\textwidth]{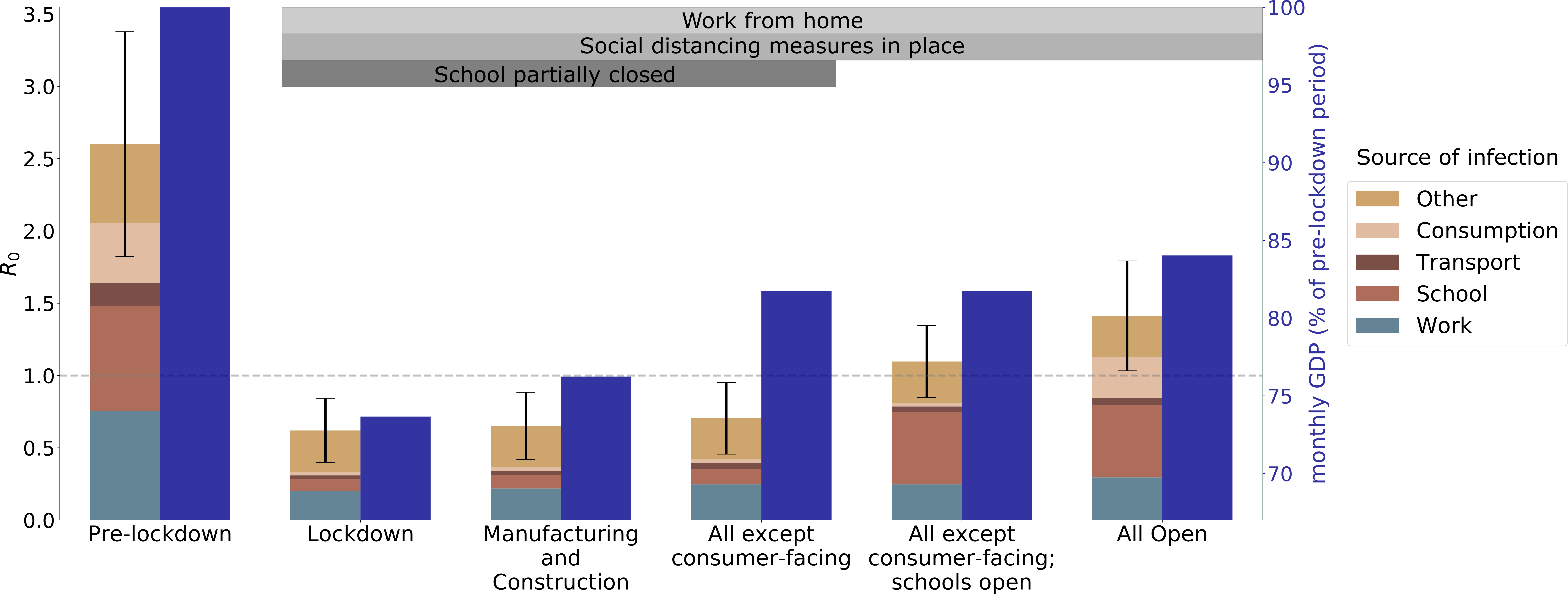}
    \caption{\textbf{How different policy scenarios affect $R_0$ and economic output.} The coloured bars show the expected reproduction number of the epidemic for each policy scenario. Different colours designate the activities that cause the epidemic to spread. The purple bars denote the percentage increase in value added relative to lockdown a month after the economy is opened under each scenario.  Black lines are two standard deviation error bars. {\it Note that we have normalized the infection rates for all scenarios so that they correspond to the \cite{jarvis2020quantifying} study during lockdown.} (Our estimate during lockdown is roughly $R_0 \approx 0.90$; their estimate is $0.62$; these agree with the error bars).  Note that Manufacturing and Construction also includes mining, agriculture, and a few others. }
        \label{fig:scenarios}
\end{figure}

Fig. \ref{fig:scenarios} summarizes our bottom line results, presenting the trade-off between increasing economic production and mitigating the spread of the pandemic under five different scenarios. The bars on the left show estimates of $R_0$ and the bars on the right show GDP as a percentage of the pre-lockdown GDP.  For comparison, pre-lockdown is shown on the left.  The scenarios are: keeping lockdown; opening Manufacturing and Construction (which is short for fully re-opening Agriculture, Mining, Manufacturing, Utilities and Construction); opening all industries except consumer-facing industries; same, but additionally opening schools; and opening all industries.   

We find that a two-month lockdown has a strong impact on the economy, with gross output, value added, and consumption decreased by $26\%$, $25\%$, and $18\%$, respectively, compared to the UK pre-lockdown levels. Compared to the economy's performance before the lockdown, an additional month of lockdown would decrease GDP from 75\% to 74\%, while re-opening only the Manufacturing and Construction sectors would increase GDP from $75\%$ to $76\%$ in a month. Re-opening all sectors except those that are consumer-facing would increase GDP to $82\%$ of its pre-lockdown value, i.e. in a month it would increase GDP by 8 percentage points with respect to the lockdown scenario. Opening all industries only adds an additional 2\% boost to GDP.  Note that the scenario with all industries open has only 84\% of pre-lockdown GDP. This is due to a combination of a persistent depression in demand for industries like restaurants (even if they are open) and the fact that consumer expectations take time to recover.

A comparison to our predictions for the increase in $R_0$ under each scenario shows that for the scenario where all industries except consumer facing industries are opened, the increase in $R_0$ relative to lockdown is small. In contrast, as soon as schools are open $R_0$ rises dramatically, and is very likely greater than one. If the economy is fully reopened, the predicted rise in $R_0$ is very likely substantially greater than one. Note that when the economy is fully reopened we find an $R_0$ still disturbingly greater than one, although much lower than the pre-lockdown value, as we assume that work from home continues and non-work related social distancing measures continue. Another reason for this is that we renormalize all our epidemiological results by the factor of $0.62/0.90$, corresponding to the ratio of our original estimate to that of \cite{jarvis2020quantifying} for the lockdown situation, thus ensuring that our estimate for the lockdown scenario corresponds to theirs. We did this because we feel that the relative values of our estimated $R_0$ across different scenarios are more reliable than the absolute values, and we defer to professional epidemiologists for estimating the absolute values. It is important to bear in mind that all these values are uncertain, and the uncertainties potentially make the difference in determining whether there will be a second wave of the epidemic in the UK. We should also stress the uncertainties in the economic results -- as we will show here, they depend rather sensitively on assumptions about the production function.

Thus, our results suggest that there is a ``sweet spot", corresponding to the scenario in which all except consumer facing industries reopen, with schools remaining closed for the children of parents who do not work or can work from home.\footnote{This understandably creates childcare problems and will require creative solutions, and our model is not designed to account for the negative effects on health and future human capital that closed school can create.} This scenario provides a good combination of a minimal predicted increase in $R_0$ and a substantial economic boost over remaining in lockdown.

The official UK government guidelines for COVID-19 recovery ``Step one'' (until June 1st)\footnote{
\url{https://www.gov.uk/government/publications/our-plan-to-rebuild-the-uk-governments-covid-19-recovery-strategy} (accessed: 2020/05/21).
} recommend that, in addition to sectors that were previously considered essential, manufacturing and construction should reopen, but that consumer-facing industries such as hospitality and non-essential retail should remain closed. Overall, this scenario corresponds to something in between our second and third scenario, depending on whether sectors such as business services fully reopen. By contrast, other countries (e.g. France) reopened personal services and non-essential retail soon after lockdown was lifted, which would correspond to something between our fourth and fifth scenario (depending on whether schools are open).

\section{Economic Model} 
\label{sec:model}
To analyse the economic benefits of staged re-opening we introduce a sectoral macroeconomic model that was inspired by the work of
\cite{battiston2007credit, hallegatte2008adaptive, henriet2012firm} and  \cite{inoue2019firm}.  We combine elements of these models and extend them to include new features.  Our model incorporates production network effects that can amplify economic shocks both upstream and downstream.

In our model producers experience supply shocks caused by a nationwide lockdown. In the lockdown workers in non-essential industries who are unable to work from home become unproductive, resulting in lowered productive capacities of industries.
At the same time demand-side shocks hit as consumers adjust their consumption preferences to avoid getting infected.
We use the first-order supply and demand shocks predicted by \cite{del2020supply} to initialise our macro model.

Our model is open-source and can be downloaded together with all relevant data\footnote{
\url{https://doi.org/10.5281/zenodo.3834116}.
}. We also provide an interactive online interface for our model, allowing the user to explore alternative scenarios and parameter ranges\footnote{
\url{https://anton-pichler.shinyapps.io/UK_COVID-19_economic_simulator/}
}.

\subsection{Timeline}

A time step $t$ in our economy corresponds to one day. 
There are $N$ industries\footnote{
See Appendix \ref{apx:notation}, Tables \ref{tab:notation_econ}-\ref{tab:notation_SIR} for a comprehensive summary of notations used.
}, one representative firm for each industry, and one representative household that owns the industries. Every day:
\begin{enumerate}
    \item Firms hire or fire workers depending on whether their workforce was insufficient or redundant to carry out production in the previous day.
    \item The representative household decides its consumption demand and industries place orders for intermediate goods.
    \item Industries produce as much as they can to satisfy demand, given that they could be limited by lack of critical inputs or lack of workers.
    \item If industries do not produce enough, they distribute their production to final consumers and other industries on a pro rata basis, that is, proportionally to demand. 
    \item Industries update their inventory levels, and profits and labor compensation are distributed to households.
\end{enumerate}
 
\begin{figure}
    \centering
\includegraphics[width = \textwidth]{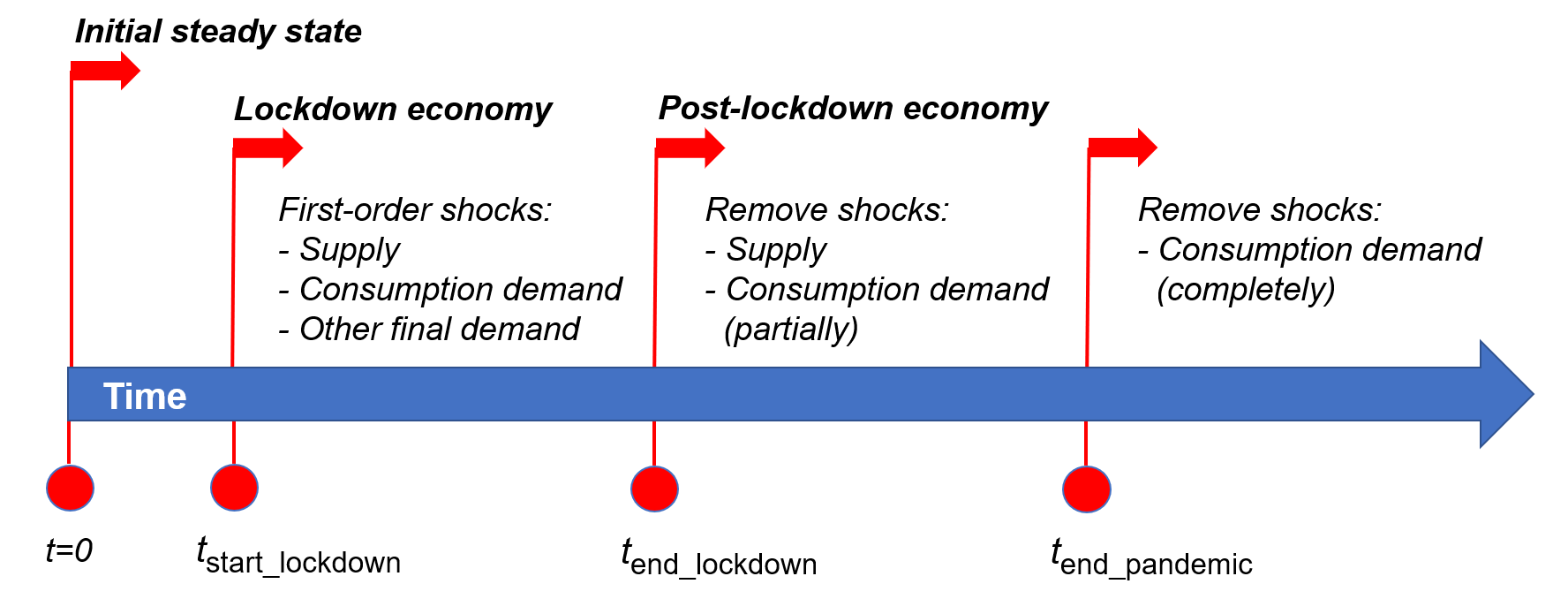}
    \caption{\textbf{Schematic model chronology.}   }
        \label{fig:model_timeline}
\end{figure}

The model is initialized at time $t=0$ in the steady state corresponding to input data. We apply the pandemic shocks on day $t_\text{start\_lockdown}$ and keep the economy in lockdown until day  $t_\text{end\_lockdown}$. At that point we remove the supply-side restrictions corresponding to the scenario.  Some consumption demand-side shocks remain in place until the pandemic is suppressed on day $t_\text{end\_pandemic}$. In our scenarios we do not lift shocks of other final demand shocks (investment, international trade). Fig. \ref{fig:model_timeline} schematically displays the overall chronology of the model.

\subsection{Accounting structure}
Let $x_{i,t}$ denote total output of industry $i$ at time $t$ and $Z_{ji,t}$ the intermediate consumption by industry $i$ of good $j$.  Industry $i$ is demand and $j$ is supply. We adopt the standard convention that in the input-output matrix columns represent demand and rows represent supply. In an economy with no ``excess'' output, i.e. in which all produced output is used up, the output of $i$ is equal to
\begin{equation} \label{eq:totout}
    x_{i,t} = \sum_{j=1}^N Z_{ij,t} + c_{i,t} + f_{i,t},
\end{equation}
where $c_{i,t}$ is household consumption of good $i$ at time step $t$ and $f_{i,t}$ is all other (exogenous) final demand, including government consumption and exports.

We let $l_{i,t}$ denote labor compensation to workers in industry $i$.  This also indicates the number of workers employed in industry $i$, under the assumption that all workers employed in the same industry earn the same wage. Profits of industry $i$ can then be written as 
\begin{equation} \label{eq:profit}
    \pi_{i,t} = x_{i,t} - \sum_{j=1}^N Z_{ji,t} - l_{i,t} - e_{i,t},
\end{equation}
where $e_{i,t}$ represents all other expenses (taxes, imports, etc.).
Note that we do not model physical capital explicitly, and we take prices as time-invariant.

For total output, total labor income, total profits and total household consumption we write
\begin{align}
    \Tilde{x}_t &\equiv \sum_{i=1}^N x_{i,t}, \label{eq:tot_x} \\
    \Tilde{l}_t &\equiv \sum_{i=1}^N l_{i,t}, \label{eq:tot_l} \\
    \Tilde{\pi}_t &\equiv \sum_{i=1}^N \pi_{i,t}, \label{eq:tot_pi}\\
    \Tilde{c}_t &\equiv \sum_{i=1}^N c_{i,t}, \label{eq:tot_c}
\end{align}
respectively.
We focus on these four variables when discussing aggregate economic impacts of the pandemic in subsequent sections.

Our analysis is based on the UK economy.
We use the latest release of the World Input-Output Database (WIOD) \citep{timmer2015illustrated} to determine the relevant values for gross output $x_{i,0}$, intermediate consumption $Z_{ij,0}$, 
household consumption $c_{i,0}$, 
other final demand $f_{i,0}$, 
labor compensation $l_{i,0}$, 
and all other expenses $e_{i,0}$ (2014 values). 
Overall, we can distinguish 55 separate industries.

\subsection{Demand}
\label{sec:demand}
It will become important to distinguish between demand, that is orders placed by customers to suppliers, and actual realized transactions. All the steps outlined above are realized transactions, which might or might not be equal to demand.

\paragraph{Industry demand.} 
The total demand faced by industry $i$ at time $t$, $d_{i,t}$, is the sum of the demand from all its customers,
\begin{equation}
    d_{i,t} = \sum_{j=1}^N O_{ij,t} + c^d_{i,t} + f^d_{i,t},
\end{equation}
where $O_{ij,t}$ (for \emph{orders}) denotes the demand from industry $j$, 
$c_{i,t}^d$ the demand from households and $f_{i,t}^d$ all other final demand.

\paragraph{Recipes.} Industries produce output according to a production recipe encoded in the technical coefficient matrix $A$, where the element $A_{ij} = Z_{ij,0}/x_{j,0}$ is the expense in input $i$ per unit of output $j$. We will relax the assumption of fixed production recipes, since not every input is critical for production in the short-run (see Appendix \ref{apx:ihs}).
Industries always demand and aim to consume inputs according to their recipe, even if lacking non-critical inputs does not cause immediate effects on its output in the short time horizon considered here.

\paragraph{Inventories.}
Due to the dynamic nature of the model, production and demand are not immediate.
Instead industries use an inventory of inputs in production. We let $S_{ij,t}$ denote the stock of material $i$ held in $j$'s inventory. Each industry $j$ aims to keep a target inventory $n_j Z_{ij,0}$ of every required input $i$ to ensure production for $n_j$ further days\footnote{
Considering an input-specific target inventory would require generalizing $n_j$ to a matrix with elements $n_{ij}$, which is easy in our computational framework but difficult to calibrate empirically.
}. We explain how we calibrate the parameters $n_j$ in Appendix \ref{apx:inventory}.

\paragraph{Intermediate demand.} 
Intermediate demand follows the dynamics originally introduced by \cite{henriet2012firm} and adopted by \cite{inoue2019firm} in the context of firm-level production network models.
To satisfy incoming demand (from $t-1$) and to reduce the gap to its target inventory, an industry $j$ makes orders to its suppliers at every time step $t$. More specifically, industry $j$ demands from industry $i$
\begin{equation} \label{eq:order_interm}
    O_{ij,t} = A_{ij} d_{j,t-1} + \frac{1}{\tau} [ n_j Z_{ij,0} - S_{ij,t} ],
\end{equation}
where $\tau$ indicates how quickly an industry adjusts its demand due to an inventory gap. Small $\tau$ corresponds to responsive industries that aim to close inventory gaps quickly. In contrast, if $\tau$ is large, intermediate demand adjusts slowly in response to inventory gaps. 
In the literature we find different choices for $\tau$, ranging from 1 \citep{henriet2012firm} to 30 \citep{hallegatte2012modeling} time steps.
In our simulations, we choose an intermediate value $\tau=10$. We present sensitivity tests with respect to $\tau$ in Appendix \ref{apx:model_sensitivity}.

\paragraph{Consumption demand.}  We let consumption demand for good $i$ be
\begin{equation}\label{eq:cd}
    c^d_{i,t}= \theta_{i,t} \Tilde{c}^d_t, 
\end{equation}
where $\theta_{i,t}$ is a preference coefficient, giving the share of goods from industry $i$ out of total consumption demand $\Tilde{c}^d_t$. The coefficients $\theta_{i,t}$ evolve exogenously, following assumptions on how consumer preferences change during the various phases of the pandemic; see Section \ref{sec:pandemic_shock}, Eq. \eqref{eq:theta}. 

Total consumption demand evolves following an adapted and simplified version of the consumption function in \cite{muellbauer2020}. 
In particular, $\Tilde{c}^d_t$ evolves according to
\begin{equation}\label{eq:consdemand}
    \log \Tilde{c}^d_t=\rho \log \Tilde{c}^d_{t-1} + \frac{1-\rho}{2} \log\left( m  \Tilde{l}_t \right) + \frac{1-\rho}{2} \log \left( m \Tilde{l}_t^p \right) + \Tilde{\epsilon}_t, 
\end{equation}
where $\Tilde{l}_t$ is current labor income, $\Tilde{l}_t^p$ is an estimation of permanent income and 
$m$ is the share of labor income that is used to consume final domestic goods, i.e. that is neither saved nor used for consumption of imported goods. From our data we find $m=0.82$. 
Consumption demand during the pandemic is affected by a change of permanent income expectations and the exogenous shock term $\Tilde{\epsilon}_t$; see Section \ref{sec:pandemic_shock}, Eqs. \eqref{eq:perm_income} and \eqref{eq:demand_shocks}.
The parameter $\rho$ indicates sluggish adjustment to new consumption levels. 
Assuming that a time step corresponds to a quarter, \cite{muellbauer2020} takes $\rho=0.6$, implying that more than 70\% of adjustment to new consumption levels occurs within two and a half quarters. 
We modify $\rho$ to account for our daily timescale: By letting $\bar{\rho}=0.6$, we take $\rho=1-(1-\bar{\rho})/90$ to obtain the same time adjustment as in \cite{muellbauer2020}\footnote{
In an autoregressive process like the one in Eq. \eqref{eq:consdemand}, about 70\% of adjustment to new levels occurs in a time $\iota$ related inversely to the persistency parameter $\rho$. Letting $Q$ denote the quarterly timescale considered by \cite{muellbauer2020}, time to adjustment $\iota^Q$ is given by $\iota^Q=1/(1-\bar{\rho})$. Since we want to keep approximately the same time to adjustment considering a daily time scale, we fix $\iota^D=90\iota^Q$. We then obtain the parameter $\rho$ in the daily timescale such that it yields $\iota^D$ as time to adjustment, namely $1/(1-\rho)=\iota^D=90\iota^Q=90/(1-\bar{\rho})$. Rearranging gives the formula that relates $\rho$ and $\bar{\rho}$.
}.
Note that, in the steady state, by definition permanent income corresponds to current income, i.e. $\Tilde{l}_t^p=\Tilde{l}_t$, and thus total consumption demand corresponds to $m \Tilde{l}_t $.\footnote{
To see this, note that in the steady state $\Tilde{c}^d_t=\Tilde{c}^d_{t-1}$. Moving the consumption terms on the left hand side and dividing by $1-\rho$ throughout yields $\log \Tilde{c}^d_t=\log\left( m  \Tilde{l}_t \right) + \Tilde{\epsilon}_t$. With no exogenous shock, we find $\Tilde{c}^d_t= m  \Tilde{l}_t$.
}

To test robustness, we present model results for alternative consumption functions in Appendix \ref{apx:model_sensitivity}. We find that our simulations are highly robust against alternative consumption models.

\paragraph{Other components of final demand.} In addition, an industry $i$ also faces demand $f_{i,t}^d$ from sources that we do not model as endogenous variables in our framework, such as government or industries in foreign countries. $f_{i,t}^d$ is not affected by the dynamics of the model. We discuss the composition and calibration of $f_{i,t}^d$ in detail in Section \ref{sec:pandemic_shock}.

\subsection{Supply}
\label{sec:supply}
Every industry aims to satisfy incoming demand by producing the required amount of output. Production is subject to the following two economic constraints:

\paragraph{Productive capacity.}
First, an industry has finite production capacity $x_{i,t}^\text{cap}$, depending on the amount of available labor input.
Initially every industry employs $l_{i,0}$ of labor and produces at full capacity 
$x_{i,0}^{\text{cap}} =  x_{i,0}$.
We assume that productive capacity depends linearly on labor inputs,
\begin{equation} \label{eq:xcap}
x_{i,t}^{\text{cap}} =  \frac{l_{i,t}}{l_{i,0}}x_{i,0}^{\text{cap}}.
\end{equation}

\paragraph{Input bottlenecks.}
Second, the production of an industry might be constrained due to an insufficient supply of critical inputs.  This can be caused by production network disruptions. While the empirical intermediate consumption at the initial time step is embodied in the technical coefficient matrix $A$, not every input is necessarily critical for production.
Modeling the severeness of intermediate input constraints realistically requires an understanding of how critical inputs are in the production of a given industry \citep{barrot2016input}.

We use the ratings of IHS Markit analysts to differentiate three types of inputs: \textit{critical}, \textit{important} and \textit{non-critical} inputs (Appendix \ref{apx:ihs}).
If an industry runs out of critical inputs, economic production halts immediately. 
Conversely, if an industry runs out of non-critical inputs, we assume that economic production is not affected. 
We also have ratings on \textit{important but not critical} inputs. As a baseline we treat important inputs as non-critical. In Section \ref{sec:prodfun_simul} we investigate in detail how alternative assumptions on the input-production relationship affect economic impacts.

For a given set of inputs if there are no limits on production capacities, industry $i$ can produce 
\begin{equation} \label{eq:xinp}
x_{i,t}^{\text{inp}} = 
\min_{j \in \mathcal{V}_i} 
\left \{ \;  \frac{ S_{ji,t} }{ A_{ji} } \right \},
\end{equation}
where $\mathcal{V}_i$ is the set of \textit{critical}
suppliers to industry $i$. 
If every input is critical, every input is binding, and this reduces to a Leontief production function.

\paragraph{Output level choice and input usage.}
Since an industry aims to satisfy incoming demand within its production constraints, realized production at time step $t$ is
\begin{equation} \label{eq:x_act}
    x_{i,t} = \min \{ x_{i,t}^{\text{cap}}, x_{i,t}^{\text{inp}}, d_{i,t}  \}.
\end{equation}
Thus production level of an industry is constrained by the smallest of three values: labor-constrained production capacity $x_{i,t}^{\text{cap}}$, intermediate input-constrained production capacity $x_{i,t}^{\text{inp}}$, or total demand $d_{i,t}$. 

The level of output then determines the actual use of inputs according to the production recipe. Industry $i$ uses an amount $A_{ji}x_{i,t}$ of input $j$,
unless $j$ is not critical and the amount of $j$ in $i$'s inventory is less than $A_{ji}x_{i,t}$. 
In this case, the quantity consumed of input $j$ by industry $i$ is equal to the remaining inventory stock of $j$-inputs $S_{ji,t} < A_{ji}x_{i,t}$.

\paragraph{Rationing.}
Without any adverse shocks, industries are always able to meet total demand, i.e. $x_{t} = d_{t}$. But in case of production capacity or/and input bottlenecks, industries may not be able to meet total demand, $x_{i,t} < d_{i,t}$, so they need to ration their output. We assume simple proportional rationing, although alternative rationing mechanisms could be considered (e.g. \cite{inoue2019firm}).

The final delivery from industry $i$ to industry $j$ then is the share of orders received
\begin{equation}
    Z_{ij,t} = O_{ij,t} \frac{x_{i,t}}{d_{i,t}}.
\end{equation}
Households receive a share of their demand
\begin{equation}
    c_{i,t} = c_{i,t}^d \frac{x_{i,t}}{d_{i,t}},
\end{equation}
and the realized final consumption of agents with exogenous final demand is 
\begin{equation}
    f_{i,t} = f_{i,t}^d \frac{x_{i,t}}{d_{i,t}}.
\end{equation}

\paragraph{Inventory updating.} The inventory of $j$ for every input $i$ is updated according to
\begin{equation}
    S_{ij,t+1} = \min \left\{ S_{ij,t} + Z_{ij,t} - A_{ij} x_{j,t}, 0 \right\}.
\end{equation}
In a Leontief production function, where every input is critical, the minimum operator would not be needed since production could never continue once inventories are run down. It is necessary here, since when inventories of non-critical inputs $i$ are depleted, industry $j$ produces output using less goods $i$ than $A_{ij} x_{j,t}$.

\paragraph{Hiring and firing.} 
Firms adjust their labor force depending on which production constraints in Eq. \eqref{eq:x_act} are binding. If the capacity constraint $x_{i,t}^{\text{cap}}$ is binding, industry $i$ decides to hire as many workers as necessary to make the capacity constraint no longer binding. Conversely, if either input constraints $x_{i,t}^{\text{inp}}$ or demand constraints $d_{i,t}$ are binding, industry $i$ lays off workers until capacity constraints become binding.  
More formally, at time $t$ labor demand by industry $i$ is given by $l^d_{i,t}=l_{i,t-1}+\Delta l_{i,t}$, with 
\begin{equation}
    \Delta l_{i,t} = \frac{l_{i,0}}{x_{i,0}}\left[ \min\{x_{i,t}^{\text{inp}},d_{i,t}\} - x_{i,t}^{\text{cap}}\right].
\end{equation} 
Note that the term $l_{i,0}/x_{i,0}$ reflects the assumption that the labor share in production is constant over the considered period. We assume frictions in the labor market in a sense that it takes time for firms to adjust their labor inputs.
Specifically, we assume that industries can increase their labor force only by a fraction $\gamma_{\text{H}}$ in direction of their target. Similarly, industries can decrease their labor force only by a fraction $\gamma_{\text{F}}$ in the direction of their target. In the absence of additional policies we usually have $\gamma_{\text{F}}>\gamma_{\text{H}}$, indicating that it is easier for firms to lay off employed than hire new workers.
Industry-specific employment evolves then according to
\begin{equation} \label{eq:labor_evolution}
    l_{i,t} = 
    \begin{cases} 
    l_{i,t-1} + \gamma_{\text{H}} \Delta l_{i,t} &\mbox{if } \; \Delta l_{i,t} \ge 0, \\
    l_{i,t-1} + \; \gamma_{\text{F}} \Delta l_{i,t} &\mbox{if }  \Delta l_{i,t} < 0.
    \end{cases}
\end{equation}
As we discuss further in Section \ref{sec:econpolicy}, $\gamma_{\text{H}}$ and $\gamma_{\text{F}}$ can be interpreted as policy variables. 
For example, the implementation of a furloughing scheme makes re-hiring of employees easier, corresponding to an increase in $\gamma_{\text{H}}$.
In our baseline simulations we choose $\gamma_{\text{H}}=1/30$ and $\gamma_{\text{F}} = 2 \gamma_{\text{H}}$. 
Given our daily time scale, this is a rather rapid adjustment of the labor force. We present sensitivity tests for these parameters in Appendix \ref{apx:model_sensitivity}.

\subsection{Pandemic shock} \label{sec:pandemic_shock}

\paragraph{Timeline}
The simulation starts in the steady state. For simplicity we let the pandemic shock hit at the same time as the lockdown starts, i.e. we do not take into account reduced demand beforehand.  We let the lockdown last for two months (60 time units), and then lift it according to the specifications below.

\paragraph{Supply shocks}
At every time step during the lockdown an industry $i$ experiences an (exogenous) first-order labor supply shock $\epsilon^S_{i,t} \in [0,1]$ that quantifies labor reductions. These reductions
are caused by the lack of labor that was previously provided by workers in non-essential industries \citep{del2020supply,fana2020covid,galasso2020working} who cannot work remotely \citep{del2020supply,dingel2020,gottlieb2020working,koren2020business}. 
For instance, if an industry is non-essential, and none of its employees can work from home, it faces a labor supply reduction of 100\% during lockdown i.e., $\epsilon^S_{i,t}=1, \ \forall t \in [t_\text{start\_lockdown}, t_\text{end\_lockdown})$. Instead, if an industry is classified as fully essential, it faces no labor supply shock and $\epsilon^S_{i,t}=0 \ \forall t$.

Letting $l_{i,0}$ be the initial labor supply before the lockdown, the maximum amount of labor available to industry $i$ at time $t$ is given as
\begin{equation}
    l_{i,t}^\text{max} = (1- \epsilon^S_{i,t}) l_{i,0}.
\end{equation}
If $\epsilon^S_{i,t} > 0$, the productive capacity of industry $i$ will be smaller than in the initial state of the economy. 
We assume that the reduction of total output is proportional to the loss of labor. In that case the productive capacity of industry $i$ at time $t$ is
\begin{equation}
x_{i,t}^{\text{cap}} = \frac{l_{i,t}}{l_{i,0}} x_{i,0}^{\text{cap}} \le (1-\epsilon^S_{i,t}) x_{i,0}.
\end{equation}
Recall from Section \ref{sec:supply} that firms can hire and fire to adjust their productive capacity to demand and supply constraints. Thus, productive capacity can be lower than the initial supply shock. 
However, during lockdown they can never hire more than $(1-\epsilon^S_{i,t}) l_{i,0}$ workers. If the lockdown is unwound for an industry $i$, first-order supply shocks are removed, i.e. we set $\epsilon^S_{i,t} = 0$, for $t\geq t_\text{end\_lockdown}$.

\paragraph{Supply shock calibration}
To initialise the economic model with first-order supply shocks from the pandemic we use the shock predictions of the recent study by \cite{del2020supply}. 
In \cite{del2020supply} supply shocks of the pandemic are derived by quantifying which work activities of different occupations can be performed from home (Remote-Labor-Index) and by using the occupational compositions of industries. Moreover, the predictions also take into account whether an industry is essential in the sense that it needs to continue 
operating during a lockdown. 
The predictions of first-order shocks are based on the US economy using a different industrial classification system. These predictions therefore need to be adopted for the UK economy and the WIOD industry classification as we outline in detail in Appendix \ref{apx:shocks}.

For the UK we estimate that $67\%$ of the work force has an essential job. However, much of this essential work can be done remotely (e.g. government and financial services).
In total we estimate that $44\%$ of workers can work remotely and that $37\%$ of workers are currently going to work, assuming that people work from home whenever possible.

\paragraph{Consumption demand shocks}

A first shock to consumption demand occurs through reductions in current income and expectations for permanent income. Expectations for permanent income depend on whether households expect a V-shaped vs. L-shaped recovery, that is, whether they expect that the economy will quickly bounce back to normal or there will be a prolonged recession. Let expectations for permanent income $\Tilde{l}_t^p$ be specified by
\begin{equation} \label{eq:perm_income}
\Tilde{l}_t^p=\xi_t  \Tilde{l}_0    
\end{equation}
In this equation, the parameter $\xi_t$ captures the fraction of pre-pandemic labor income $\Tilde{l}_0$ that households expect to retain in the long run. We first give a formula for $\xi_t$ and then explain the various cases.
\begin{equation}{\label{eq:xit}}
\xi_t =
\begin{cases}
    1, & t<t_\text{start\_lockdown}, \\
    \xi^L = 1-\frac{1}{2}\frac{\tilde{l}_0-\tilde{l}_{t_\text{start\_lockdown}}}{\tilde{l}_0}, & t_\text{start\_lockdown}\leq t \leq t_\text{end\_lockdown},\\
    1-\rho + \rho\xi_{t-1} + \nu_{t-1}, & t > t_\text{end\_lockdown}.
\end{cases}
\end{equation}
Before lockdown, we let $\xi_t \equiv 1$. 
During lockdown, following \cite{muellbauer2020} we assume that $\xi_t$ is equal to one minus half the relative reduction in labor income that households experience due to the direct labor supply shock, and denote that value by $\xi^L$. (For example, given a relative reduction in labor income of 16\%, $\xi^L=0.92$.)\footnote{During lockdown, labor income may be further reduced due to firing. For simplicity, we choose not to model the effect of these further firings on permanent income.} After lockdown, we assume that 50\% of households believe in a V-shaped recovery, while 50\% believe in an L-shaped recovery. We model these expectations by letting $\xi_t$ evolve according to an autoregressive process of order one, where the shock term $\nu_t$ is a permanent shock that reflects beliefs in an L-shaped recovery. With 50\% of households believing in such a recovery pattern, it is $\nu_t\equiv-(1-\rho)(1-\xi^L)/2$.\footnote{The specification in Eq. \eqref{eq:xit} reflects the following assumptions: (i) time to adjustment is the same as for consumption demand, Eq. \eqref{eq:consdemand}; (ii) absent permanent shocks, $\nu_t=0$ after some $t$, $\xi_t$ returns to one, i.e. permanent income matches current income; (iii) with 50\% households believing in an L-shaped recovery, $\xi_t$ reaches a steady state given by $1-(1-\xi^L)/2$: with $\xi^L=0.92$ as in the example above, $\xi_t$ reaches a steady state at 0.96, so that permanent income remains stuck four percentage points below pre-lockdown current income.}

In addition to the income effect, during a pandemic consumption/saving decisions and consumer preferences over the consumption basket are changing, leading to first-order demand shocks \citep{CBO2006, del2020supply}.
For example, consumers are likely to demand less services from the hospitality industry, even if it is able to supply these services. Transport is very likely to face substantial demand reductions, despite being classified as an essential industry in many countries. A key question is whether reductions in demand for ``risky'' goods and services is compensated by an increase in demand for other goods and services, or if lower demand for risky goods translates into higher savings.

We consider a demand shock vector ${\epsilon}_t$, whose components $\epsilon_{i,t}$ are the relative changes in demand for goods of industry $i$. These components evolve in the various phases of the pandemic, as defined in the following equations:
 \begin{equation}\label{eq:demand_shocks}
    \epsilon_{i,t} = 
     \begin{cases}
     0, &  \text{if } t<t_\text{start\_lockdown}, \\
     \epsilon^D_{i}, &  \text{if } t_\text{start\_lockdown}\leq t<t_\text{end\_lockdown}, \\
     0, & \text{if no on-site consumption of $i$ \&}\\
     & t\geq t_\text{end\_lockdown}, \\
     \frac{\epsilon_i^D}{\log 100} \log \left( 100 - \frac{99t}{t_{\text{end\_pandemic}}} \right), & \text{if on-site consumption of $i$ \& }\\
     & t_\text{end\_lockdown} \leq t < t_\text{end\_pandemic},\\
     0, &  \text{if } t\geq t_\text{end\_pandemic}.
     \end{cases}
 \end{equation}
We use the estimates by the \cite{CBO2006, del2020supply}, which we denote by $\epsilon^D_i$, to parameterize $\epsilon_{i,t}$ during lockdown. Roughly speaking, these shocks are massive for restaurants and transport, mild for manufacturing, null for utilities, and positive for health (see Appendix \ref{apx:shocks}).

\begin{figure}
    \centering
\includegraphics[width = 0.8\textwidth]{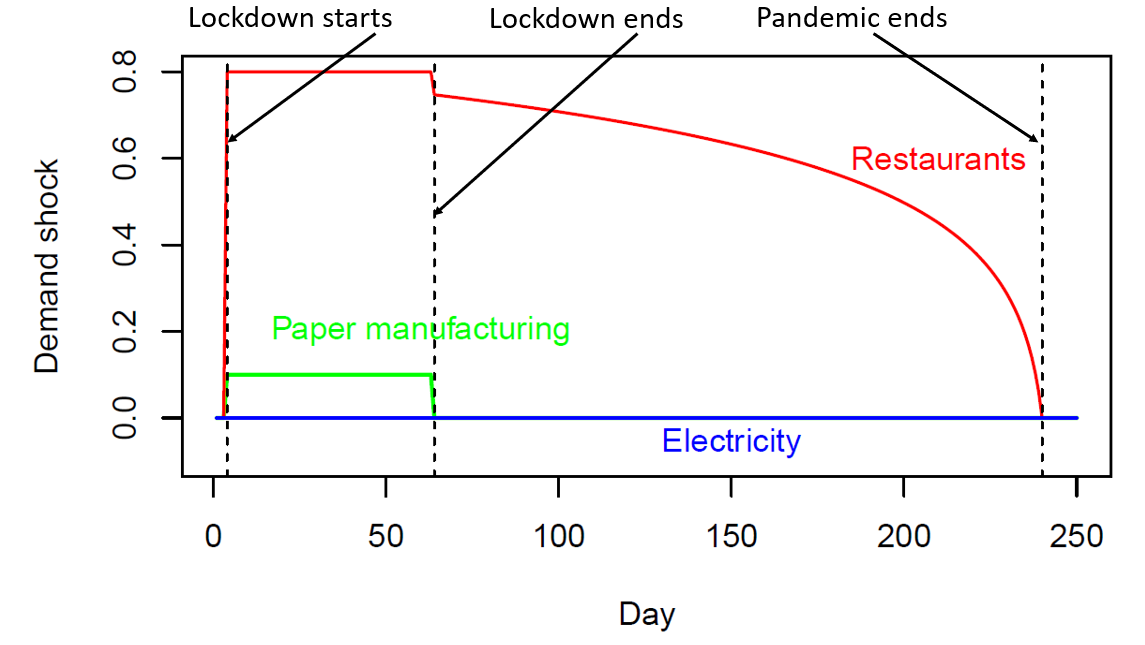}
    \caption{\textbf{Illustration of demand shocks.} Electricity (blue line) does not experience any demand shock; paper manufacturing (green line) only experiences a moderate demand shock during lockdown; restaurants (red line) experience a strong demand shock during lockdown, and then the demand shock remains persistent in the initial phase of reopening, disappearing only as the end of the pandemic approaches.  }
        \label{fig:demand_shocks}
\end{figure}

When the lockdown is lifted, demand shocks for industries that do not involve on-site consumption are removed; in contrast, demand for industries that involve on-site consumption (restaurants, theatres, etc.)\footnote{
For deciding whether an industry faces on-site consumption we use the same list that we compiled for the epidemic model, supplementing it with industries that are not very infectious collectively, but that individually could be perceived as risky. For example, infections while buying a car are a negligible share of all infections, but visiting a car seller might be perceived as risky. Specifically, we classify as industries involving on-site consumption those with the following codes: G45, G47, H49, H50, H51, H52, H53, I, L68, M69\_M70, O84, P85, R\_S, T.
}
goes back to normal more slowly, and in a non-linear way. The specification in Eq. \eqref{eq:demand_shocks} captures the idea that demand for on-site consumption industries is likely to resume very slowly after lockdown and to accelerate towards its pre-pandemic level as the pandemic approaches an end (or at least is perceived to come to a conclusion).\footnote{
Note that the specification in Eq. \eqref{eq:demand_shocks} also allows for a small bump in consumption demand at the time the lockdown is lifted.
} Recent evidence from transaction data in China \citep{chen2020impact} backs the assumption that demand for these industries resumes more slowly than for industries that do not face on-site consumption. An illustration for three industries that either do not experience any demand shock, experience a demand shock only during lockdown or experience a demand shock throughout the pandemic is given in Fig. \ref{fig:demand_shocks}.

We now explain how the demand shock vector affects consumption demand. Recall from Eq. \eqref{eq:cd}, $c^d_{i,t}= \theta_{i,t} \Tilde{c}^d_t$, that consumption demand is the product of the total consumption scalar $\Tilde{c}^d_t$ and the preference vector $\theta_t$, whose components $\theta_{i,t}$ represent the share of total demand for good $i$. 
We initialize the preference vector by considering the initial consumption shares, that is $\theta_{i,0}=c_{i,0}/\sum_j c_{j,0}$. 
By definition, the initial preference vector $\theta_{0}$ sums to one, and we keep this normalization at all following time steps. To do so, we consider an auxiliary preference vector $\bar{\theta}_t$, whose components $\bar{\theta}_{i,t}$ are obtained by applying the shock vector $\epsilon_{i,t}$. That is, we define $\bar{\theta}_{i,t}=\theta_{i,0} (1-\epsilon_{i,t})$ and define $\theta_{i,t}$ as
\begin{equation} \label{eq:theta}
    \theta_{i,t}= 
    \frac{\bar{\theta}_{i,t}}{\sum_j \bar{\theta}_{j,t}} = 
    \frac{ (1-\epsilon_{i,t}) \theta_{i,0} }{\sum_j  (1-\epsilon_{j,t}) \theta_{j,0} } .
\end{equation} 

The difference $1-\sum_i \bar{\theta}_{i,t}$ is the aggregate reduction in consumption demand due to the demand shock, which would lead to an equivalent increase in the saving rate. However, households may not want to save all the money that they are not spending. For example, they most likely want to spend on food the money that they are saving on restaurants. Therefore, we define the aggregate demand shock $\Tilde{\epsilon}_t$ in Eq. \eqref{eq:consdemand} as 
\begin{equation} \label{eq:epsilon}
\Tilde{\epsilon}_t=\Delta s \left(1- \sum_{i=1}^N \bar{\theta}_{i,t} \right) (1-\rho),
\end{equation}
where $\Delta s$ is the change in the savings rate. When $\Delta s=1$, households save all the money that they are not planning to spend on industries affected by demand shocks; when $\Delta s=0$, they spend all that money on goods and services from industries that are affected less. For our simulations, we take an intermediate value $\Delta s=0.5$. Finally, the term $(1-\rho)$ is needed to account for the autoregressive process in Eq. \eqref{eq:consdemand}.\footnote{If $\Tilde{\epsilon}_t$ was constant, in the steady state $\log \Tilde{c}^d_t$ would be reduced by $\Delta s \left(1- \sum_{i=1}^N \bar{\theta}_{i,t} \right)$.  }

\paragraph{Demand shock calibration.}
Note that WIOD distinguishes five types of final demand: 
(I) \textit{Final consumption expenditure by households},	(II) \textit{Final consumption expenditure by non-profit organisations serving households}, 
(III) \textit{Final consumption expenditure by government}	(IV) \textit{Gross fixed capital formation} and
(V) \textit{Changes in inventories and valuables}.
Additionally, all final demand variables are available for every country.
The endogenous consumption variable $c_{i,0}$ corresponds to (I), but only for domestic consumption. All other final demand categories, including all types of exports, are absorbed into $f_{i,0}$.

We apply different initial shocks to the different demand categories presented above. For domestic final demand variables we assume the following initial shocks:
We use the estimates from \cite{CBO2006} and \cite{del2020supply} to calibrate the consumption demand shock variable $\epsilon_i^D$ which we
apply to the final consumption variables (I) and (II).  We assume that investment (IV) is reduced by 5.6\%, in line with the US Bureau of Economic Analysis (BEA) estimates for the reduction in investment in the US from 2019Q4 to 2020Q1.
We do not apply any exogenous shocks to categories (III) \textit{Final consumption expenditure by government}	
and
(V) \textit{Changes in inventories and valuables}.

To initialise the model with foreign demand shocks, we use the recent estimates on trade by the World Trade Organisation.
In their recent forecast international trade is predicted to decline between 12-33\% for European countries \citep{bekkers2020wto}. 
We follow the pessimistic scenario of the WTO and assume a drop of 33\% in foreign intermediate and final demand.

A summary of all shocks is provided in Appendix \ref{apx:shocks}, Table \ref{tab:FO_shocks}. There is considerable uncertainty in our estimates of first-order demand shocks, which we aim to reduce in the future by collecting additional data. However, sensitivity tests shown in Appendix \ref{apx:shock_uncertainty} suggest that our model predictions are fairly robust against uncertainties in the shock estimates.

\subsection{Policy intervention} \label{sec:econpolicy}

An exogenous policymaker -- the government -- can influence economic outcomes in three possible ways. 
First, the key policy which we are considering is the implementation and withdrawal of a lockdown.
While the implementation of a lockdown affects all industries simultaneously according to the exogenous first-order supply and demand shocks, the lockdown can be unwound for different sets of industries. We experiment with different re-opening scenarios which we also evaluate with respect to their impact on infectious contagion (Section \ref{sec:epidemic_model}).

Second, the government can also pay out additional social benefits to workers to compensate income losses. During the pandemic only a fraction of the initial labor force is employed, due both to direct shocks and subsequent firing/furloughing, resulting in lower labor compensation, i.e. $\tilde{l}_t < \tilde{l}_0$, for $t\geq t_\text{start\_lockdown}$.
The government can reimburse a fraction $b$ of the income loss $\tilde{l}_t - \tilde{l}_0$ as social benefits, increasing disposable income of households to
\begin{equation}
    \tilde{l}_{t}^* = \tilde{l}_{t} + b ( \tilde{l}_{0} - \tilde{l}_{t}).
\end{equation}
Following the current UK policy on furloughing, we set $b=0.8$ in our default simulations.

As a third policy dimension we consider labor force adjustment parameters $\gamma_\text{H}$ and $\gamma_\text{F}$. Recall from Eq. \eqref{eq:labor_evolution} that the larger these parameters, the quicker firms can adjust their labor inputs. Hiring and firing of employees can be costly without further support by the government.
We assume that a furloughing policy scheme increases the flexibility of adding and removing labor inputs to an industry.

We explore in a somewhat stylized way the effect of furloughing on the economy by varying parameters $b$, $\gamma_\text{H}$ and $\gamma_\text{L}$. Setting these parameters to larger values represents a regime where furloughing is encouraged by the government, whereas smaller values indicate the business-as-usual scenario without furloughing.


\section{Economic impact of COVID-19 on the UK economy}
\label{sec:econimpact}

We now show results of the economic model and compare model predictions to data. We focus on the baseline calibration discussed above. 
For convenience all model parameters are reported again in Table \ref{tab:econmodelpars}.

\begin{table}[htbp]
	\centering
	\caption{Parameters of the economic model for our baseline simulations.}
		\begin{tabular}{|l|c|c|}
			\hline
			 Name & Symbol & Value \\
			 \hline
			 Consumption function & & Eq. \eqref{eq:consdemand} \\
			 Production function & & Eqs. \eqref{eq:xcap}-\eqref{eq:x_act} \\
			 Inventory adjustment & $\tau$ & 10 \\
			 Upward labor adjustment & $\gamma_H$ & 1/30 \\
			 Downward labor adjustment & $\gamma_F$ & 1/15 \\
			 Consumption adjustment & $\rho$ & 0.987 \\
			 Government benefits & b & 0.8 \\
			 Change in savings rate & $\Delta s$ & 0.5 \\
			\hline						
		\end{tabular}
	\label{tab:econmodelpars}
\end{table}

We let the model start in the steady state at the beginning of 2020. The economy rests in steady state until March 23$^\text{rd}$, at which point we apply the pandemic shock. For this simulation we assume that lockdown lasts two months, until May 23$^\text{rd}$, at which point all supply-side restrictions are unwound. We show this specific scenario for illustration purposes, while we consider other reopening scenarios in Sections \ref{sec:reopen_singles} and \ref{sec:reopening}.
We let the model run for another month and a half, until the end of June, to analyse its recovery path. We do not run the model further in the future, both because of the great uncertainties involved and because our assumptions on non-critical inputs are only valid for a limited time span.

\begin{figure}[hbt]
    \centering
\includegraphics[width = 1.0\textwidth]{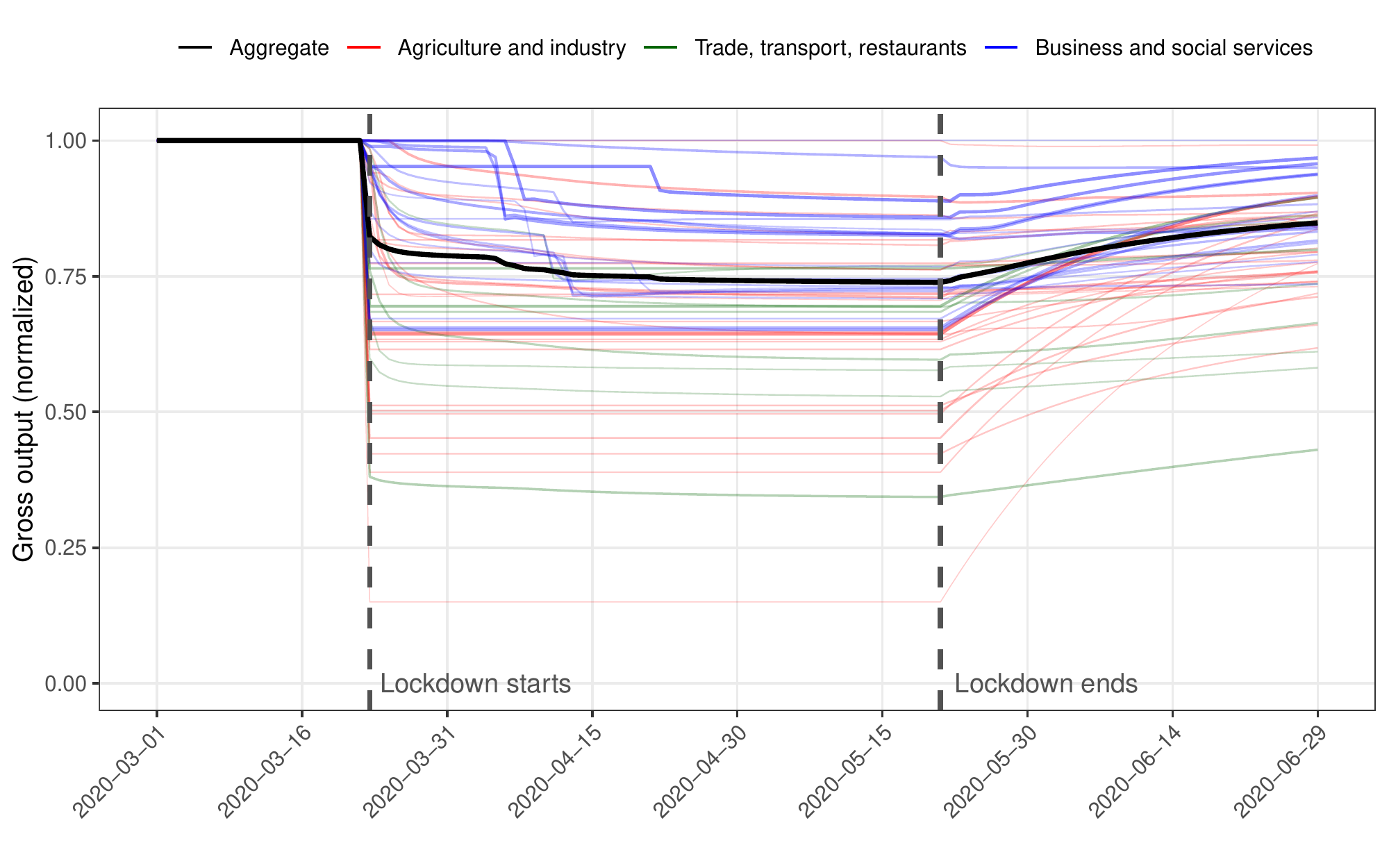}
    \caption{\textbf{Economic production as a function of time.} We plot production (gross output) as a function of time for each of the 55 industries.  Aggregate production is a thick black line and each sector is colored. Agricultural and industrial sectors are colored red; trade, transport, and restaurants are colored green; service sectors are colored blue. All sectoral productions are normalized to their pre-lockdown levels, and each line size is proportional to the steady-state gross output of the corresponding sector.}
        \label{fig:aggregatevssectoral}
\end{figure}

Fig. \ref{fig:aggregatevssectoral} shows model results for production (gross output); results for other important variables, such as profits, consumption and labor compensation (net of government benefits) are similar. When the lockdown starts, there is a sudden drop in economic activity, shown by a sharp decrease in production. A second smaller drop in production occurs at the beginning of April, due to some service sectors further reducing production. Throughout the simulation, however, service sectors tend to perform better than manufacturing, trade, transport and accommodation sectors. The main reason is that most service sectors face both lower supply and demand shocks, as a high share of workers can effectively work from home, and there is no on-site consumption for most business and professional services. In fact, consumption even increases for several industries (consumption of health is an example).

When the lockdown is lifted, the economy starts approaching its previous level, but this is not all achieved by the end of June. While some sectors quickly return close to pre-lockdown levels, recovery for other sectors is much slower. For example, Restaurants and Transport (green lines) recover very slowly, due to the assumption that consumers are cautious towards industries that involve on-site consumption (see Section \ref{sec:pandemic_shock} and Fig. \ref{fig:demand_shocks}). The aggregate level of consumption also does not return to pre-lockdown levels, due to a reduction in expectations of permanent income associated with beliefs in an L-shaped recovery (Section \ref{sec:demand}), and due to the fact that we do not remove shocks to investment and exports (see Section \ref{sec:pandemic_shock}).

Considering both the lockdown period and the post-lockdown partial recovery, our forecast for GDP in the second quarter of 2020 compared to the last quarter of 2019 is -21.5\%. This estimate is more pessimistic than the majority of forecasts for the UK economy done by economic institutions and consulting firms,\footnote{\url{https://www.gov.uk/government/collections/data-forecasts}} which, on average, are around -15\%. However, it is more optimistic than the estimate by the Bank of England, which predicts a -25\% reduction in aggregate GDP.\footnote{See the monetary policy report of May 2020 (\url{https://www.bankofengland.co.uk/-/media/boe/files/monetary-policy-report/2020/may/monetary-policy-report-may-2020}), section 2 ``Current economic conditions''.}

To test how realistic the results of our model are we compare as many model predictions to data as possible. For aggregate data, we focus on the UK, also considering Spain, France, and Italy. 

The UK recently released early estimates of national accounts in the first quarter of 2020.\footnote{\url{https://www.ons.gov.uk/economy/grossdomesticproductgdp/bulletins/gdpfirstquarterlyestimateuk/januarytomarch2020}} Because lockdown started in the UK only on March 23, the impact on the UK economy over all the first quarter was modest. Indeed, GDP reduced by about 2\%, and consumption decreased by 1.7\%. Running our model until the end of March, we find a GDP reduction of 1.7\% and a consumption reduction of 1.3\%. Labor compensation actually increased in the UK by 0.8\%, while in our model it decreased by 0.3\%. While data for labor compensation are probably not affected by only a week of lockdown, it is interesting that both in data and in our model consumption decreased more than income, a peculiar feature of this pandemic-induced recession \citep{muellbauer2020}.

Spain, France, and Italy have reported a larger effect. Remarkably, recently released data from statistical offices reveals that all these countries expect a decline in GDP of around 5\% in 2020-Q1, which is substantially larger than the 2\% reduction in the UK. Since this is mostly due to these countries starting widespread lockdowns between one and two weeks earlier than the UK, we rerun our model starting lockdown on March 15. In this case, we find that quarterly GDP decreases by 3.3\%. This is somewhat off the 5\% mark, but part of the error probably comes from the fact that we calibrate our model on UK data (e.g. IO tables), and we assume no reduction in GDP at all before March 15 (in contrast to the evidence of disruptions in supply chains, reduced international travel and early reaction by some consumers prior to this date).

So far, we considered testing data aggregated at the national level, and only pertaining at most to the first 15 days of lockdown. However, we want to test the predictions of our model at a detailed sectoral level, and also explore how our model fares deeper into the lockdown period. The best sectoral data that we could find were released by the States of Washington and Texas, which released weekly unemployment claims data at a high level of industrial disaggregation. In particular, Washington released data up to 6-digit NAICS, while Texas released data for 17 broad industries.\footnote{We had to clean the data imputing some missing information and to do a crosswalk from NAICS to WIOD sector. Details are available upon request.}  To compare with the predictions of our model, we make the hypothesis that relative reductions in employment across different sectors are similar in the UK as in Washington and Texas; for example, restaurant workers are more likely to be fired or furloughed than workers in food manufacturing in all these places. We consider unemployment claims from March 14 through April 25 for Washington, and through April 18 for Texas. While March 14 was not the official start of lockdown for either State, unemployment claims started to spike during that week, making it the ideal starting point to compare the predictions of our model to data. To run the model for the same time span as the data, we run it for 42 days after imposing lockdown to compare to Washington data, and for 35 days for Texas data. 

\begin{figure}
    \centering
\includegraphics[width = 0.49\textwidth]{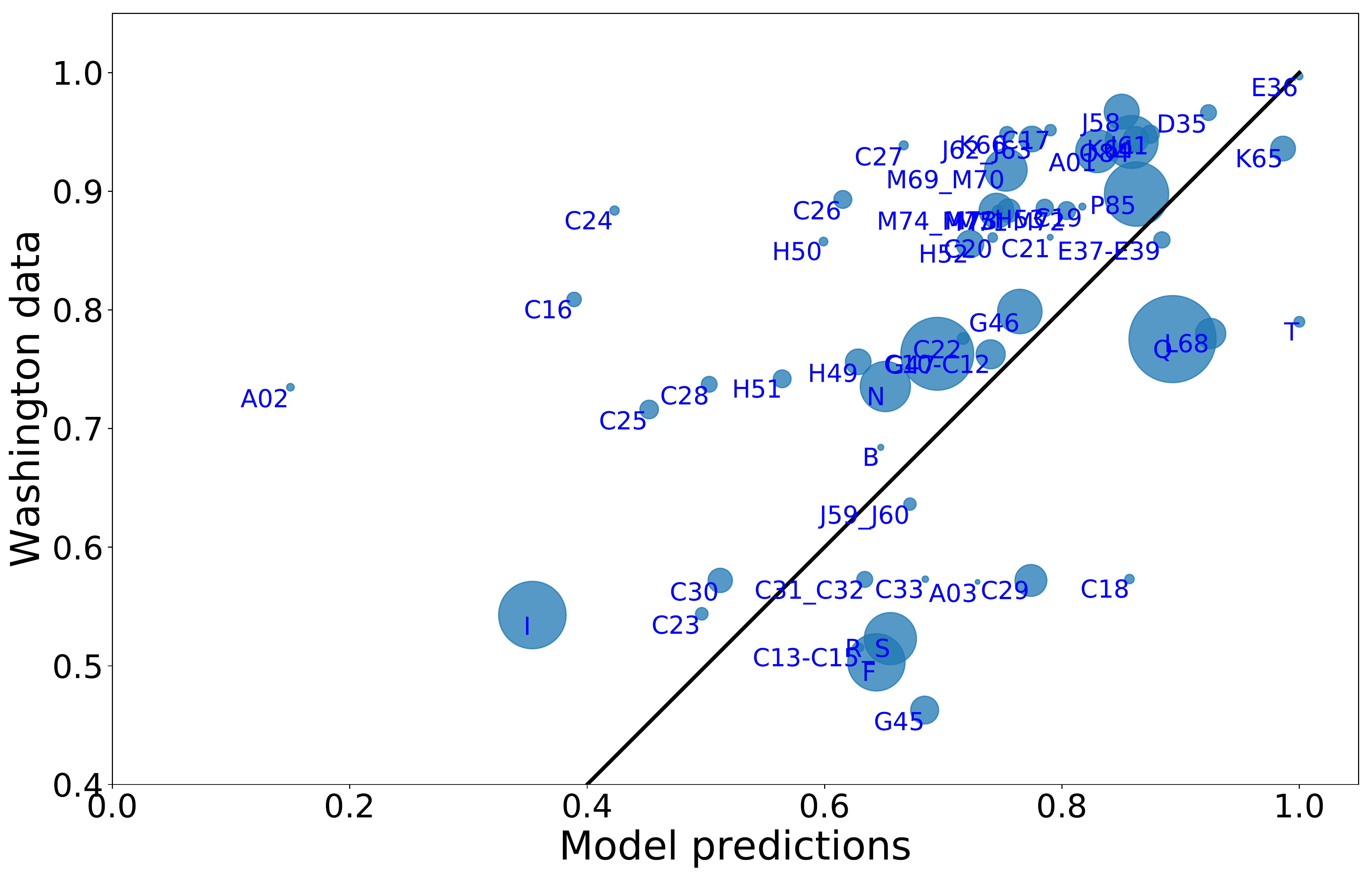}
\includegraphics[width = 0.49\textwidth]{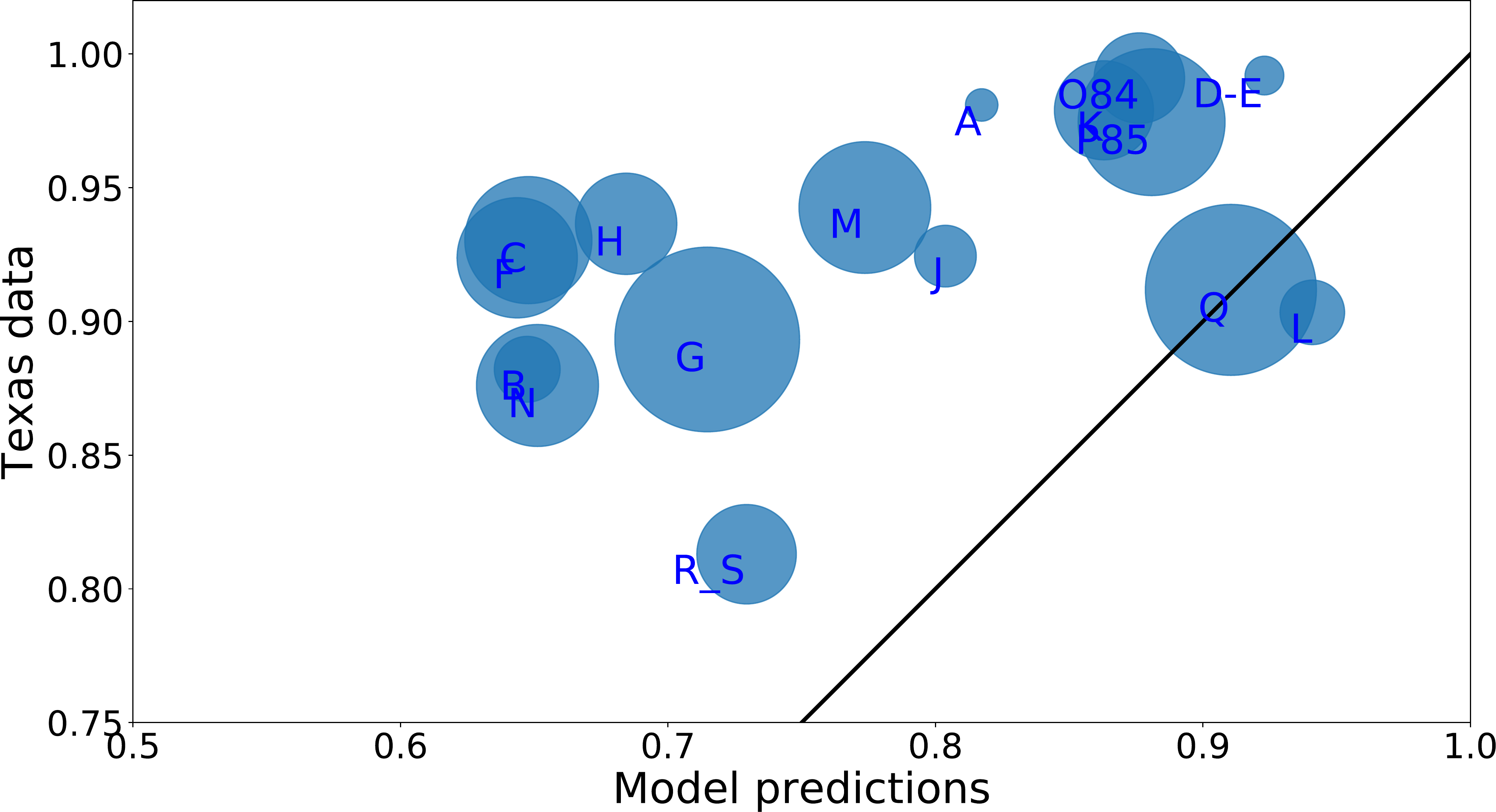}
    \caption{\textbf{Comparison of model predictions to data.} In both panels, on both axes, we show the ratio between employment 42 (Washington) or 35 (Texas) days after the shock and employment before the shock. Dots are scaled according to employment shares. The dots labels refer to industry codes, see Appendix for the corresponding names. The identity line is shown for reference.}
        \label{fig:washington_model_predictionsAessential2}
\end{figure}

Figure \ref{fig:washington_model_predictionsAessential2} shows the ratio of employment levels on April 25 to employment levels on March 14, both in the model and in the Washington data, across all sectors. The Pearson correlation between the model's predictions and the data is $0.44$, and the correlation weighted by the employment share of each sector is $0.66$. This indicates that predictions for the largest sectors are more accurate.  In most cases the model somewhat overestimates the reduction in employment. However, in a few cases employment reduction is actually underestimated, for example in health (Q, large dot on the right).

Comparison with Texas data yields similar results, except that in this case the model vastly overestimates the number of firings. However, correlations are higher in this case, as the Pearson correlation coefficient is 0.68 and the weighted correlation coefficient is 0.72.

We perform the same comparison between model predictions and empirical data for alternative specifications in Appendix \ref{apx:sensitivity_correlation}. Considering all sensitivity cases studied in Appendix \ref{apx:senstivity}, the correlation coefficients listed above are very robust. The only exception is the case in which we consider a Leontief production function, where the correlation between model predictions and data even becomes negative. The correlation is also low if we consider important inputs as critical or half-critical.

Overall, we take both the aggregate and sectoral results as an indication that the outcomes of the model are in good qualitative agreement with the reality. There remains some significant quantitative differences, but also a substantial margin for improvement as we have not fitted many parameters, demand shocks can be improved, and we have to compare a UK model to state-level US data. We also take the result as a clear indication that the Leontief production function produces predictions at odds with empirical data, supporting our modeling choice of considering non-critical inputs.


\section{Supply, demand and network effects}
\label{sec:supply_demand_network}

\subsection{Production and input criticality} \label{sec:prodfun_simul}

A key innovation of our model is that it uses a production function that distinguishes between critical and non-critical inputs, based on the IHS Markit analyst ratings. 
To better understand how different assumptions on input criticality influence outcomes, we implemented various specifications. 

The most rigid case is the Leontief production function, in which every positive entry in the technical coefficient matrix $A$ is a binding input to an industry, reducing ``input bottlenecks'', Eq. (\ref{eq:xinp}), to 
\begin{equation} \label{eq:xinp_leo}
x_{i,t}^{\text{inp}} = 
\min_{ \{ j: \; A_{ji} > 0 \} } 
\left \{ \;  \frac{ S_{ji,t} }{ A_{ji} } \right \},
\end{equation}
In this case an industry would halt production immediately if inventories of any input are run down, even for small and potentially negligible inputs .

At the other extreme, we implement a linear production function for which all inputs are perfectly substitutable. Here, production in an industry can still continue if inputs cannot be provided, as long as there is sufficient supply of alternative inputs. Thus, there is no input bottlenecks for individual inputs, however, production can be constrained if the input inventories are insufficient. In this case we have 
\begin{equation} \label{eq:xinp_linear}
    x_{i,t}^{\text{inp}} =   \frac{ \sum_j S_{ji,t} }{ \sum_j A_{ji} } .     
\end{equation}
Note that intermediate inputs are perfectly substitutable in this case, but a lack of labor supply cannot be compensated by other inputs.

We also implemented three different production function specifications based on the criticality ratings of IHS Markit analysts. Recall that we define \emph{critical} inputs as those that were rated 1, \emph{non critical} those that were rated 0, and \emph{important} those that were rated 0.5.
First, we use the baseline production function of the main text where we set all input ratings of 0.5 equal to 0 (this makes them non-critical inputs). Second, we set all 0.5 input ratings equal to 1 to make them critical inputs. This moves us closer to the Leontief production function.
As a third case, we implement the specification where an industry's production scales proportionally to the 0.5 rating of important inputs. Thus, we have
\begin{equation} \label{eq:xinp_05}
    x_{i,t}^{\text{inp}} =
    \min_{ 
    \{ 
     \forall j \in \mathcal{V}_i, \; \forall k \in \mathcal{U}_i
    \} 
} 
    \left \{ \;  \frac{ S_{ji,t} }{ A_{ji} },  
   \frac{1}{2} \left(\frac{ S_{ki,t} }{ A_{ki} } +  x_{i,0}^\text{cap}\right)
     \right \},     
\end{equation}
where $\mathcal{V}_i$ is the set of \textit{critical} inputs and 
$\mathcal{U}_i$ is the set of \textit{important} inputs to industry $i$. 
This means that if an \textit{important} input goes down by 50\% compared to initial levels, production of the industry would decrease by 25\%. In case the stock of this input is fully depleted, production drops to 50\% of initial levels.

\begin{figure}[H]
\centering
\includegraphics[trim = {0cm 0cm 0cm 0cm}, clip,width=\textwidth]{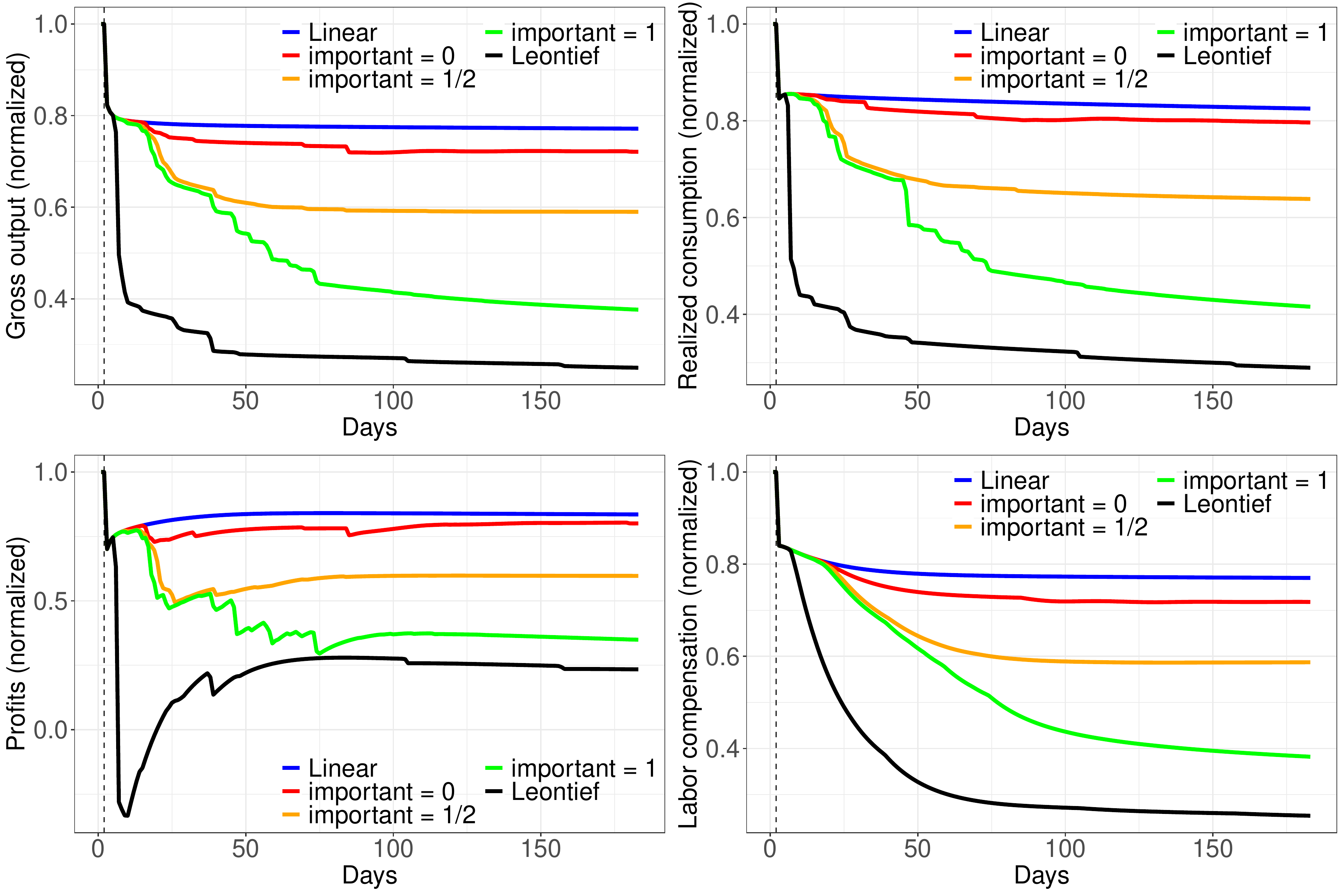}
    \caption{ {\bf Comparison of production functions for indefinite lockdown.}
    Total output, realized consumption, profits and labor compensation for five production function specifications after the lockdown is imposed at $t=2$ (dashed vertical line). All values are normalized to the initial no-lockdown steady state and there is no re-opening of industries.
    Linear (blue) denotes a linear production. The red line represents the baseline production function with all important (0.5) inputs being set to be non-critical. The orange line is the case where production scales with a factor of 1/2 with important inputs. The green line (important=1) is the case where all important inputs are treated as critical. The black line the Leontief production function where all inputs are critical.
    }
    \label{fig:prodfun_base}
\end{figure}

In Fig. \ref{fig:prodfun_base} we show simulation results on total production $\tilde{x}_t$,  
labor compensation $\tilde{l}_t$, 
profits $\tilde{\pi}_t$ and
household consumption $\tilde{c}_t$; Eq. \eqref{eq:tot_x}-\eqref{eq:tot_c}. 
Note that value added is the sum of profits and labor compensation. In the simulation, $t_\text{start\_lockdown}=2$,\footnote{Choosing a time step for applying the pandemic shock is arbitrary, as the model rests in steady state beforehand. In Section \ref{sec:econimpact} we chose March $23^{\text{rd}}$ as the starting date of lockdown to easily compare to UK quarterly data, in the following we apply the pandemic shock at $t=2$, with no loss in generality.} and there is no re-opening in the following six months (180 days).

As expected we find the largest drop for all economic variables for the Leontief production function, where every input can potentially become binding (black line). 
For the Leontief economy our model predicts roughly a 70\% drop in gross output, consumption and labor compensation over the next six months if the lockdown continues.

Again in line with intuition, we obtain the mildest economic impacts for the linear production function (blue). There is still a substantial drop in all variables due to first-order shocks, but little further adjustment resulting from network effects.

The production function specifications using the results of the IHS Markit survey lie somewhere in between these two extremes. Treating all important inputs as critical also yields a severe drop in economic production and all other variables (green). The red line indicates the scenario where all important inputs are considered to be non-critical, which is the baseline specification of our model. 
The results for this case are more similar to the linear production function, although slightly more severe due to the higher risk of lacking critical inputs. Treating important inputs as half-critical such that output scales with the inputs by the factor of 1/2 is the `median' scenario in these simulations, as indicated by the orange line.

In Appendix \ref{apx:prodfun} we show the recovery paths after re-opening the entire economy. Although the production functions can yield substantially different economic impacts in lockdown, they converge in the long-run after the economy is re-opened.

\subsection{First-order shocks, shock propagation and total impact}
\label{sec:prodnet}

\begin{figure}[H]
\centering
\includegraphics[width = .92\textwidth]{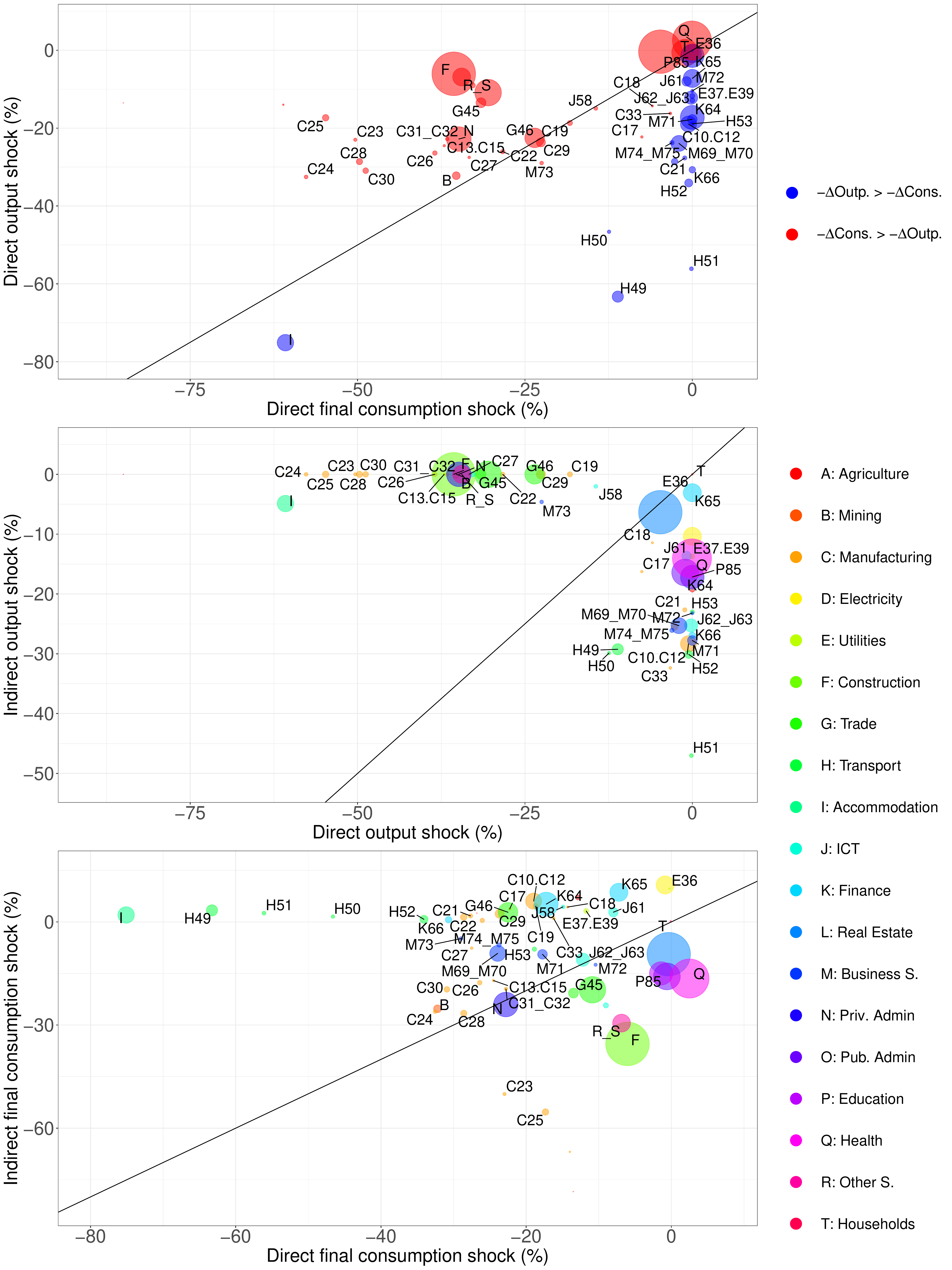}
    \caption{
    \emph{Upper panel:}
    {\bf Sectoral first-order shocks in the UK economy} on the supply and demand sides. The axes give \%-reductions in gross output (y-axis) and in final consumption (x-axis). A blue disk indicates that the monetary shock in absolute terms is larger on the supply side than on the demand side (blue disks can thus lie above the identity line and vice versa). Disk size corresponds to initial gross output of industries. Details on the shocks and industry labels can be found in Appendix \ref{apx:shocks}, Table \ref{tab:FO_shocks}.
    \emph{Center panel:}
    {\bf Comparison of direct and indirect output shocks.}
    Direct supply shocks shown as reduction in sectoral output (x-axis) plotted against indirect impacts on sectoral production (y-axis). Industries that face a large initial supply shock tend to experience smaller higher-order impacts, while higher-order effects can be large for industries that experienced little or no initial supply shock.  
    \emph{Bottom panel:}
    {\bf Comparison of direct and indirect final consumption shocks.}
    Direct final demand shocks (x-axis) plotted against indirect impacts on final demand (y-axis). Disk size corresponds to initial level of final demand satisfied per industry.
    }
        \label{fig:in_vs_direct}
\end{figure}

Our model is initialised with the first-order supply and demand shocks discussed in Section \ref{sec:pandemic_shock} and simulates how the economic system translates these shocks into overall economic impacts. Recall that exogenous supply shocks lead to an immediate reduction in gross output. 
We let the direct output shock be
\begin{equation}
    \text{OS}_i^{\text{direct}} \equiv \frac{x_{i,0}^\text{shocked} - x_{i,0}}{x_{i,0}} = \frac{(1-\epsilon^S_{i,0}) x_{i,0} - x_{i,0}}{x_{i,0}} = -\epsilon^S_{i,0},
\end{equation}
which is equivalent to first-order supply shocks.
Similarly, exogenous demand shocks instantaneously decrease final consumption.
More formally, we define direct final consumption shocks as
\begin{equation}
    \text{CS}_i^{\text{direct}} \equiv \frac{c_{i,0}^\text{shocked} + f_{i,0}^\text{shocked}- c_{i,0} - f_{i,0} }{c_{i,0} + f_{i,0}}.
\end{equation}
The upper panel of Fig. \ref{fig:in_vs_direct}  visualises for each sector of the UK economy the reduction in gross output $\text{OS}_i^{\text{direct}}$ and final consumption $\text{CS}_i^{\text{direct}}$ as a result of first-order shocks. While some industries such as Forestry and Logging (A02) face larger immediate reductions in gross output than final consumption, transport industries (H49-51) experience much larger negative shocks in final consumption. Reductions in both final consumption and gross output are enormous for Accommodation and Food Service Activities (I).

We also quantify higher-order impacts on gross output (supply-side) and final consumption (demand-side). 
These indirect effects are time-dependent since overall economic performance changes in time as can be seen from the simulations above.
Note that higher-order impacts in gross output do not necessarily need to be caused by supply-side shocks but could also result from a lack of demand. Conversely, final consumption reductions can stem from lowered production levels.
We let the total output shock at any time denote 
\begin{equation*}
\text{OS}_{i,t}^{\text{total}} =  x_{i,t} /x_{i,0} - 1.
\end{equation*}
The indirect output shock is then computed as the residual of the total output shock $\text{OS}_{i,t}^{\text{total}}$ and direct output shock $\text{OS}_i^{\text{direct}}$.
Thus, for the indirect output shock we have
\begin{equation}
    \text{OS}_{i,t}^{\text{indirect}} = \text{OS}_{i,t}^{\text{total}} - \text{OS}_i^{\text{direct}}.
\end{equation}
Similarly, we compute indirect final consumption shocks as
\begin{equation}
    \text{CS}_{i,t}^{\text{indirect}} = \text{CS}_{i,t}^{\text{total}} - \text{CS}_i^{\text{direct}},
\end{equation}
where $\text{CS}_{i,t}^\text{total} = (c_{i,t} + f_{i,t} ) / (c_{i,0} + f_{i,0}) - 1$ is the overall final consumption shock.

The center panel of Fig. \ref{fig:in_vs_direct} shows a scatter plot where the x-axis denotes direct output shocks $\text{OS}_i^\text{direct}$ and the y-axis indirect output shocks $\text{OS}_{i,t}^\text{direct}$ for $t=60$, i.e. two months into lockdown. 
Points scatter in an inverted L-shape indicating that industries that experience large direct output shocks do not reduce production much more in the course of the lockdown. In contrast, many industries that experience little or no direct shocks to their productive capacities downsize economic production substantially after two months.

The bottom panel of Fig. \ref{fig:in_vs_direct} is the same but for final consumption instead of output. Although similar patterns can be observed, there are also a few larger differences. For the majority of industries there are less extreme direct and indirect effects on final consumption. These industries thus lie closer to the identity line. While all higher-order supply-side effects are non-positive, almost half of the industries face positive higher-order consumption effects, although they tend to be very small. Note that the vast majority (92\%) still face negative total consumption effects. Positive values on the y-axis indicate that higher-order impacts on final consumption are slightly mitigating initial shocks. For example, Transport and Warehousing industries (H49-52) are substantially hit by direct demand shocks. The total impact after two months of lockdown is somewhat below these levels.

We also used our first-order shocks to calibrate simpler traditional input-output models (see \cite{miller2009input} for an excellent overview). We show in Appendix \ref{apx:io_models} that under much simplified model parametrizations we can recover the classic Leontief model and get similar sectoral predictions as the Gosh model in steady state. The problem with these models is that they are not able to take supply and demand shocks into account at the same time. While alternative IO models such as the mixed endogenous/exogenous model \citep{dietzenbacher2015reflections} can be used with simultaneous supply and demand constraints, the model does not necessarily yield feasible solutions corresponding to positive output and consumption values. This is indeed what we find for the UK economy. Calibrating the mixed endogenous/exogenous IO model to our first-order shocks results in infeasible economic allocations.

We now demonstrate how the first-order shocks shown in Fig. \ref{fig:in_vs_direct} are amplified through the production network and how they affect overall economic impact. 
To evaluate the relative contributions of supply and demand shocks to overall outcomes, we run the following simulations: First, we run the baseline model setup where both initial supply and demand shocks are present as discussed above.
Second, we run the model considering only supply shocks, i.e. we set all consumer demand shocks to zero, ${\epsilon}^D_i=0$, as well as remove all shocks to other final demand categories, i.e. $f^d_{i,t} = f^d_{i,0}$

As a third simulation scenario we run the model with all initial supply shocks switched off and include only initial demand shocks.

\begin{figure}[H]
    \centering
\includegraphics[width = .95\textwidth]{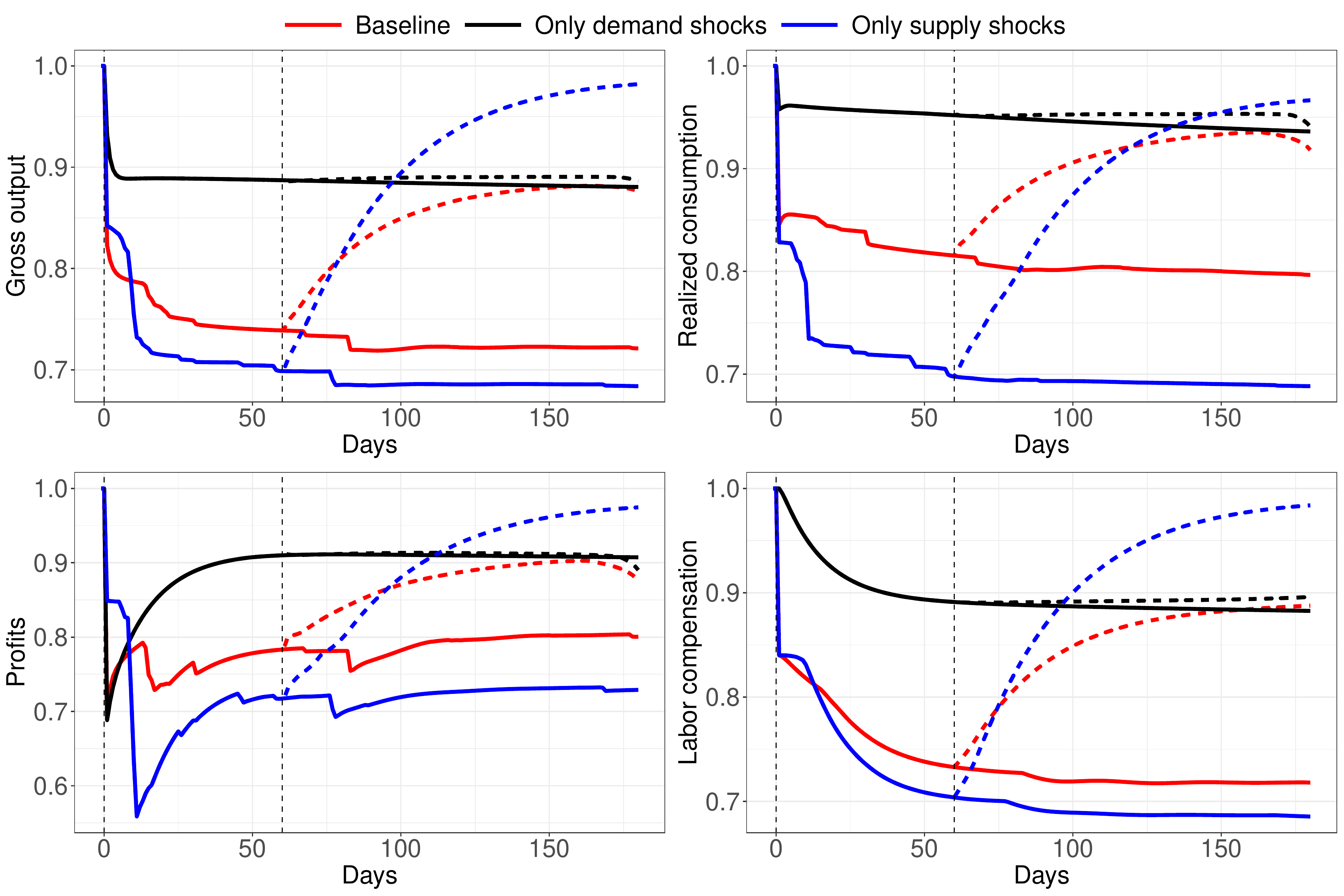}
    \caption{
    {\bf Dynamic effect of supply shocks vs. demand shocks.} 
    Normalized values of gross output, realized consumption, profits and labor compensation for different shock scenarios. Baseline (red) denotes the model default setup where both supply and demand shocks are used. The black/red line shows the case where only demand/supply shocks are switched on. The lockdown starts at $t=0$ and ends for all industries after two months at $t=60$ (vertical dashed lines). }
        \label{fig:shockimpact}
\includegraphics[width = .46\textwidth]{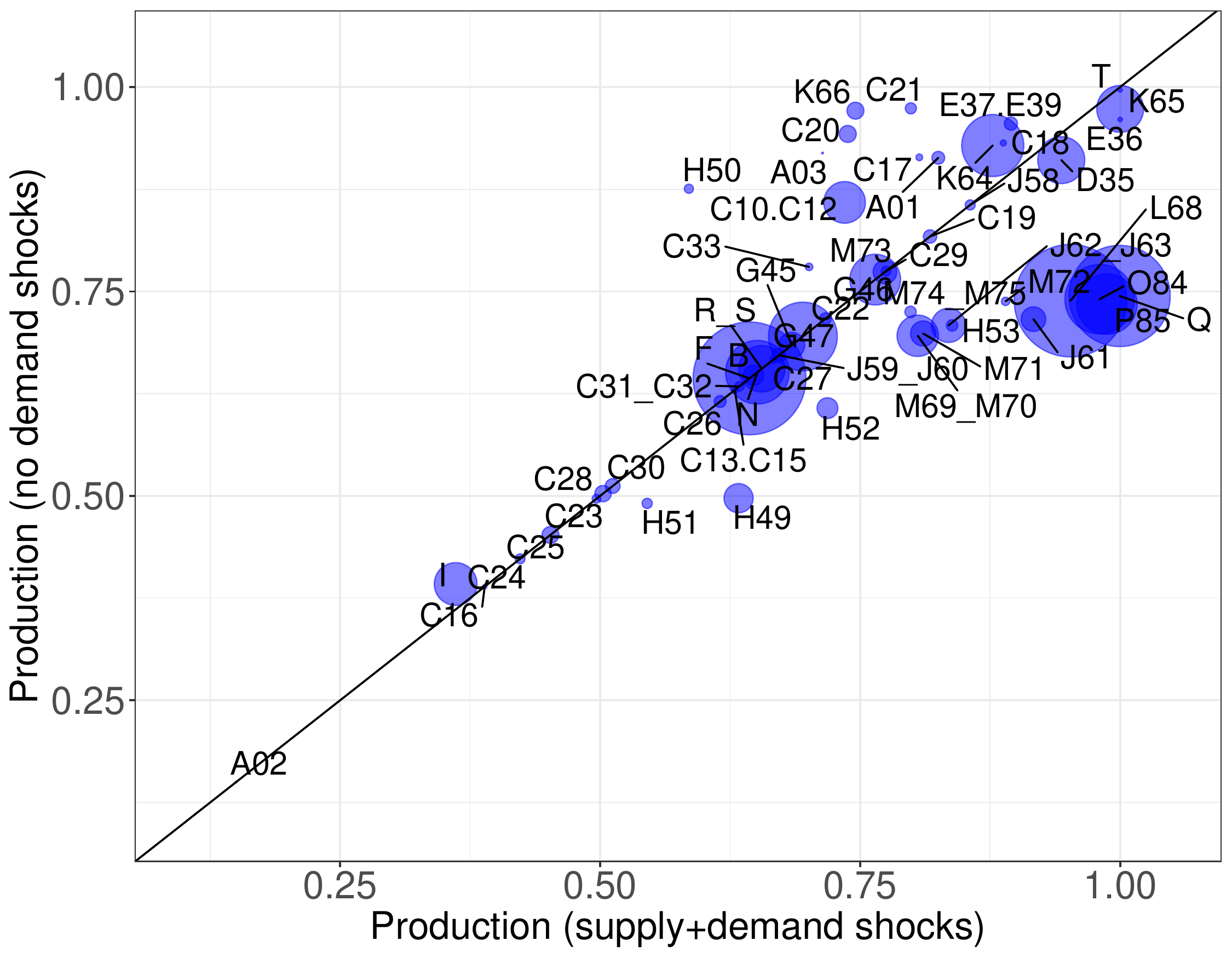}
\includegraphics[width = .53\textwidth]{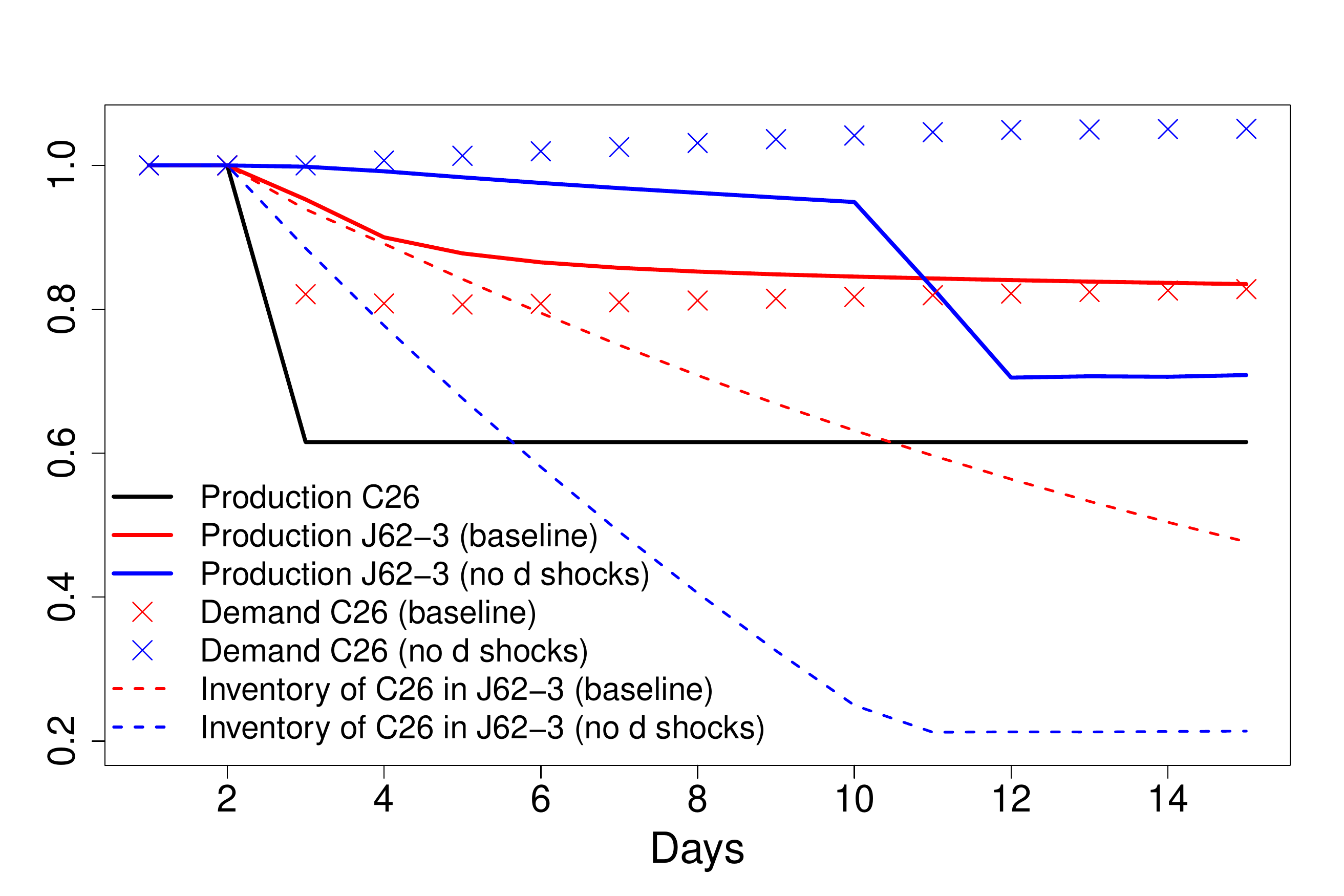}
    \caption{{\bf How can production decrease when demand shocks are removed?}
    \emph{Left panel:} Comparing normalized production at the sectoral level two weeks after lockdown when supply and demand shocks are turned on (x-axis) and with demand shocks turned off (y-axis). A few large sectors achieve substantially less production if there are no initial demand shocks in the economy. 
    \emph{Right panel:} A sectoral example of downstream shock propagation of the ``Baseline'' and ``Only supply shocks'' scenarios in Fig. \ref{fig:shockimpact}. Sector IT (ISIC J62-3) produces less after 11 time steps if there are no demand shocks in the economy (blue solid line) compared to both supply and demand shocks being present (red solid), since it quickly runs out of critical input C26, Manuf. Electronic, inventories (blue vs. red dashed line). C26 produces the same in both cases (black line) due to binding capacity constraints. If there are no initial demand shocks, C26 faces higher aggregate demand (blue vs. red crosses). Due to higher demand for C26 goods and lower production of C26 goods, IT receives less C26 if there are no demand shocks in the economy.
    }
        \label{fig:shockimpact_example}
\end{figure}

Fig. \ref{fig:shockimpact} shows four key macroeconomic variables (total output $\tilde{x}_t$, consumption $\tilde{c}_t$, profits $\tilde{\pi}_t$ and labor compensation $\tilde{l}_t$) for all three simulation scenarios, both when lockdown continues (solid lines) and when the lockdown is lifted for all industries after 60 days (dashed lines).
In Appendix \ref{apx:supply_demand}, we run the same simulations for alternative production functions.
It becomes clear that demand shocks lead to much smaller economic impacts than supply shocks (blue vs. red solid lines). On the other hand, we find that the economy recovers much quicker after the lockdown if there are no demand shocks, as indicated by the large positive slope of the red dashed line. When there are only demand shocks, recovery is slow. In this case unwinding the economy from lockdown brings limited positive effects due to persistence in exports and investment shocks, sluggish consumption adaptation and a portion of consumers believing in an L-shaped recovery.  

Strikingly, we observe that overall negative economic impact in lockdown is larger if the model economy faces only supply shocks instead of being exposed to supply and demand shocks simultaneously (red vs. black solid line). If demand shocks are absent, total output lies roughly 5\% below the baseline scenario where both types of shocks are present, except for the first eleven days.

Why is it that turning off demand shocks leads to larger adverse overall impacts? The reason is that in case of large supply constraints and no reduction in final demand, there is higher competition for relatively few goods. If producers cannot satisfy aggregate demand, they need to ration their output to customers (recall that we use proportional rationing). In case of large aggregate demand every customer receives only a relatively small share, which could be even less for some industries compared to the scenario where demand shocks are turned on.
If these goods are critical inputs, production in concerned industries will come down once inventories of these inputs are run down. Thus, removing demand shocks can increase the risk of input bottlenecks in production. Put simply, decreasing final demand of some key intermediate goods ensures continued supply of these intermediate goods to other intermediate industries.

In the particular case considered here, it turns out that several large industries have to reduce production as a consequence of fierce competition for critical inputs, as can be seen in Fig. \ref{fig:shockimpact_example} a). 
Without demand shocks industries such as Health (Q), Education (P85), Real Estate (L68) or IT (J62-3) produce up to 25\% less when compared to the baseline case two weeks after lockdown. Although several other industries such as Manufacturing Chemicals (C20) and Pharmaceuticals (C21) produce substantially more without demand shocks, this does not offset the overall adverse effect on the economy as a whole.  This can be thought of as a coordination failure.

To better understand how this happens we zoom into the first 15 days of Fig. \ref{fig:shockimpact} a). As an illustrative example in Fig. \ref{fig:shockimpact_example} b), we show how production constraints in Manufacturing Electronic (C26) lead to larger input bottlenecks in IT (J62-3) in the absence of initial demand shocks.
There is no difference in Manufacturing Electronic production (black line) for both scenarios due to binding capacity constraints, but it faces larger aggregate demand if there are no demand shocks (blue crosses) compared to both shocks being present (red crosses). Thus, sector IT for which Manufacturing Electronic goods are critical inputs has to run down its input inventories quicker if there are no demand shocks (blue dashed line), since sector Manufacturing Electronic can deliver less goods in this case. This is reflected in total output of IT. If there are no demand shocks present, IT production is higher for the first few time steps (blue solid line), but drops below the baseline production of both supply and demand shocks (red solid line). We observe similar dynamics for other industries as well.

The case of Manufacturing Electronic production and the coupled output of the IT sector exemplifies the complexity of shock propagation through production networks. 
We highlight several striking features of this analysis which tend to be entirely neglected in most macroeconomic studies, even if they incorporate industry-specific effects.

First, the specification of production functions and input criticality plays an important role. Most economic analyses use some form of CES production functions with non-zero substitution coefficients. Under this approach, while it is in principle possible to construct elaborated CES nests where different degrees of substitutability are allowed between different inputs of a given sector \citep{baqaee2020nonlinear}, in practice it is so hard to calibrate the parameters that only one \citep{barrot2016input} or a few \citep{baqaee2020nonlinear,bonadio2020global} parameters are specified. None of the recent IO papers have considered different degrees of substitution between groups of intermediate inputs. The survey considered here (Appendix \ref{apx:ihs}) instead introduces a distinction between critical and non-critical inputs, for each separate industry, allowing us to keep the Leontief assumption of a strong lack of substitutability for critical inputs, which is arguably a key feature of short-run dynamics after large shocks, while at the same time not allowing some non-critical inputs to prevent production. This is a step toward more realism, as in exceptional circumstances like a pandemic, we believe that it is likely that firms can still operate even if several inputs that they usually use are not available.
Of course, this is admittedly imperfect and could be improved, and we have made the strong assumption that the lack of use a non-critical input simply does not decrease production and translate into higher profits. Assuming a drop in productivity in this case would change the quantitative results, but would not, however, fundamentally change the dynamics.

Second, the size of inventories held by industries is crucial. Similar to equity buffers in financial distress models, inventories act as buffers against production shocks originating upstream and propagating downstream.  
Inventory effects are not present in most macroeconomic studies \citep{favero2020restarting, bodenstein2020social, eichenbaum2020macroeconomics, mckibbin2020global} and IO models (see Table \ref{tab:lit_table}), and only appear in a very stylized manner in other empirical work \citep{mandel2020economic, inoue2020propagation}.
Detailed information on input-specific inventories on industry and firm levels, as well as on behaviour and inventory management rules, could vastly improve our understanding of shock propagation in production networks.

Third, dynamics really matter over the short time horizons relevant for the pandemic lockdown and its immediate aftermath. It can take days to weeks for shocks to cascade through several layers of a large production network and our simulations suggest that it can take months until the dynamics reach a steady state. Moreover, the presence of input bottlenecks due to the lack of critical inputs can amplify initial shocks in highly nonlinear ways (see Appendix \ref{apx:shock_uncertainty}). The propagation of shocks is path-dependent.  This is due to the fact that different industries can have very different customers, resulting in heterogeneous contagion dynamics that depend on ``who gets hit first''. 
General equilibrium models and most input-output models implicitly assume zero adjustment time and compare pre- and post-shock equilibrium states of the economy to quantify overall impacts. Our analysis suggests that the shock propagation dynamics play an important role in the short time horizons of pandemic lockdowns. In other words, equilibrium comparative static is warranted only when adjustment is faster than the arrival of new shock \citep{ando1963essays}. This is not the case currently, where the lifting of the lockdown happens before the system has had a chance to reach the lockdown steady-state.

\subsection{Re-opening a network economy} \label{sec:reopen_singles}

We next investigate how unwinding social distancing measures in certain sectors affects overall economic output. We consider more realistic re-opening scenarios in Section \ref{sec:reopening} together with their impact on infection. Here, we focus on more stylized, theoretically interesting, examples under different production function assumptions to better understand the driving forces behind overall impacts of staged re-opening.

We study the following simulations: As before we represent the economy in lockdown by initialising the model as usual with first-order shocks. We then consider two cases. First, the re-opening scenario where lockdown is relieved after two months for a given set of industries, i.e. for those industries we set $\epsilon^S_{i,t} = 0$ and demand adjusts as discussed in Section \ref{sec:pandemic_shock}. Second, the lockdown scenario, where the lockdown continues and no shocks are removed. We then compare the two scenarios to quantify the boost in economic activity of re-opening a given industry compared to the lockdown.

\begin{figure}[H]
    \centering
\includegraphics[trim = {3cm 0cm 3cm 0cm}, clip, width = 1\textwidth]{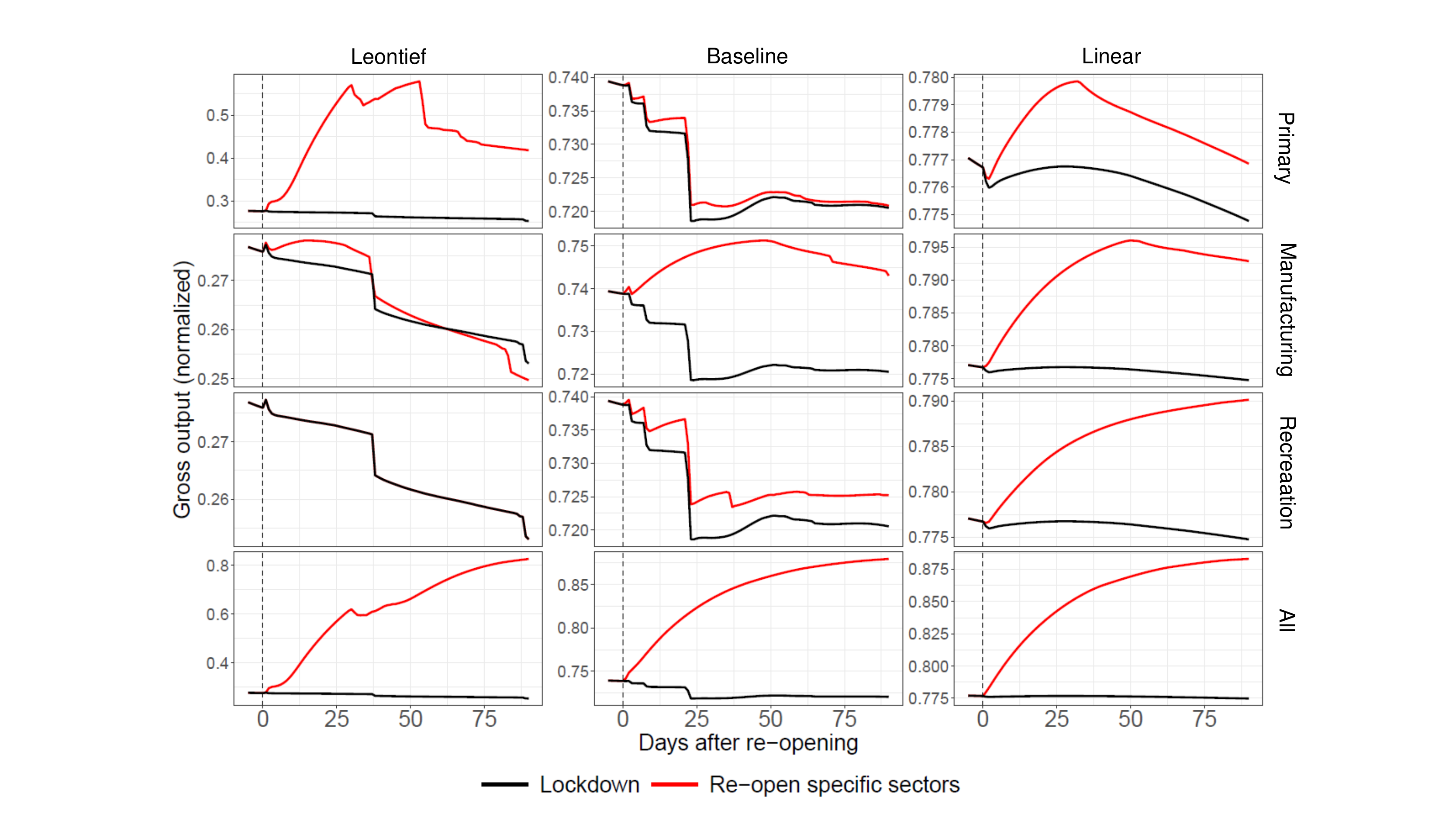}
    \caption{ 
    {\bf Economic performance of indefinite lockdown and re-opening single sectors for different production functions.}
    Black lines indicate total output if the lockdown continues and red lines if only one given sector is re-opened at $t=0$.
    Columns represent different production functions, rows denote the sectors which are re-opened. 
    The left column show the effect of re-opening for different sectors in a Leontief economy. The center and right columns show the same for the baseline and linear production functions, respectively.
    Row 1 shows economic effects of re-opening only primary sectors (ISIC A and B). Row 2 shows the same for opening Manufacturing (ISIC C), row 3 for ``Recreation'' (ISIC I, R, S) and row 4 for opening all sectors simultaneously. Note that the scale of the y-axes varies between panels.
    }
        \label{fig:reopen_analysis}
\end{figure}

Fig. \ref{fig:reopen_analysis} summarises our findings. Each panel shows total production normalized by pre-shock output on the y-axes for both scenarios (re-open sectors in red, continued lockdown in black). The x-axes shows the number of days where day zero is when the lockdown is lifted in the re-opening scenario. Thus, the economy was already two months in lockdown before day zero which is not shown since production is identical for both scenarios during that period.
Panel columns represent simulation results for different production function specifications. Panel rows indicate the industries which are re-opened if the lockdown is relaxed.

The economic boost of re-opening varies largely between different sectors and also depends strongly on the production mode assumed.
Let us first consider the Leontief production function (left panels). Here, we find a huge increase in economic activity if the highly upstream primary sectors (Agriculture and Mining) are re-opened. Note that primary sectors only account for 2\% of UK's total economic output. Opening primary sectors has much smaller effects when using the baseline production function, where inputs are only partially critical, and the linear production function, where inputs are are not critical at all.

When re-opening the much larger manufacturing sectors (15\% of total output), we obtain a completely different impact on economic output. Strikingly, we find for a Leontief production function that economic output can be lower if manufacturing sectors are opened. The reason is similar to why smaller aggregate shocks can lead to larger overall impacts as discussed in Fig \ref{fig:shockimpact}. Manufacturing is a large sector relying on many inputs which are critical inputs for other sectors too. Production constraints in other industries might render it impossible to provide larger amounts of those inputs. If manufacturing sectors are re-opened, competition for those scarce inputs increases, resulting in less intermediate consumption for non-manufacturing industries which might face input bottlenecks as a consequence. We do not observe this for the alternative production function setups which relax the strong Leontief assumption. Here, economic output increases by 2-3 percentage points.

We find again very different results when reopening Other Services and Food and Accommodation, here for brevity called recreation. These sectors are large (6\% of total output) and heavily affected by the lockdown since they include theaters, hotels, restaurants and other social activities. It is interesting that opening these industries has no impact on overall economic production when assuming a Leontief production function. This is because these are highly downstream industries and their economic output is of little significance for the intermediate consumption of other industries. Thus, opening recreational sectors has mostly demand-side effects, and given the capacity constraints of upstream sectors this extra demand cannot be satisfied, resulting in no change in overall production. We find positive effects of opening recreational sectors under the alternative production function assumptions where total production increases slightly above lockdown levels.

The bottom row of panels compares the restart of the economy when the lockdown is lifted for all industries simultaneously. The largest economic boost is given in the Leontief economy which re-starts at very low levels after lockdown. The recovery paths are similar of our baseline and the linear production function. Note that recovery is not instantaneous, but takes a considerable amount of time. A month after re-opening, the economy still operates well below initial production levels.

\section{The effect of reopening on the reproduction number}
\label{sec:epidemic_model}

We attempt to quantify the effects of the re-opening scenarios on $R_0$, the reproduction number of the epidemic that ultimately determines the overall number of deaths.

Rather than coupling a complete epidemiological model with the economic model, we focus on determining $R_0$ by modelling only the infection rate as it varies across economic scenarios. It is illustrative to consider a simple SIR model,
\begin{eqnarray}
    \dot{S} & = & -\beta S I/M,\\
    \nonumber
    \dot{I} & = & \beta S I/M - \gamma I,\\\nonumber
    \dot{R} & = & \gamma I,
\end{eqnarray}
where dots over a variable denote its time derivative, $S$ is the number of people who are susceptible, $I$ is the number who are infected, and $R$ is the number who have recovered, and $M = S+I+R$ the total population, which we assume constant.

The model has two parameters, the transmission rate $\beta$ and the recovery rate $\gamma$. We focus on the early stage of the epidemic, that is, when the number of recovered individuals is small with respect to the rest of the population and there is no herd immunity \footnote{To give an upper bound, \citet{vollmer2020} recently reported an attack rate of around 13\% for Lombardy, the most affected region of Italy.}. The rate of exponential growth in the early stages of the epidemic is $R_0 = \beta/\gamma$, and is the key parameter determining the outcome of the outbreak. When $R_0<1$, the outbreak is minimal, but when $R_0\geq 1$, the outbreak reaches a finite proportion of the population, and larger $R_0$ implies a larger final number of individuals infected.  

While $\gamma$ is largely unaffected by public health measures absent any treatments for the disease, $\beta$ depends strongly on public policies and individual behavior. Since we are interested in $R_0$,
and $\gamma$ is relatively constant, we focus on modelling $\beta$. To compute $R_0$, we use
\begin{equation}
R_{0}(t)= \beta_t / \gamma \propto \beta_t.
   \label{eq:r0_prop}
\end{equation}
We now discuss how we compute $\beta_t$, and come back to computing $R_0$ at the end of the section.

\subsection{Decomposing infection across activities}
The parameter $\beta$ encompasses two factors: the number of contacts and the risk of infection during a contact. Usually epidemiologists try to distinguish between contacts at home, school, work, and other places \citep{mossong2008social,mikolajczyk2008collecting,mccreesh2019estimating,stromgren2017place,ferguson2020,Texas_covid_2020}. While we also want to distinguish home-based and work-based contacts, in the context of reopening industries a key question is that of consumption-based infections \citep{eichenbaum2020macroeconomics}. For instance, reopening cinemas would pose a threat to people working in cinemas, but the number of consumers attending cinemas is vastly larger than the number of workers. Thus, data on the number of contacts at work would fail to capture this risk. 

We decompose $\beta$ into five sources of infection: infections at the workplace ($\beta_w$), infections at school ($\beta_s$), infections during consumption activities ($\beta_c$), infections during commuting ($\beta_T$) and infections that are not influenced by whether or not an industry is open (for short, ``home'' infections $\beta_h$). We write
\begin{equation}
   \beta(t) = \beta^* \Big( \beta_w(t)+\beta_s(t)+\beta_c(t)+\beta_T(t)+\beta_h(t) \Big),
   \label{eq:beta}
\end{equation}
where $\beta^*$ is a disease specific parameter, and the components sum up to one when the economy is fully open ($t=0$),
\begin{equation}
\beta_w(0)+\beta_s(0)+\beta_c(0)+\beta_T(0)+\beta_h(0) = 1.
   \label{eq:beta_0}
\end{equation}
In Appendix \ref{apx:epidemic_modelling} we show how we can derive this equation and the functional form for each $\beta_x(t)$ where $x = w, s, c, T, h.$
With this formulation we can measure $\beta^*$ using data on the speed of diffusion of the virus in a fully open economy. Since there are estimates of $R_0$ before the lockdown, and considering Eq. \eqref{eq:r0_prop}, to evaluate $R_0$ during the lockdown and for various scenarios we only need to evaluate the changes to each of the terms within the parenthesis in Eq. \eqref{eq:beta}. To do this we rely on social contact surveys that estimate the intensity-weighted number of contacts of an average person across each activity (see Table \ref{table:sharesall} and
Appendix \ref{apx:epidemiology_data} for details).

\begin{table}[H]
\centering
\begin{tabular}{|l|l|p{3cm}|}
  \hline
Source of infection & Symbol  & Share of intensity-weighted contacts \\ 
  \hline
 Work & $\beta_w$& 0.29 \\ 
 School & $\beta_s$& 0.28 \\ 
  Consume & $\beta_c$& 0.16 \\ 
  Transport& $\beta_T$& 0.06 \\ 
  Home-related & $\beta_h$& 0.21 \\ 
   \hline
   Sum & & 1 \\ 
   \hline
\end{tabular}
\caption{Calibration of values for Eqs. \eqref{eq:beta}-\eqref{eq:beta_0}, using our intensity-weighted share of contacts per activity derived from \citet{stromgren2017place}, see Appendix \ref{apx:epidemiology_data}.}
\label{table:sharesall}
\end{table}

\subsection{Decomposing work and consumption infection across industries}
Our goal is to focus on what happens when schools and/or a group of industries are opened for work and/or for consumption. To do this we split the population into three categories: a fraction $\eta^s$ are students and pupils, a fraction $\eta^u$ are non-working adults, and the rest are workers, which we further split into $N$ industries, each containing a share $\eta_i$ of the population, so that we have
\begin{equation}
    \eta^s + \eta^u + \sum_{i=1}^N\eta_i = 1.
\end{equation}

The adult non-working population (including the inactive and unemployed population) do not produce output and cannot get infected through the work or study channel. Students and pupils do not produce output, but they do interact with others and can get infected at school. Workers produce economic output and can get infected through the work channel. The economic output and risk of infection of a worker is determined by the industry they work in. 

In what follows we consider policy variables that affect each of the $\beta$ terms in Eq. \eqref{eq:beta}. For instance, one of the key policy variables is
\[
\delta_{i,w} \equiv \text{Share of workers of industry $i$ that go to work physically.}
\]
We defer to the next subsection the details of how we compute this share in each scenario. We now show how each term of Eq. \eqref{eq:beta} (work, schools, consumption, transport and home) is computed, and in particular how they depend on $\delta_{i,w}$ and other policy variables (see Appendix \ref{apx:epidemic_modelling} and \ref{apx:epidemiology_data} for more details).

\paragraph{Home-related.} We assume that $\beta_h (t)$ is unaffected by whether or not a given industry opens. However, we account for reduction in household interactions during lockdown due to social distancing guidelines. We assume that
\[
\beta_h (t)=  \beta_h(0) \Big((1-\delta_h) \kappa + \delta_h \Big),
\]
where $\delta_h$ is a policy variable that takes the value 0 if social distancing for family/friends contacts is in place and 1 otherwise, and $\kappa$ is the share of social/family/friends contacts that are not avoidable by social distancing. We calibrate this by assuming that $\kappa$ is the share of household/home contacts, and ($1-\kappa$) is the share of contacts due to visiting friends and relatives, time in family cars, and contacts in public urban spaces. As discussed in Appendix \ref{apx:epidemiology_data}, we find $\kappa =0.76$.

\paragraph{Work-related.} Ideally, we would want to know the share of work-related infections that are due to workers of industry $i$. We are unable to obtain this, but from O*NET data we can estimate an index of exposure to infection. To incorporate this information, we assume
\begin{equation}
   \beta_w(t) = \beta_w(0) \sum_{i=1}^{N} \delta_{i,w} \frac{\eta_i b_{i,w}}{\sum_{k=1}^N \eta_k b_{k,w}},
\end{equation}
where $\delta_{i,w}$ is a policy variable that is equal to 1 in the pre-lockdown period (more details below), $b_{i,w}$ is an indicator of intensity weighted number of contacts in industry $i$, and $\eta_i$ is the share of population in industry $i$.
To calibrate $b_{i,w}$, we take O*NET occupation-level data on the exposure to infection and on physical proximity. We construct exposure to infection and physical proximity indexes at the industry level by using the share of occupations in each industry, and then construct $b_{i,w}$ as the average of the industry-level exposure to infection and physical proximity.

\paragraph{Schools.} We model students and pupils separately (workers in Education face a risk under the ``Work-related'' category). We assume that the school closure implies that all children above 14 (a share $1-g$) are not allowed to school, and those at 14 or below (a share $g$) are allowed if their parents work but cannot work from home. This excludes from school the children of the adult non working population, and we assume that the students and pupils do not work. Let $\delta_s$ be one if schools are open as normal, and zero if they are partially closed. Then, the fraction of the students and pupils population that are attending school is 
\begin{equation}
   \mu^s  =   \left( \delta_s + (1 - \delta_s) \left( g \sum_{i=1}^N \delta_{i,w}\eta_i \right) \right).
\end{equation}
For simplicity, we assume that the school infection rate scales linearly with the fraction of students attending schools as follows,
\begin{equation}
   \beta_s(t)  =  \beta_s(0) \mu^s.
\end{equation}

\paragraph{Consumption-related.} We proceed as for work-related infections, and write
\begin{equation}
   \beta_c(t) = \beta_c(0) \sum_{i=1}^N \delta_{i,c} b_{i,c},
\end{equation}
where $\delta_{i,c}$ is a policy variable that is equal to 1 in the pre-lockdown period (more details below), and the $b_{i,c}$ are such that $\sum_{i=1}^N b_{i,c}=1$. 
To calibrate $b_{i,c}$, we derive from \citet{stromgren2017place} a breakdown of consumption-related contacts into those related to Retail, those related to Restaurants, and those related to Sports Venues. We then map each of these three categories into a single separate WIOD classification.

\paragraph{Public transport.} Formally transport is just an industry where consumers risk catching the virus. However, in other consumption-related industries, the number of consumers depends only whether this particular industry is open. Transport is different because the number of people taking transport depends on how many \emph{other} industries are open \-- if all industries are open, trains are packed and there are more contacts/infections. Therefore we treat Transport separately, and we assume that all transport-related infections are between commuters. If we assume that the number of contacts of one commuter is proportional to the number of other commuters, infections are proportional to the square of the proportion of usual commuters that do commute in a given scenario. Thus, we have 
\begin{equation}
  \beta_T(t) = \beta_T(0) \left(\frac{ \mu^s\eta^s +\sum_{i=1}^N \delta_{i,w}\eta_i   }{\eta^s + \sum_{i=1}^N\eta_i}\right)^2 
\end{equation}
where the left term inside the parenthesis corresponds to student commuters and the right term to work commuters. The term in the denominator is a normalizing factor that guarantees consistency at time $t=0.$ (See Appendix \ref{apx:epidemic_modelling} for details on the derivation).

\subsection{Policy scenarios: reopening selected industries}
\label{sec:policyscenarios}

A policy is a set $\Lambda \equiv \Big\{ \{\delta_{i,c}\}_{i=1\dots N}, \{\delta_{i,w}\}_{i=1\dots N}, \delta_s, \delta_h\Big\} $. It is helpful to first note the values of $\Lambda$ before and after the lockdown. 

\paragraph{Pre-Lockdown.} All industries are open for workers and consumers. All schools are open and there is no friends and family social distancing.
\[
\begin{split}
\delta_{i,w}(\text{Pre-Lockdown}) &= 1 \\
\delta_{i,c}(\text{Pre-Lockdown}) &= 1 \\
\delta_{s}(\text{Pre-Lockdown}) &= 1\\
\delta_{h}(\text{Pre-Lockdown}) &= 1
\end{split}
\]
In this case, $\beta(t)=\beta^*$.

\paragraph{Scenario I: Full lockdown.} Workers go to work physically if and only if they are essential and they cannot work from home, which happens for a share of workers equal to $e_{iw}(1 - r_i)$, where $e_{iw}$ is the degree to which the industry is essential, and $r_i$ is the Remote Labor Index (see Appendix \ref{apx:shocks} for details on how we estimate $e_{iw}$ and $r_i$ for each industry). Consumers consume physically only what they can consume physically.
\[
\begin{split}
\delta_{i,w}(\text{Lockdown}) &= e_{i,w}(1 - r_i)\\
\delta_{i,c}(\text{Lockdown}) &= e_{i,c}\\
\delta_{s}(\text{Lockdown}) &= 0 \\
\delta_{h}(\text{Lockdown}) &= 0
\end{split}
\]
where $e_{i,c}$ is an ``essential consumption index''. For each industry (in fact, retail is the only relevant one), it indicates how much of infection-related consumption is still open during lockdown. In practice, we assume $e_{i,c}=0$ for all industries except retail,  and $e_{i,c}=e_{i,w}$ for retail, that is, we assume that the share of the retail-based infections that continue during lockdown can be proxied by the share of retail workers who are essential.

\hspace{5mm}

We consider four degrees of reopening: roughly speaking, reopening only manufacturing and construction, reopening everything except consumer-facing industries (with or without fully re-opening schools), and reopening everything. For all scenarios we assume that friends and family social distancing remains in place, $\delta_h=0$, and and that everyone that can work from home continues to work remotely.
Table \ref{table:scenario_summary} shows the main scenarios and the key dimensions in which they differ.

\begin{table}[H]
\footnotesize
\centering
\begin{tabular}{|p{4.4cm}|p{1.4cm}|p{2cm}|p{1.8cm}|p{1.8cm}|p{1cm}|p{1.4cm}|}
\hline
 & I & II & III & IV & V & VI\\
\textbf{Scenario} & Lockdown & Manufacturing and Construction & All except consumer-facing. & All except consumer-facing, Schools opened & Open & Pre-lockdown\\
\hline
\textbf{Work from home?} & Yes & Yes &Yes & Yes & Yes & No\\
\textbf{Schools open?} & Partially  & Partially  & Partially &  Fully &  Fully & Fully \\ 
\textbf{On-site consumption open?} &No &No &No &  No & Yes & Yes \\ 
\textbf{Home-related distancing?} &Yes &Yes &Yes &  Yes & Yes & No \\ 
 \hline
\end{tabular}
\caption{Six scenarios for reopening the economy.}
\label{table:scenario_summary}
\end{table}
\normalsize

\paragraph{Scenario II: Manufacturing and Construction.} We label this scenario ``Manufacturing and Construction'' for short, but we mean opening all A-F industries: Agriculture, Mining, Manufacturing, Utilities and Construction.  Because all workers who can work from home do so, we have
\[
\begin{split}
\delta_{i,w}(\text{Manufacturing and Construction})  &= 
\begin{cases}
(1 - r_i) & \quad \forall i \in \text{A-F}, \\
e_{i,w}(1 - r_i) &  \quad \forall i \in \text{G-T}. \\
\end{cases}\\
\delta_{i,c}(\text{Manufacturing and Construction}) &= e_{i,c} \hspace{20mm} \quad \forall i.
\end{split}
\]

\paragraph{Scenario III: All except consumer-facing.} This means opening all industries (A-T), except the three ``consumer-facing'' industries G47 (Retail), I (Accommodation and Food), and RS (Other Services, which includes recreation and personal services).
\[
\begin{split}
\delta_{i,w}(\text{All except consumer-facing.})  &= 
\begin{cases}
e_{iw}(1 - r_i) & \quad \forall i \in \Big\{\text{G47, I, RS}\Big\},\\
(1 - r_i) &  \quad \text{otherwise}. \\
\end{cases}\\
\delta_{i,c}(\text{All except consumer-facing.}) &= e_{i,c} \hspace{20mm} \quad \forall i.
\end{split}
\]

\paragraph{Scenario IV: All except consumer-facing. + Schools.} This is the same as Scenario III, but we assume that schools are fully open instead of receiving only children of workers who cannot work from home, that is $\delta_s=1$

\paragraph{Scenario V: Open.} We reopen all industries for work and consumption.
\[
\begin{split}
\delta_{i,w}(\text{All open})  &= (1 - r_i)  \quad \forall i \\
\delta_{i,c}(\text{All open}) &= 1 \quad \forall i.
\end{split}
\]

\subsection{Computing $R_0$}
We assume that before the lockdown, $R_0(\text{pre-lockdown})=2.6$ \citep{jarvis2020quantifying}. Then we compute how $\beta$ is reduced by social distancing measures during the lockdown. This gives us an estimate of the lockdown $R_0$ which we denote $\tilde{R_0}$
\begin{equation}
\tilde{R_0}= R_0 (\text{pre-lockdown}) \times \frac{\beta(\text{lockdown})}{\beta(\text{pre-lockdown})} \end{equation}
We find $\tilde{R_0}\approx 0.90$. However, the recent study by \citet{jarvis2020quantifying} finds that in the UK during lockdown, $R_0 \approx 0.62$. It is not surprising that we overestimate the lockdown $R_0$, as our model does not incorporate all the basic sanitary measures that would apply to the contacts that have not been reduced. For instance, we consider supermarket infections to stay the same because supermarkets are open, but social distancing applies in supermarkets and there are extra cleaning procedures in place. Another example would be essential workers. In our model, essential workers contribute to infections now just as they did before, while in reality it is likely that their conditions have been made at least a bit safer. A final example includes the effect of information campaigns on hand washing. 
To take this into account, we rescale all our estimates for the scenarios so that they start from a lockdown value at 0.62.

\begin{equation}
R_0(\Lambda) = \frac{0.62}{\tilde{R_0} } \times \frac{\beta(\Lambda)}{\beta(\text{pre-lockdown})} 
\end{equation}

This rescaling implies that fully re-opening the economy back to the pre-lockdown situation (that is, removing all limitations to work, consumption, school and social contact, while keeping the extra sanitary precautions) would bring $R_0$ to $(0.62/1.04) \times 2.6=1.55$.

We obtain standard errors for $R_0$ as follows. \citet{jarvis2020quantifying} report a pre-lockdown mean $R_0$ of 2.6 with a standard error of 0.54, that is, a standard error of $0.54/2.6=21\%$ of the mean. Their post-lockdown estimate is $0.62$ with a 95\% confidence interval (0.37 - 0.89), that is, a standard error of $((0.62-0.37)/2)/0.62=20\%$ of the mean. In view of this, we assume that for all our estimates, one standard error always equal 20\% of the mean $R_0$ estimates. We report confidence intervals as two standard errors around the mean.


\subsection{Economic performance vs. infections: sector-specific re-opening}
\label{sec:reopening}

We now show simulations for the four economic scenarios outlined in Section \ref{sec:policyscenarios}: lockdown, open manufacturing and construction, open all industries except consumer-facing ones, open all industries (school opening does not affect economic scenario). 

\begin{figure}[H]
    \centering
\includegraphics[width = \textwidth]{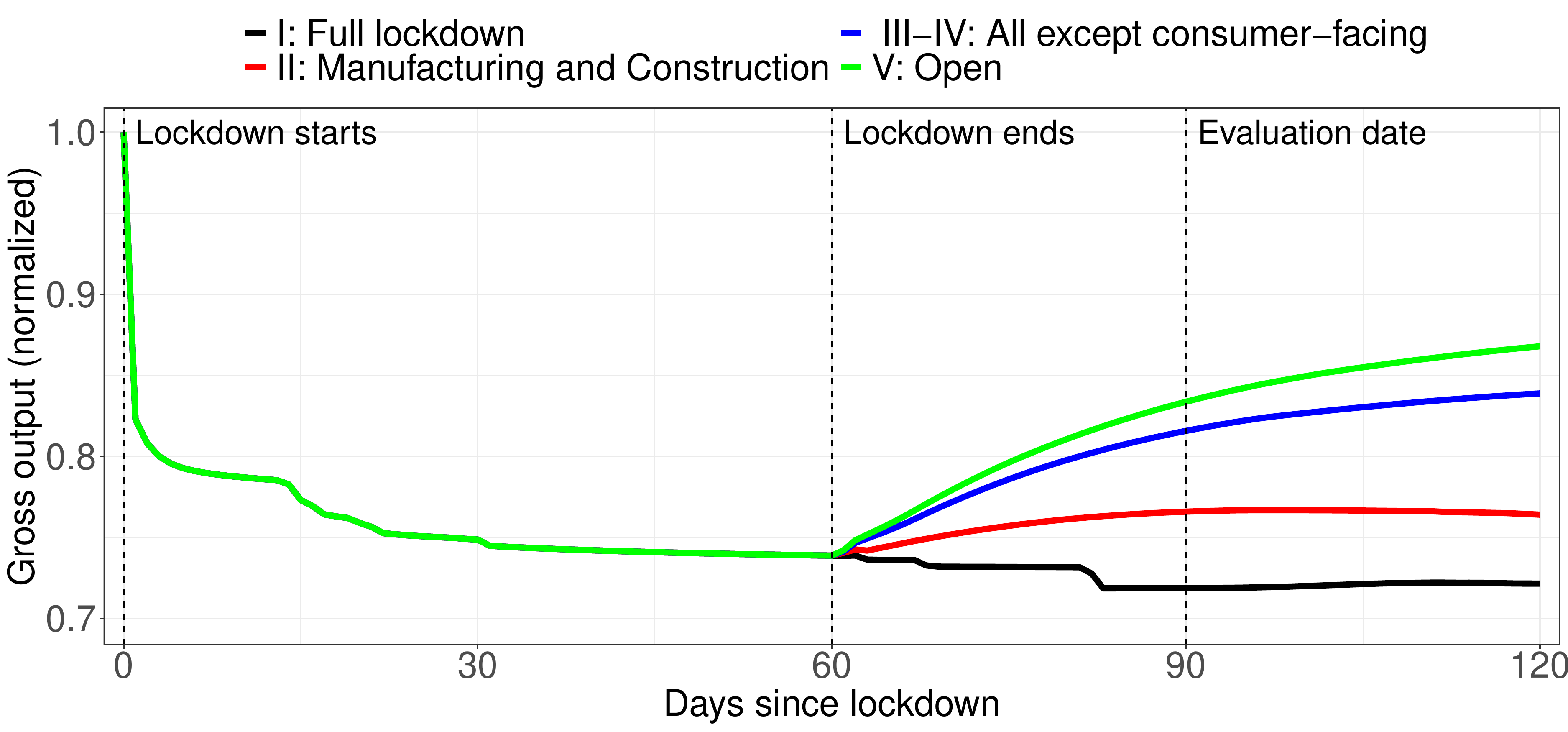}
    \caption{\textbf{Reopening scenarios.} We show production normalized to pre-lockdown levels, for four months since lockdown starts, assuming lockdown is lifted . We consider the four economic scenarios outlined in Section \ref{sec:policyscenarios}.
   }
    \label{fig:scenarios_output}
\end{figure}

As can be seen in Figure \ref{fig:scenarios_output}, keeping the lockdown leads to a further reduction in production, due to firms exhausting their inventories. Re-opening only all A-F industries (Agriculture, Mining, Manufacturing, Utilities and Construction) leads to an increase in production, however the jump is much bigger when opening all industries except consumer-facing ones. Conversely, there is not a large benefit in terms of production from opening consumer-facing industries. To compute the increases in production in Figure \ref{fig:scenarios}, we consider the first 30 days after lockdown is lifted; as it is easy to see, the economy is far from completely recovering after 30 days, as discussed in Section \ref{sec:econimpact}.

There is a trade-off between an increase in production and mitigating the epidemic spread when opening industries. In the bar plot in Fig. \ref{fig:scenarios} we illustrate this trade-off. The bars on the left show our estimates of $R_0$, the higher the bar, the faster the epidemic spread. The bars on the right show the GDP (as a percentage of the pre-lockdown GDP) of each scenario. In Scenario II, where Manufacturing and Construction open, the effect on $R_0$ is negligible, but has a 3 percentage points higher GDP than the lockdown scenario. The negligible increase in $R_0$ is due to the low percentage of the labor force that resumes work. $15\%$ of the employed people work in $A-F$ industries, and only $10\%$ cannot work from home. Furthermore, $6\%$ were already working on-site due to the essential nature of their work. Thus, under scenario II, only $4\%$ of the employed labor force resume on-site work.

Scenario III presents a slight increase in $R_0$ with respect to the lockdown scenario. The slight increase is mostly because non-consumer-facing industries from $G-R\_S$ have a high remote labor index. Therefore, assuming that all the workers that can telework stay home, the number of people returning to work is small. Scenario III has a 8 percentage points higher GDP than the lockdown scenario. This is due to a direct effect as well as a indirect effect where business services now resume work in tandem with primary and secondary sector, lifting key bottlenecks in supply chains.

Scenario IV, where we include opening schools for all children (not only for the below 14 children of people working on-site), increases $R_0$ substantially. With $17\%$ of the population being $14$ years old or younger, it is not surprising that opening schools increases the speed of the epidemic spread. Since we do not consider productivity decrease due to childcare work, the GDP of Scenario IV is equal to the GDP in Scenario III.
Scenario V, where consumer-facing industries open, increases substantially epidemic spread due to the share of contacts that happen in restaurants, hotels, gyms, etc. However, there is only a 2 percentage point increase in GDP compared to the previous scenario. This is mostly due to a lack of recovery in demand for the hospitality and recreation industries.

\section{Discussion} \label{sec:discuss}

In this paper we have investigated how locking down and re-opening the economy as a policy response to the COVID-19 pandemic affects economic performance and contagion.
We introduced a novel economic model specifically designed to address the unique features of the current pandemic. 
The model is industry-specific, incorporating the production network and inventory dynamics. We use survey results by industry experts to model how critical different inputs are in the production of a specific industry. 
We calibrate the model to the UK economy and find that two months after lockdown gross output and consumption are down by 27\% when compared to pre-lockdown levels.

We find that industries are affected by direct demand and supply shocks in highly heterogeneous ways. While many manufacturing industries face large supply shocks, transport industries experience mostly demand-side shocks. Other industries including hotels and restaurants are substantially exposed to both shock types simultaneously. 
We find similar industrial heterogeneity for higher-order impacts.

We analysed how shocks propagate through the production network, resulting in non-trivial economic impacts.
First, we have shown that input criticality plays an important role in the downstream amplification of shocks. Second, we found that inventory levels can act as buffers against production shocks and are crucial for understanding economic impacts -- an aspect usually neglected in other studies.
Third, it has become evident that time scales matter as shock propagation is not immediate but takes time. Overall, we find that first-order shocks can be translated into overall impacts in highly nonlinear ways. We even find cases where smaller aggregate shocks can lead to larger economic impacts as a result of unbalanced supply and demand dynamics. This `coordination failure' suggests that it could be dangerous to re-open single sectors of the economy by themselves without understanding how they are embedded in the production network. Our results suggest that the economic boost from opening an industry depends on the up-/downstream location of that industry as well as how severe the economy suffers from input bottlenecks. In case the economy faces serious productive constraints, re-opening a single sector can even have adverse effects on economic output.

There is a trade-off between re-opening the economy and facing an increase in epidemic spreading. To help understand this we develop an epidemic model where the infection rate is divided between different economic-related activities: work, schools, consumption, transport, and others. Within work and consumption, we consider the relative risk of infection between industries. We find that there is little variance between the risk of infection workers face in different industries. In contrast, the risk of infection due to consumption is concentrated in three consumer-facing industries: retail, restaurants and hotels, and other services (including gym and entertainment events). Our results show that keeping consumer-facing industries and schools closed, and having people who can telework work from home can significantly increase the economic output while having a relatively small increase in the spread of the epidemic.

\FloatBarrier

\small
\bibliographystyle{agsm}
\bibliography{tech_ref}
\FloatBarrier
\normalsize

\newpage
\appendix
\section*{Appendix}
\label{sec:Appendix}

\section{First-order economic shocks and work context industry variables} 
\label{apx:shocks}

\subsection{NAICS-WIOD mapping of shocks}

Due to the COVID-19 pandemic industries experience supply-side reductions due to the closure of non-essential industries and workers not being able to perform their activities at home. Many industries also face substantial reductions in demand. \cite{del2020supply} provide quantitative predictions of these first-order supply and demand shocks for the US economy. To calculate supply-side predictions, \cite{del2020supply} classified industries as essential or non-essential and constructed a Remote Labor Index, which measures the ability of different occupations to work from home. Under the assumption that the distribution of occupations across industries and that the percentage of essential workers within an industry is the same for the US and the UK, we can map the supply-shocks estimated by \cite{del2020supply} into the UK economy as follows.

First, we build a crosswalk from the NAICS 4-digit industry classification to the classification system used in WIOD, which is a mix of ISIC 2-digit and 1-digit codes. We build this crosswalk using the NAICS to ISIC 2-digit crosswalk from the European Commission and then aggregating the 2-digit codes that are presented as 1-digit in the WIOD classification system. We then do an employment-weighted aggregation of the supply shocks from \cite{del2020supply} for the 277 industries at the NAICS 4-digit classification level to the 55 industries in the WIOD classification. Some of the 4-digit NAICS industries map into more than one WIOD industry classification. When this happens we assume employment is split uniformly among the WIOD industries the NAICS industry maps into. Finally, we make one modification to deal with imputed rents for the Real Estate Sector. Imputed rents account for $69\%$ of the monetary value of the sector\footnote{Calculation obtained from monetary values reported in \url{https://www.ons.gov.uk/economy/grossvalueaddedgva/datasets/nominalandrealregionalgrossvalueaddedbalancedbyindustry}}. We assume that the supply shock does not affect imputed rents for the Real State Sector and thus consider that the supply shock only affects $31\%$ of the sector. With this modification the final supply shock to the Real Estate Sector is $15\%.$

For calibrating consumption demand shocks, we use the same data as \cite{del2020supply} which are based on the \cite{CBO2006} estimates. These estimates are available only on the more aggregate 2-digit NAICS level which are straightforward to map into WIOD ISIC categories. Table \ref{tab:FO_shocks} gives an overview of all first-order shocks applied to WIOD industries.

\subsection{Essential score, remote labor index, and industries' work context} 
Using the same methodology as before, i.e., doing a crosswalk from NAICS-4 digit to the classification system used in WIOD and using employment shares to aggregate, we map the essential score and remote labor index computed in \cite{del2020supply} into the WIOD list of industries. We use these industry remote labor index and essential score at the WIOD industry classification level to estimate the number of people working in each industry for each scenario i.e., to estimate $\delta_{iw}(t)$.

\begin{figure}
    \centering
\includegraphics[width = 0.5\textwidth]{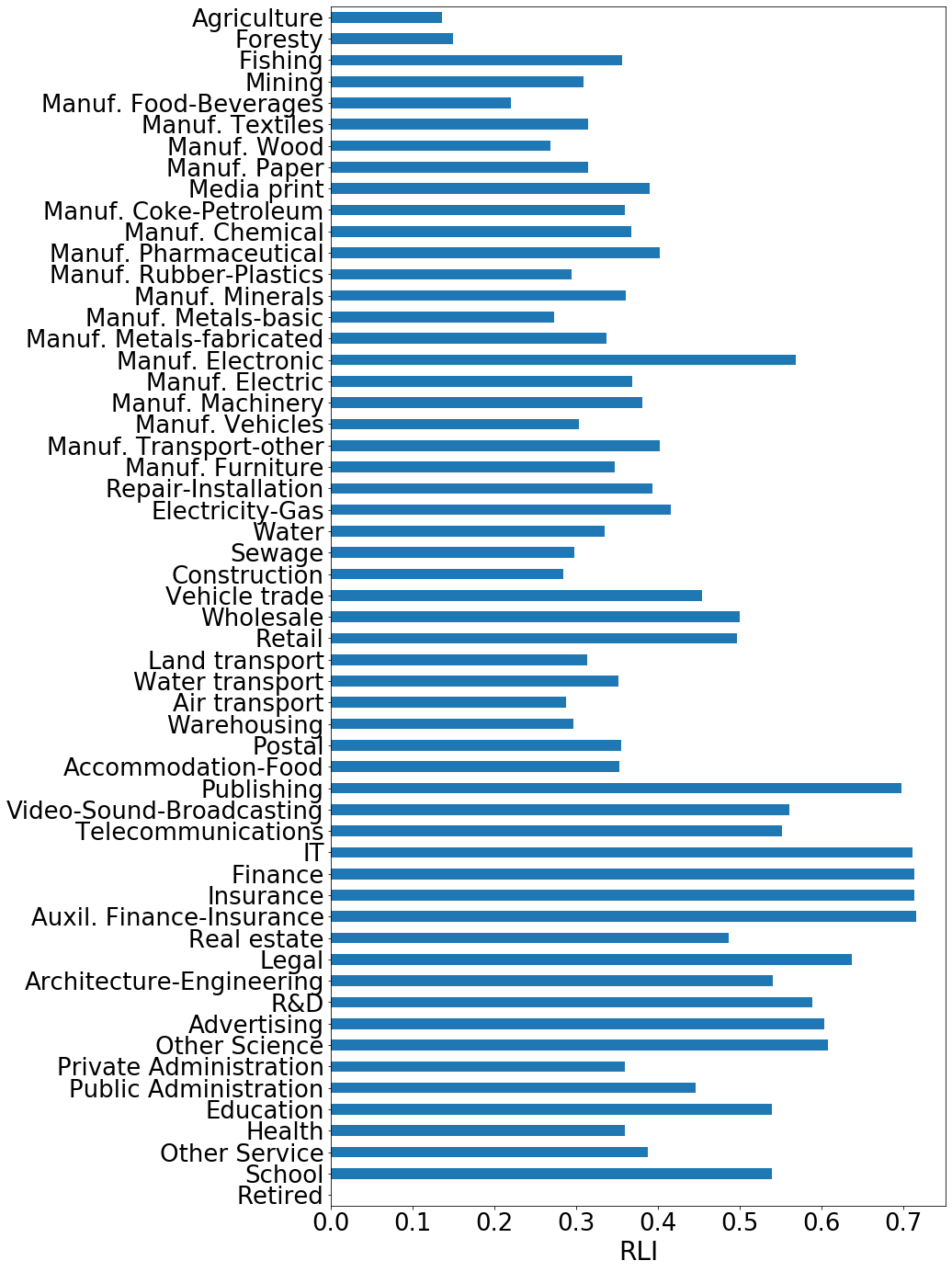}
    \caption{\textbf{Remote Labor Index of industries.} Remote labor index of the WIOD industry classification. See Table \ref{tab:FO_shocks} for code-industry name.}
        \label{fig:RLI}
\end{figure}

O*NET provides different Work Context\footnote{
\url{https://www.onetonline.org/find/descriptor/browse/Work_Context/}
}
 indices for occupations, including ``Exposure to disease and infection" and ``Physical proximity", for brevity we refer to these indexes as exposure to infection and physical proximity. Using the same methodology than \cite{del2020supply} we map these occupation indexes into the NAICS 4-digit industry classification. In particular, we use the data from the BLS, which indicates the occupational composition of each industry, and take the employment weighted average of the occupation's work context employed in each industry. After computing the exposure to infection, physical proximity and outdoors work for the industries at the NAICS 4-digit industry classification we map them into the WIOD classification system with the above mentioned crosswalk methodology. As we explain in Appendix \ref{apx:epidemiology_data} we use the  exposure to infection and physical proximity index of each industry to estimate the relative risk of contracting COVID-19 workers in each industry face.

\begin{table}[ht]
\scriptsize
\centering
\begin{tabular}{|cl|cccc|cc|cc|}
  \hline
    & & 
  \multicolumn{4}{c|}{ Supply  } &
  \multicolumn{2}{c|}{ Cons. demand} & \multicolumn{2}{c|}{ Other demand} \\
ISIC & Sector & $x$ & $\epsilon^S$ & RLI & ess. & $c$ & $\epsilon^D$ & $f$ & $f$ shock \\ 
  \hline
A01 & Agriculture & 0.8 & 0.0 & 14 & 100 & 0.9 & -10 & 0.3 & -20 \\ 
  A02 & Foresty & 0.0 & -85.0 & 15 & 0 & 0.0 & -10 & 0.0 & -23 \\ 
  A03 & Fishing & 0.1 & 0.0 & 36 & 100 & 0.0 & -10 & 0.1 & -32 \\ 
  B & Mining & 1.3 & -35.3 & 31 & 51 & 0.1 & -10 & 1.4 & -33 \\ 
  C10-C12 & Manuf. Food-Beverages & 2.8 & -0.6 & 22 & 99 & 2.5 & -10 & 1.3 & -32 \\ 
  C13-C15 & Manuf. Textiles & 0.4 & -37.1 & 31 & 47 & 0.1 & -10 & 0.5 & -27 \\ 
  C16 & Manuf. Wood & 0.2 & -61.1 & 27 & 18 & 0.1 & -10 & 0.1 & -16 \\ 
  C17 & Manuf. Paper & 0.4 & -7.5 & 31 & 89 & 0.1 & -10 & 0.2 & -30 \\ 
  C18 & Media print & 0.3 & -6.0 & 39 & 90 & 0.1 & -10 & 0.1 & -18 \\ 
  C19 & Manuf. Coke-Petroleum & 0.9 & -18.3 & 36 & 71 & 1.5 & -10 & 0.7 & -33 \\ 
  C20 & Manuf. Chemical & 1.1 & -2.6 & 37 & 96 & 0.3 & -10 & 1.6 & -32 \\ 
  C21 & Manuf. Pharmaceutical & 0.7 & -1.1 & 40 & 98 & 0.3 & -10 & 1.2 & -31 \\ 
  C22 & Manuf. Rubber-Plastics & 0.7 & -28.3 & 29 & 60 & 0.1 & -10 & 0.6 & -28 \\ 
  C23 & Manuf. Minerals & 0.5 & -50.3 & 36 & 20 & 0.1 & -10 & 0.2 & -27 \\ 
  C24 & Manuf. Metals-basic & 0.6 & -57.7 & 27 & 20 & 0.0 & -10 & 1.7 & -33 \\ 
  C25 & Manuf. Metals-fabricated & 1.1 & -54.8 & 34 & 18 & 0.1 & -10 & 0.7 & -18 \\ 
  C26 & Manuf. Electronic & 0.8 & -38.5 & 57 & 10 & 0.2 & -10 & 1.4 & -28 \\ 
  C27 & Manuf. Electric & 0.4 & -33.3 & 37 & 46 & 0.1 & -10 & 0.8 & -30 \\ 
  C28 & Manuf. Machinery & 1.1 & -49.7 & 38 & 20 & 0.2 & -10 & 2.1 & -30 \\ 
  C29 & Manuf. Vehicles & 1.6 & -22.6 & 30 & 65 & 1.3 & -10 & 2.6 & -29 \\ 
  C30 & Manuf. Transport-other & 1.0 & -48.8 & 40 & 17 & 0.1 & -10 & 2.5 & -31 \\ 
  C31\_C32 & Manuf. Furniture & 0.6 & -36.6 & 35 & 43 & 0.2 & -10 & 0.8 & -25 \\ 
  C33 & Repair-Installation & 0.4 & -3.3 & 39 & 95 & 0.0 & -10 & 0.0 & -17 \\ 
  D35 & Electricity-Gas & 3.2 & 0.0 & 42 & 100 & 3.6 & 0 & 0.1 & -20 \\ 
  E36 & Water & 0.2 & 0.0 & 33 & 100 & 0.6 & 0 & 0.0 & -11 \\ 
  E37-E39 & Sewage & 0.8 & 0.0 & 30 & 100 & 0.6 & 0 & 1.1 & -16 \\ 
  F & Construction & 7.9 & -35.6 & 28 & 51 & 0.3 & -10 & 11.5 & -6 \\ 
  G45 & Vehicle trade & 1.7 & -31.6 & 45 & 41 & 2.1 & -10 & 0.6 & -23 \\ 
  G46 & Wholesale & 3.5 & -23.6 & 50 & 51 & 3.3 & -10 & 4.3 & -30 \\ 
  G47 & Retail & 4.7 & -30.5 & 50 & 37 & 16.4 & -10 & 0.6 & -28 \\ 
  H49 & Land transport & 2.0 & -11.1 & 31 & 83 & 2.7 & -67 & 0.2 & -28 \\ 
  H50 & Water transport & 0.6 & -12.4 & 35 & 81 & 0.6 & -67 & 0.7 & -32 \\ 
  H51 & Air transport & 0.6 & -0.1 & 29 & 100 & 1.2 & -67 & 0.4 & -32 \\ 
  H52 & Warehousing & 1.4 & -0.5 & 30 & 99 & 0.1 & -67 & 0.4 & -26 \\ 
  H53 & Postal & 0.7 & 0.0 & 36 & 100 & 0.1 & 0 & 0.1 & -30 \\ 
  I & Accommodation-Food & 2.9 & -60.8 & 35 & 6 & 8.4 & -80 & 0.7 & -32 \\ 
  J58 & Publishing & 0.6 & -14.4 & 70 & 48 & 0.5 & 0 & 0.6 & -24 \\ 
  J59\_J60 & Video-Sound-Broadcasting & 0.9 & -32.8 & 56 & 34 & 1.1 & 0 & 1.2 & -15 \\ 
  J61 & Telecommunications & 1.6 & -0.9 & 55 & 99 & 1.9 & 0 & 0.8 & -22 \\ 
  J62\_J63 & IT & 2.3 & -0.2 & 71 & 99 & 0.2 & 0 & 2.6 & -13 \\ 
  K64 & Finance & 4.3 & 0.0 & 71 & 100 & 3.2 & 0 & 2.9 & -32 \\ 
  K65 & Insurance & 3.2 & 0.0 & 71 & 100 & 6.3 & 0 & 1.5 & -31 \\ 
  K66 & Auxil. Finance-Insurance & 1.1 & 0.0 & 72 & 100 & 0.2 & 0 & 2.0 & -33 \\ 
  L68 & Real estate & 7.8 & -4.8 & 49 & 6 & 25.3 & 0 & 1.0 & -8 \\ 
  M69\_M70 & Legal & 2.8 & -2.0 & 64 & 94 & 0.0 & 0 & 1.3 & -25 \\ 
  M71 & Architecture-Engineering & 1.7 & 0.0 & 54 & 100 & 0.1 & 0 & 1.4 & -19 \\ 
  M72 & R\&D & 0.5 & 0.0 & 59 & 100 & 0.0 & 0 & 1.1 & -11 \\ 
  M73 & Advertising & 0.6 & -22.5 & 60 & 36 & 0.0 & 0 & 0.3 & -30 \\ 
  M74\_M75 & Other Science & 0.7 & -3.0 & 61 & 94 & 0.3 & 0 & 1.0 & -30 \\ 
  N & Private Administration & 4.4 & -34.9 & 36 & 42 & 1.0 & 0 & 2.7 & -29 \\ 
  O84 & Public Administration & 4.8 & -1.1 & 45 & 97 & 0.6 & 0 & 11.9 & -1 \\ 
  P85 & Education & 4.2 & 0.0 & 54 & 100 & 1.6 & 0 & 8.7 & -2 \\ 
  Q & Health & 7.0 & -0.1 & 36 & 100 & 2.8 & 15 & 15.0 & 1 \\ 
  R\_S & Other Service & 3.2 & -34.5 & 39 & 47 & 5.6 & -5 & 2.5 & -10 \\ 
  T & Household activities & 0.2 & 0.0 &  0 & 0 & 0.8 & 0 & 0.0 & -32 \\ 
   \hline
\end{tabular}
\caption{
{\bf Industry-specific first-order shocks. }
Column $x$ denotes relative shares of gross output, $\epsilon^S$ the supply shock, RLI the Remote Labor Index and ess. the essential score of industries.
Column $c$ represents relative shares of consumer consumption and $\epsilon^D$ the demand shock to consumption.
Column $f$ denotes relative shares of other final consumption (exports, gross capital formation, inventory changes, government) and $f$ shock the shock to other final demand. All values are in \%.
} 
\label{tab:FO_shocks}
\end{table}
\FloatBarrier

\section{Inventory data and calibration} 
\label{apx:inventory}

We could not find UK data to calibrate inventory target parameters $n_j$ in Eq. \eqref{eq:order_interm}. The only reliable data that we could find are those from the US National Income and Product Accounts (NIPA). We describe these data and how we processed them in this appendix. We obtain ratios of level of inventories to monthly sales across industries, with a high level of disaggregation for manufacturing industries, and more uniform information for other industries. Among other things, we show that these ratios are remarkably stable over time. We take this evidence as supporting the idea that inventory to sales ratios are fundamental concepts that have to do with the nature of production rather than with specificities of the US economy. Therefore, we consider these ratios as proxies of $n_j$, after multiplying them by 30 to take into account that we are considering a daily rather than a monthly time scale.

\subsection{Data sources}

All inventory data have been sourced from the Bureau of Economic Analysis (BEA)'s National Income and Product Accounts (NIPA). In particular, we used data from the following tables:

\begin{itemize}
\item Table 5.8.6B. Real Private Inventories and Real Domestic Final Sales by Industry, Chained Dollars - LastRevised: March 26, 2020 [BEA/NIPA-T50806B]
\item Table 208. Real Gross Output by industry, Chained Dollars - LastRevised: April 6, 2020 [BEA/GDPbyIndustry-208Q]
\item Table 1BU. Real Manufacturing and Trade Inventories, Seasonally Adjusted, End of Period [Chained 2012 dollars, 1997 forward, NAICS] - LastRevised: March 26, 2020 [BEA/NIUnderlyingDetail-U001B]
\item Table 2BU. Real Manufacturing and Trade Sales, Seasonally Adjusted at Monthly Rate [Chained 2012 dollars, 1997 forward, NAICS] - LastRevised: March 26, 2020 [BEA/NIUnderlyingDetail-U002BU]
\item Table 4BU1. Real Manufacturing Inventories, by Stage of Fabrication (Materials and supplies), Seasonally Adjusted, End of Period [Chained 2012 dollars, 1997 forward, NAICS] - LastRevised: March 26, 2020 [BEA/NIUnderlyingDetail-U004B1]
\item Table 4BU2. Real Manufacturing Inventories, by Stage of Fabrication (Work-in-process), Seasonally Adjusted, End of Period [Chained 2012 dollars, 1997 forward, NAICS] - LastRevised: March 26, 2020 [BEA/NIUnderlyingDetail-U004B2]
\item Table 4BU3. Real Manufacturing Inventories, by Stage of Fabrication (Finished goods), Seasonally Adjusted, End of Period [Chained 2012 dollars, 1997 forward, NAICS] - LastRevised: March 26, 2020 [BEA/NIUnderlyingDetail-U004B3]
\end{itemize}

For inventory data, we started considering tables 4BU1, 4BU2 and 4BU3, focusing on all 3-digit NAICS manufacturing sectors. We then added information on inventories in trade sectors from table 1BU, focusing on the following industries: Merchant wholesale industries (NAICS 42 except 4251), motor vehicles, parts, and supplies wholesalers (4231), retail trade industries (44-45), motor vehicle and parts dealers (441). We finally added information on inventories in all other industries from table 5.8.6B.\footnote{We also complemented merchant wholesale industries by adding non-merchant wholesale (4251), so as to recover the wholesale sector as a whole.} 

We then used tables 208 and 2BU to extract information on gross output by industry. Information about gross output in wholesale and retail trade is not consistent in the two tables: in table 208 (and several other sources), yearly gross output for these sectors is around 1800 billion dollars; in table 2BU, it is around 6000 billion dollars. We use information from table 208, as, according to the BEA, ``underlying detail'' tables such as 2BU may be less accurate.\footnote{We do not find any other discrepancy, for example information provided in tables 4BU1-4BU2-4BU3 was consistent with more aggregate information in table 1BU, and that information was in turn consistent with data from table 5.8.6B.}

The latest available data (2019Q4) are reported in Table \ref{tab:datainvtab1}, where we also report monthly gross output and the ratio between the level of inventories and monthly gross output. For presentation purposes, we split inventories in certain aggregate sectors according to output shares of subsectors within these sectors --- for example, we disambiguate between mining, utilities and construction, although these are given together in table 5.8.6B.

\begin{figure}[h!]
\includegraphics[width=1\textwidth]{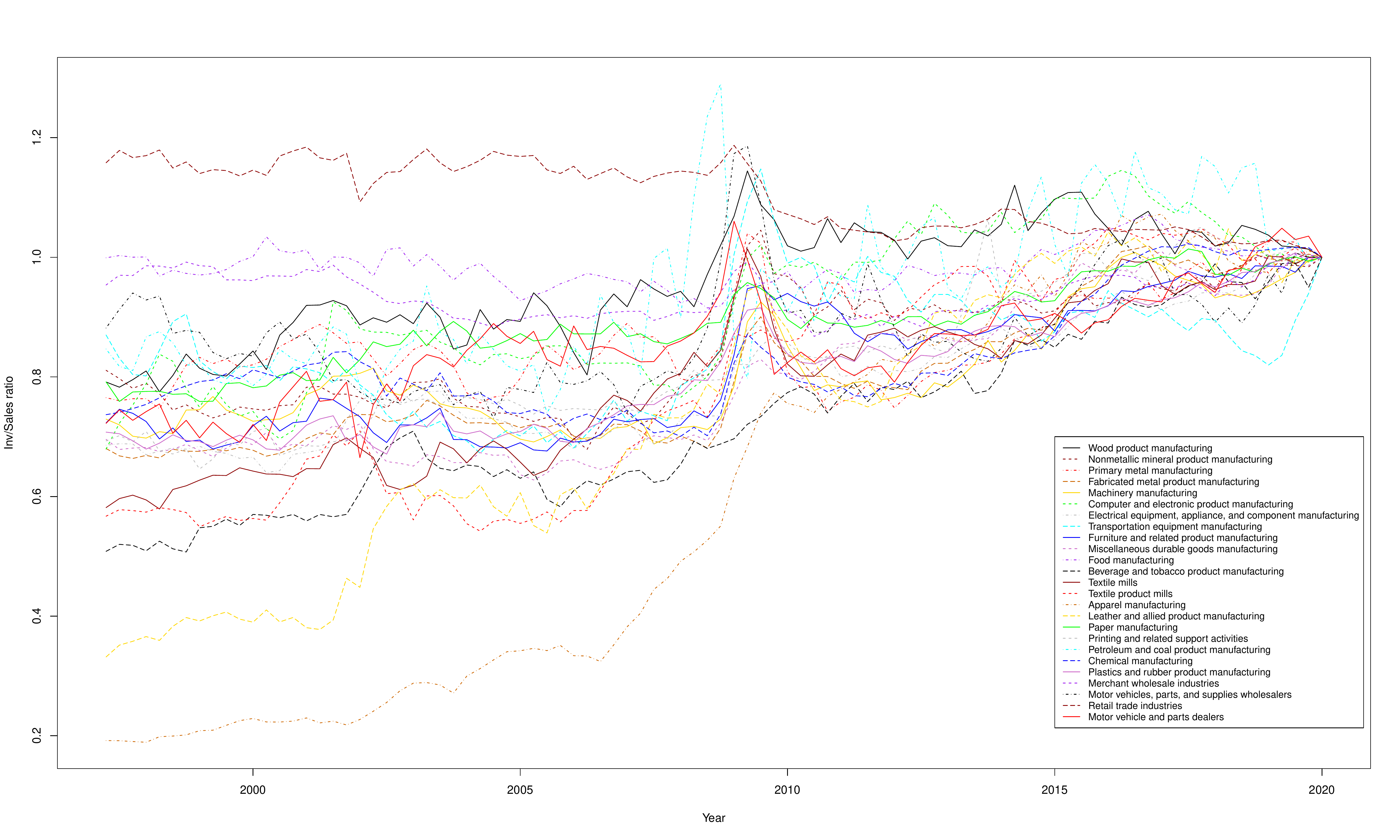}
\centering
\caption{  }
\label{fig:datainvfig1}
\end{figure}

As shown in Fig. \ref{fig:datainvfig1}, the ratios between the level of inventories and monthly gross output are remarkably stable (ratios are normalized to their values in 2019Q4), varying by no more than 20\% in the last 10 years. There appears to be an upward trend from 1997, where the average ratio was around 80\% of the 2019Q4 value, and the only industries whose ratios increased substantially are apparel manufacturing and leather and allied products manufacturing. We view the temporal stability of these ratios as supporting the idea that they can be used for other countries.

\subsection{Mapping to WIOD codes}

We next map data from NAICS to the industrial classification used in the World Input Output Database (WIOD), which is an aggregation of 2-digit International Standard Industrial Classification (ISIC) sectors. Using official concordance tables, which are valid for 4-digit NAICS and ISIC codes, is not the best option, as our data are not available to that level of disaggregation. We resort instead to manual mapping between the NAICS sectors for which we have data and the WIOD sectors. In particular, we use the crosswalk available in Table \ref{tab:datainvcrosswalk}. When a NAICS sector maps uniquely to a WIOD sector, we directly attribute inventory and gross output data. When multiple NAICS sectors map to (one or more) WIOD sectors, we aggregate data for all relevant NAICS sectors. When one or more NAICS sector map to multiple WIOD sectors, we attribute data using as weights sectoral gross outputs from the 2014 WIOD table for the U.S. 

\begin{table}[ht]

\resizebox{1\textwidth}{!}{
{\scriptsize
\centering

\begin{tabular}{|r|L{10cm}|}
  \hline
NAICS & WIOD \\ 
  \hline
321 & C16 \\ 
  327 & C23 \\ 
  331 & C24 \\ 
  332 & C25 \\ 
  333 & C28 \\ 
  334 & C26 \\ 
  335 & C27 \\ 
  336 & C29, C30 \\ 
  337, 339 & C31\_C32, C33 \\ 
  311, 312 & C10-C12 \\ 
  313, 314, 315, 316 & C13-C15 \\ 
  322 & C17 \\ 
  323 & C18 \\ 
  324 & C19 \\ 
  325 & C20, C21 \\ 
  326 & C22 \\ 
  4231\_441 & G45 \\ 
  42-4231 & G46 \\ 
  44\_45-441 & G47 \\ 
  11 & A01, A02, A03 \\ 
  \hline
  21\_\_23 & B, D35, E36, E37-E39, F \\ 
  \hline
  48\_\_92 & H49, H50, H51, H52, H53, I, J58, J59\_J60, J61, J62\_J63, K64, K65, K66, L68, M69\_M70, M71, M72, M73, M74\_M75, N, O84, P85, Q, R\_S, T \\ 
   \hline
\end{tabular}
}
}
\caption{Crosswalk NAICS to WIOD}
\label{tab:datainvcrosswalk}
\end{table}

The results are shown in Table \ref{tab:datainvtab2}. They make sense. As an example, consider the ratio between level of inventories and monthly gross output in the NAICS sectors 313, 314, 315, 316 and in the WIOD sector C13-C15. It is clear that the latter is a weighted average of the former ratios, weighted by the size of the NAICS subsectors.

\newpage

\begin{landscape}

\begin{table}[ht]

\tiny
\centering

\caption{Data are from 2019Q4. Everything in millions of 2012 chained dollars.}
\label{tab:datainvtab1}
\begin{tabular}{llrrrrrrr}
  \hline
naics\_code & naics\_name & stock\_mat\_sup & work\_in\_process & finished\_goods & total\_inv & go\_yearly & go\_monthly & ratio\_inv\_go\_monthly \\ 
  \hline
321 & Wood product manufacturing & 4975 & 2072 & 4638 & 11685 & 93672 & 7806 & 1.50 \\ 
  327 & Nonmetallic mineral product manufacturing & 6497 & 1728 & 8456 & 16681 & 110484 & 9207 & 1.81 \\ 
  331 & Primary metal manufacturing & 15409 & 12215 & 15307 & 42931 & 249984 & 20832 & 2.06 \\ 
  332 & Fabricated metal product manufacturing & 25870 & 20066 & 25938 & 71874 & 368076 & 30673 & 2.34 \\ 
  333 & Machinery manufacturing & 29528 & 18621 & 20667 & 68816 & 353964 & 29497 & 2.33 \\ 
  334 & Computer and electronic product manufacturing & 24361 & 19743 & 16948 & 61052 & 335520 & 27960 & 2.18 \\ 
  335 & Electrical equipment, appliance, and component manufacturing & 10743 & 7907 & 7457 & 26107 & 132144 & 11012 & 2.37 \\ 
  336 & Transportation equipment manufacturing & 34067 & 98210 & 17703 & 149980 & 936336 & 78028 & 1.92 \\ 
  337 & Furniture and related product manufacturing & 5057 & 1883 & 3538 & 10478 & 71112 & 5926 & 1.77 \\ 
  339 & Miscellaneous durable goods manufacturing & 11069 & 6560 & 15186 & 32815 & 151440 & 12620 & 2.60 \\ 
  311 & Food manufacturing & 21590 & 9238 & 38713 & 69541 & 801432 & 66786 & 1.04 \\ 
  312 & Beverage and tobacco product manufacturing & 8981 & 5681 & 7141 & 21803 & 124788 & 10399 & 2.10 \\ 
  313 & Textile mills & 2056 & 1179 & 2721 & 5956 & 28332 & 2361 & 2.52 \\ 
  314 & Textile product mills & 1419 & 683 & 2296 & 4398 & 23880 & 1990 & 2.21 \\ 
  315 & Apparel manufacturing & 2127 & 1496 & 5193 & 8816 & 10692 & 891 & 9.89 \\ 
  316 & Leather and allied product manufacturing & 645 & 311 & 1288 & 2244 & 4572 & 381 & 5.89 \\ 
  322 & Paper manufacturing & 11145 & 2110 & 8814 & 22069 & 174264 & 14522 & 1.52 \\ 
  323 & Printing and related support activities & 2820 & 1163 & 3137 & 7120 & 76968 & 6414 & 1.11 \\ 
  324 & Petroleum and coal product manufacturing & 18917 & 20958 & 32313 & 72188 & 1033320 & 86110 & 0.84 \\ 
  325 & Chemical manufacturing & 41896 & 23577 & 52226 & 117699 & 734484 & 61207 & 1.92 \\ 
  326 & Plastics and rubber product manufacturing & 14955 & 3936 & 15273 & 34164 & 223452 & 18621 & 1.83 \\ 
  42 & Wholesale industries & 823900 & 0 & 0 & 823900 & 1861400 & 155117 & 5.31 \\ 
  4231 & Motor vehicles, parts, and supplies wholesalers & 63705 & 0 & 0 & 63705 & 138677 & 11556 & 5.51 \\ 
  44-45 & Retail trade industries & 674155 & 0 & 0 & 674155 & 1844000 & 153667 & 4.39 \\ 
  441 & Motor vehicle and parts dealers & 257911 & 0 & 0 & 257911 & 363100 & 30258 & 8.52 \\ 
  111-113 & Farms & 0 & 0 & 174416 & 174416 & 487200 & 40600 & 4.30 \\ 
  114-115 & Forestry, fishing, and related activities & 0 & 0 & 19368 & 19368 & 54100 & 4508 & 4.30 \\ 
  21 & Mining & 0 & 0 & 32662 & 32662 & 732800 & 61067 & 0.53 \\ 
  22 & Utilities & 0 & 0 & 20316 & 20316 & 455800 & 37983 & 0.53 \\ 
  23 & Construction & 60363 & 0 & 0 & 60363 & 1354300 & 112858 & 0.53 \\ 
  48-49 & Transportation and warehousing & 10737 & 0 & 0 & 10737 & 1163000 & 96917 & 0.11 \\ 
  51 & Information & 18463 & 0 & 0 & 18463 & 1999800 & 166650 & 0.11 \\ 
  52-53 & Finance, insurance, real estate, rental, and leasing & 55949 & 0 & 0 & 55949 & 6060000 & 505000 & 0.11 \\ 
  54-56 & Professional and business services & 37009 & 0 & 0 & 37009 & 4008500 & 334042 & 0.11 \\ 
  61-62 & Educational services, health care, and social assistance & 25173 & 0 & 0 & 25173 & 2726500 & 227208 & 0.11 \\ 
  71-72-81 & Arts, entertainment, recreation, accommodation, and food services & 12343 & 0 & 0 & 12343 & 1336900 & 111408 & 0.11 \\ 
  92 & Government & 33553 & 0 & 0 & 33553 & 3634200 & 302850 & 0.11 \\ 
   \hline
\end{tabular}

\end{table}

\begin{table}[ht]
\tiny
\centering
\caption{Data are from 2019Q4. Everything in millions of 2012 chained dollars.}
\label{tab:datainvtab2}
\begin{tabular}{lL{9cm}rrrrrrr}
  \hline
wiod\_code & wiod\_name & stock\_mat\_sup & work\_in\_process & finished\_goods & go\_yearly & tot\_inv & go\_monthly & ratio\_inv\_go\_monthly \\ 
  \hline
A01 & Crop and animal production, hunting and related service activities & 0 & 0 & 172754 & 482557 & 172754 & 40213 & 4.30 \\ 
  A02 & Forestry and logging & 0 & 0 & 13056 & 36470 & 13056 & 3039 & 4.30 \\ 
  A03 & Fishing and aquaculture & 0 & 0 & 7974 & 22273 & 7974 & 1856 & 4.30 \\ 
  B & Mining and quarrying & 0 & 0 & 31740 & 712120 & 31740 & 59343 & 0.53 \\ 
  C10-C12 & Manufacture of food products, beverages and tobacco products & 30571 & 14919 & 45854 & 926220 & 91344 & 77185 & 1.18 \\ 
  C13-C15 & Manufacture of textiles, wearing apparel and leather products & 6247 & 3669 & 11498 & 67476 & 21414 & 5623 & 3.81 \\ 
  C16 & Manufacture of wood and of products of wood and cork, except furniture; manufacture of articles of straw and plaiting materials & 4975 & 2072 & 4638 & 93672 & 11685 & 7806 & 1.50 \\ 
  C17 & Manufacture of paper and paper products & 11145 & 2110 & 8814 & 174264 & 22069 & 14522 & 1.52 \\ 
  C18 & Printing and reproduction of recorded media & 2820 & 1163 & 3137 & 76968 & 7120 & 6414 & 1.11 \\ 
  C19 & Manufacture of coke and refined petroleum products  & 18917 & 20958 & 32313 & 1033320 & 72188 & 86110 & 0.84 \\ 
  C20 & Manufacture of chemicals and chemical products  & 30865 & 17369 & 38475 & 541095 & 86709 & 45091 & 1.92 \\ 
  C21 & Manufacture of basic pharmaceutical products and pharmaceutical preparations & 11031 & 6208 & 13751 & 193389 & 30990 & 16116 & 1.92 \\ 
  C22 & Manufacture of rubber and plastic products & 14955 & 3936 & 15273 & 223452 & 34164 & 18621 & 1.83 \\ 
  C23 & Manufacture of other non-metallic mineral products & 6497 & 1728 & 8456 & 110484 & 16681 & 9207 & 1.81 \\ 
  C24 & Manufacture of basic metals & 15409 & 12215 & 15307 & 249984 & 42931 & 20832 & 2.06 \\ 
  C25 & Manufacture of fabricated metal products, except machinery and equipment & 25870 & 20066 & 25938 & 368076 & 71874 & 30673 & 2.34 \\ 
  C26 & Manufacture of computer, electronic and optical products & 24361 & 19743 & 16948 & 335520 & 61052 & 27960 & 2.18 \\ 
  C27 & Manufacture of electrical equipment & 10743 & 7907 & 7457 & 132144 & 26107 & 11012 & 2.37 \\ 
  C28 & Manufacture of machinery and equipment n.e.c. & 29528 & 18621 & 20667 & 353964 & 68816 & 29497 & 2.33 \\ 
  C29 & Manufacture of motor vehicles, trailers and semi-trailers & 21551 & 62129 & 11199 & 592339 & 94879 & 49362 & 1.92 \\ 
  C30 & Manufacture of other transport equipment & 12516 & 36081 & 6504 & 343997 & 55101 & 28666 & 1.92 \\ 
  C31\_C32 & Manufacture of furniture; other manufacturing & 13924 & 7290 & 16168 & 192166 & 37382 & 16014 & 2.33 \\ 
  C33 & Repair and installation of machinery and equipment & 2202 & 1153 & 2556 & 30386 & 5911 & 2532 & 2.33 \\ 
  D35 & Electricity, gas, steam and air conditioning supply & 0 & 0 & 19237 & 431602 & 19237 & 35967 & 0.53 \\ 
  E36 & Water collection, treatment and supply & 0 & 0 & 657 & 14746 & 657 & 1229 & 0.53 \\ 
  E37-E39 & Sewerage; waste collection, treatment and disposal activities; materials recovery; remediation activities and other waste management services  & 0 & 0 & 4371 & 98070 & 4371 & 8173 & 0.53 \\ 
  F & Construction & 57335 & 0 & 0 & 1286362 & 57335 & 107197 & 0.53 \\ 
  G45 & Wholesale and retail trade and repair of motor vehicles and motorcycles & 321616 & 0 & 0 & 501777 & 321616 & 41815 & 7.69 \\ 
  G46 & Wholesale trade, except of motor vehicles and motorcycles & 760195 & 0 & 0 & 1722723 & 760195 & 143560 & 5.30 \\ 
  G47 & Retail trade, except of motor vehicles and motorcycles & 416244 & 0 & 0 & 1480900 & 416244 & 123408 & 3.37 \\ 
  H49 & Land transport and transport via pipelines & 5299 & 0 & 0 & 573956 & 5299 & 47830 & 0.11 \\ 
  H50 & Water transport & 653 & 0 & 0 & 70739 & 653 & 5895 & 0.11 \\ 
  H51 & Air transport & 1957 & 0 & 0 & 211979 & 1957 & 17665 & 0.11 \\ 
  H52 & Warehousing and support activities for transportation & 2031 & 0 & 0 & 219937 & 2031 & 18328 & 0.11 \\ 
  H53 & Postal and courier activities & 1144 & 0 & 0 & 123877 & 1144 & 10323 & 0.11 \\ 
  I & Accommodation and food service activities & 9256 & 0 & 0 & 1002493 & 9256 & 83541 & 0.11 \\ 
  J58 & Publishing activities & 3427 & 0 & 0 & 371174 & 3427 & 30931 & 0.11 \\ 
  J59\_J60 & Motion picture, video and television programme production, sound recording and music publishing activities; programming and broadcasting activities & 3351 & 0 & 0 & 362983 & 3351 & 30249 & 0.11 \\ 
  J61 & Telecommunications & 6735 & 0 & 0 & 729434 & 6735 & 60786 & 0.11 \\ 
  J62\_J63 & Computer programming, consultancy and related activities; information service activities & 5816 & 0 & 0 & 629926 & 5816 & 52494 & 0.11 \\ 
  K64 & Financial service activities, except insurance and pension funding & 7307 & 0 & 0 & 791410 & 7307 & 65951 & 0.11 \\ 
  K65 & Insurance, reinsurance and pension funding, except compulsory social security & 10347 & 0 & 0 & 1120663 & 10347 & 93389 & 0.11 \\ 
  K66 & Activities auxiliary to financial services and insurance activities & 4967 & 0 & 0 & 537998 & 4967 & 44833 & 0.11 \\ 
  L68 & Real estate activities & 28803 & 0 & 0 & 3119741 & 28803 & 259978 & 0.11 \\ 
  M69\_M70 & Legal and accounting activities; activities of head offices; management consultancy activities & 11781 & 0 & 0 & 1276047 & 11781 & 106337 & 0.11 \\ 
  M71 & Architectural and engineering activities; technical testing and analysis & 4626 & 0 & 0 & 501018 & 4626 & 41752 & 0.11 \\ 
  M72 & Scientific research and development & 2492 & 0 & 0 & 269945 & 2492 & 22495 & 0.11 \\ 
  M73 & Advertising and market research & 2510 & 0 & 0 & 271811 & 2510 & 22651 & 0.11 \\ 
  M74\_M75 & Other professional, scientific and technical activities; veterinary activities & 819 & 0 & 0 & 88716 & 819 & 7393 & 0.11 \\ 
  N & Administrative and support service activities & 11268 & 0 & 0 & 1220478 & 11268 & 101706 & 0.11 \\ 
  O84 & Public administration and defence; compulsory social security & 35648 & 0 & 0 & 3861125 & 35648 & 321760 & 0.11 \\ 
  P85 & Education & 3317 & 0 & 0 & 359267 & 3317 & 29939 & 0.11 \\ 
  Q & Human health and social work activities & 21640 & 0 & 0 & 2343861 & 21640 & 195322 & 0.11 \\ 
  R\_S & Other service activities & 7819 & 0 & 0 & 846949 & 7819 & 70579 & 0.11 \\ 
  T & Activities of households as employers; undifferentiated goods- and services-producing activities of households for own use & 216 & 0 & 0 & 23374 & 216 & 1948 & 0.11 \\ 
  U & Activities of extraterritorial organizations and bodies &  &  &  &  &  &  &  \\ 
   \hline
\end{tabular}
\end{table}

\end{landscape}
\FloatBarrier

\section{Critical vs. non-critical inputs} 
\label{apx:ihs}

A survey was designed to address the question when production can continue during a lockdown.
For each industry, IHS Markit analysts were asked to rate every input of a given industry. The exact formulation of the question was as follows:
``For each industry in WIOD, please rate whether each of its inputs are essential. We will present you with an industry X and ask you to rate each input Y. The key question is: Can production continue in industry X if input Y is not available for two months?''
Analysts could rate each input according to the following allowed answers:
\begin{itemize}
    \item \textbf{0} -- This input is \textit{not} essential
    \item \textbf{1} -- This input is essential
    \item \textbf{0.5} -- This input is important but not essential
    \item \textbf{NA} -- I have no idea
\end{itemize}
To avoid confusion with the unrelated definition of essential industries which we used to calibrate first-order supply shocks, we refer to inputs as \textit{critical} and \textit{non-critical} instead of \textit{essential} and \textit{non-essential.}

Analysts were provided with the share of each input in the expenses of the industry. It was also made explicit that the ratings assume no inventories such that a rating captures the effect on production if the input is not available.

Every industry was rated by one analyst, except for industries Mining and Quarrying (B) and Manufacture of Basic Metals (C24) which were rated by three analysts. To improve input ratings, we aim to increase the sample size of analyst ratings for every industry in the next few weeks. In case there are several ratings we took the average of the ratings and rounded it to 1 if the average was at least $2/3$ and 0 if the average was at most $1/3$. Average input ratings lying between these boundaries are assigned the value 0.5.

\begin{figure}[!h]
\centering
\includegraphics[trim = {0cm 0cm 0cm 0cm}, clip,width=\textwidth]{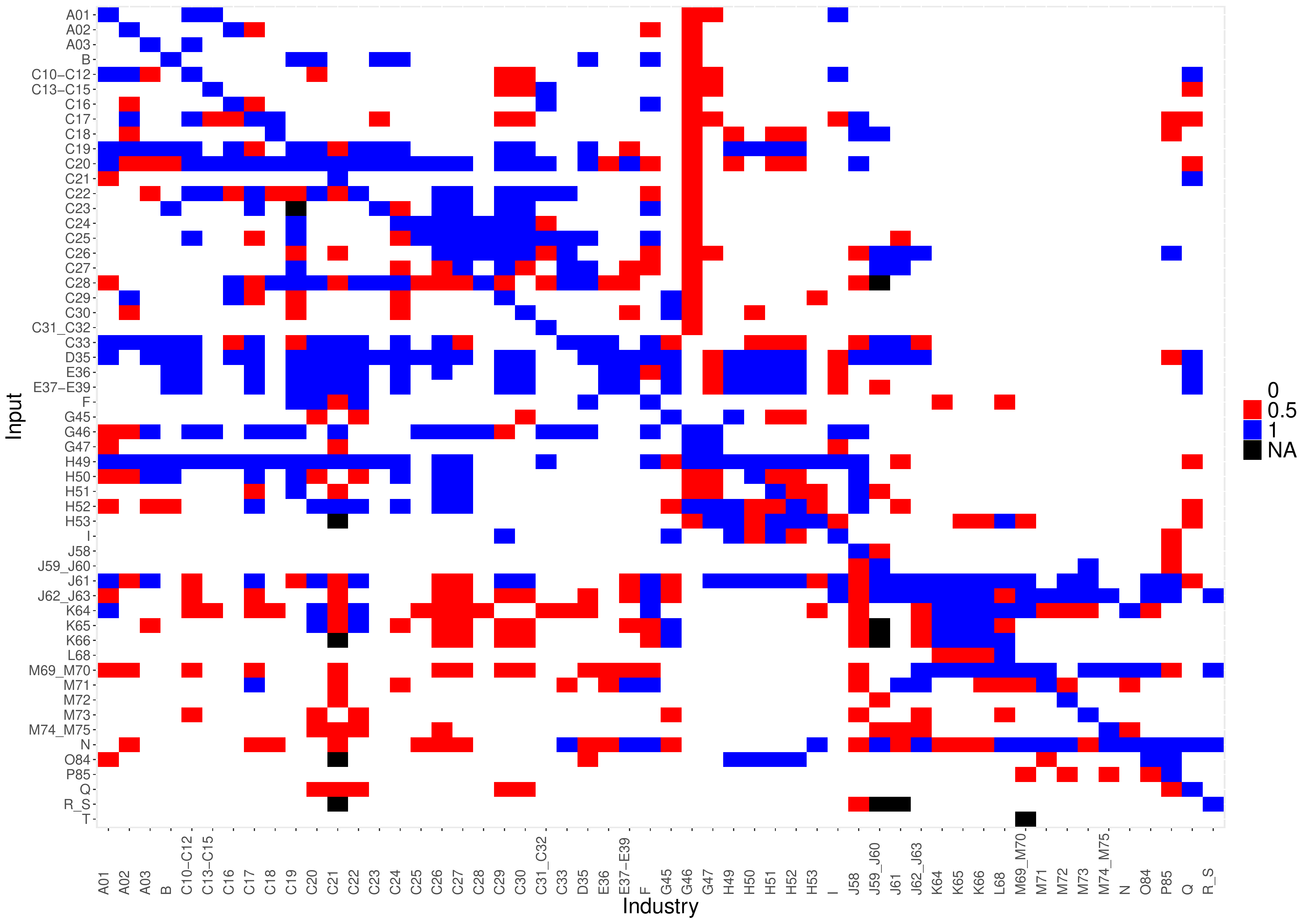}
    \caption{Criticality scores of IHS Markit analysts. 
    Rows are inputs (supply) and columns industries using these inputs (demand).
    The blue color indicates critical (score=1), red important (score=0.5) and white non-critical (score=0) inputs. Black denotes inputs which have been rated with NA. The diagonal elements are considered to be critical by definition. For industries with multiple input-ratings we took the average of all ratings and assigned a score=1 if the averaged score was at least $2/3$ and a score=0 if the average was smaller or equal to $1/3$. 
    }
    \label{fig:ihs_matrix}
\end{figure}

The ratings for each industry and input are depicted in Fig. \ref{fig:ihs_matrix}. A column denotes an industry and the corresponding rows its inputs. 
Blue colors indicate \textit{critical}, red \textit{important, but not critical} and white \textit{non-critical} inputs. Note that under the assumption of a Leontief production function every element would be considered to be critical, yielding a completely blue-colored matrix. The results shown here indicate that the majority of elements are non-critical inputs (2,338 ratings with score = 0), whereas only 477 industry-inputs are rates as critical. 365 inputs are rates as important, although not critical (score = 0.5) and \textit{NA} was assigned eleven times.

\begin{figure}[!h]
\centering
\includegraphics[trim = {0cm 0cm 0cm 0cm}, clip,width=1\textwidth]{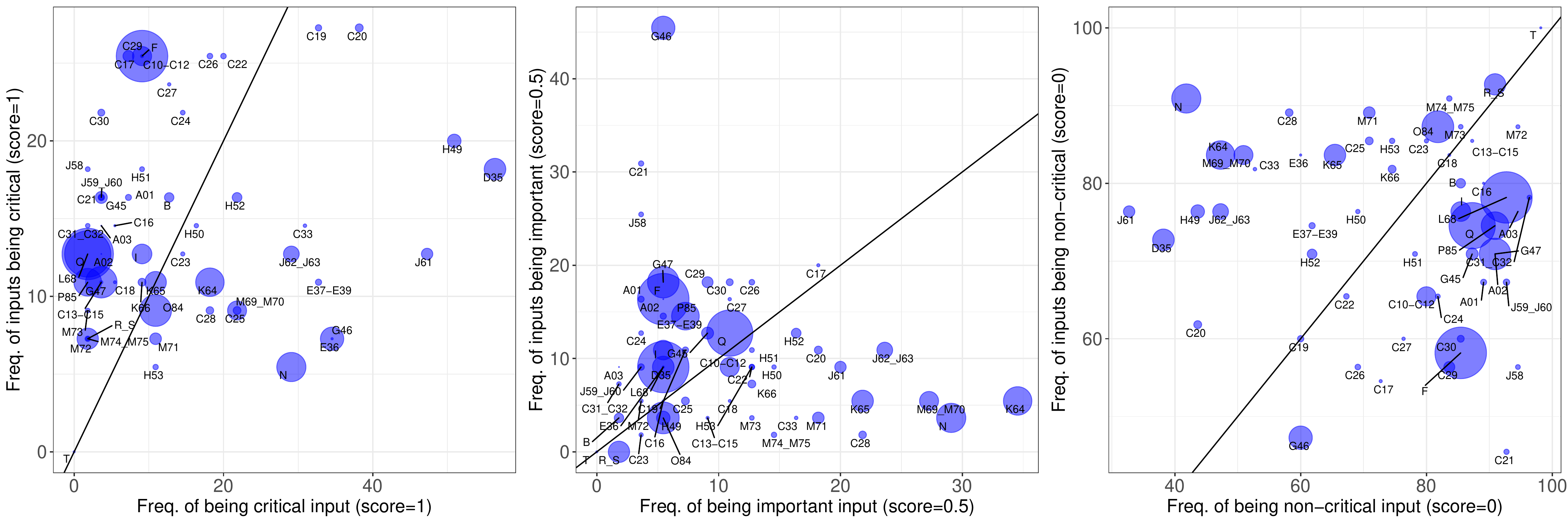}
    \caption{(Left panel) Plotting how often an industry is rated as an critical input to other industries (x-axis) against the share of critical inputs this industry is using.
    The center and right panel are the same as the left panel, except for using half-critical and non-critical scores, respectively. In each plot the identity line is shown. Point sizes are proportional to gross output.
    }
    \label{fig:ihs_scatter}
\end{figure}

The left panel of Fig. \ref{fig:ihs_scatter} shows for each industry how often it was rated as critical input to other industries (x-axis) and how many critical inputs this industry relies on in its own production (y-axis). Electricity and Gas (D35) are rated most frequently as critical inputs in the production of other industries (score=1 for almost 60\% of industries). Also frequently rated as critical are Land Transport (H49) and Telecommunications (J61). 
On the other hand, many manufacturing industries (ISIC codes starting with C) stand out as relying on a large number of critical inputs. For example, around 27\% of inputs to Manufacture of Coke and Refined Petroleum Products (C19) as well as to Manufacture of Chemicals (C20) are rated as critical.

The center panel of Fig. \ref{fig:ihs_scatter} shows the equivalent plot for 0.5 ratings (important, but not critical inputs). Financial Services (K64) are most frequently rated as important inputs which do not necessarily stop the production of an industry if not available.
Conversely, the industry relying on many important, but not binding inputs is Wholesale and Retail Trade (G46) of which almost half of its inputs got rated with a score = 0.5.
This makes sense given that this industry heavily relies on all these inputs, but lacking one of these does not halt economic production. This case also illustrates that a Leontief production function could starkly overestimated input bottlenecks as Wholesale and Retail Trade would most likely still be able to realize output even if a several inputs would not be available.

In the right panel of Fig. \ref{fig:ihs_scatter} we show the same scatter plot but for non-critical inputs. 25 industries are rated to be non-critical inputs to other industries in 80\% of all cases, with Household Activities (T) and Manufacture of Furniture (C31-32) being rated as non-critical in at least 96\%.
Industries like Other Services (R-S), Other Professional, Scientific and Technical Activities (M74-75) and Administrative Activities (N) rely on mostly non-critical inputs ($>$90\%).

A detailed breakdown of the input- and industry-specific ratings are given in Table \ref{tab:ihs_results}.

\begin{table}[!h]
\tiny
\centering
\resizebox{1\textwidth}{!}{

\begin{tabular}{|cl|cccc|cccc|c|}
  \hline
  & & 
  \multicolumn{4}{c|}{ Input-based rankings  } &
  \multicolumn{4}{c|}{ Industry-based rankings} & \\
ISIC & Sector (abbreviated) & $1$ & $0.5$ & $0$ & NA & $1$ & $0.5$ & $0$ & NA & n \\ 
  \hline
A01 & Agriculture & 4 & 2 & 49 & 0 & 9 & 9 & 37 & 0 & 1 \\ 
  A02 & Foresty & 2 & 3 & 50 & 0 & 7 & 9 & 39 & 0 & 1 \\ 
  A03 & Fishing & 2 & 1 & 52 & 0 & 8 & 5 & 42 & 0 & 1 \\ 
  B & Mining & 7 & 1 & 47 & 0 & 9 & 2 & 44 & 0 & 3 \\ 
  C10-C12 & Manuf. Food-Beverages & 5 & 6 & 44 & 0 & 14 & 5 & 36 & 0 & 1 \\ 
  C13-C15 & Manuf. Textiles & 2 & 5 & 48 & 0 & 6 & 2 & 47 & 0 & 1 \\ 
  C16 & Manuf. Wood & 3 & 3 & 49 & 0 & 8 & 3 & 44 & 0 & 1 \\ 
  C17 & Manuf. Paper & 5 & 10 & 40 & 0 & 14 & 11 & 30 & 0 & 1 \\ 
  C18 & Media print & 3 & 6 & 46 & 0 & 6 & 3 & 46 & 0 & 1 \\ 
  C19 & Manuf. Coke-Petroleum & 18 & 4 & 33 & 0 & 15 & 6 & 33 & 2 & 1 \\ 
  C20 & Manuf. Chemical & 21 & 10 & 24 & 0 & 15 & 6 & 34 & 0 & 1 \\ 
  C21 & Manuf. Pharmaceutical & 2 & 2 & 51 & 0 & 9 & 17 & 25 & 7 & 1 \\ 
  C22 & Manuf. Rubber-Plastics & 11 & 7 & 37 & 0 & 14 & 5 & 36 & 0 & 1 \\ 
  C23 & Manuf. Minerals & 8 & 2 & 44 & 2 & 7 & 1 & 47 & 0 & 1 \\ 
  C24 & Manuf. Metals-basic & 8 & 2 & 45 & 0 & 12 & 7 & 36 & 0 & 3 \\ 
  C25 & Manuf. Metals-fabricated & 12 & 4 & 39 & 0 & 5 & 3 & 47 & 0 & 1 \\ 
  C26 & Manuf. Electronic & 10 & 7 & 38 & 0 & 14 & 10 & 31 & 0 & 1 \\ 
  C27 & Manuf. Electric & 7 & 6 & 42 & 0 & 13 & 9 & 33 & 0 & 1 \\ 
  C28 & Manuf. Machinery & 10 & 12 & 32 & 2 & 5 & 1 & 49 & 0 & 1 \\ 
  C29 & Manuf. Vehicles & 4 & 5 & 46 & 0 & 14 & 10 & 31 & 0 & 1 \\ 
  C30 & Manuf. Transport-other & 2 & 6 & 47 & 0 & 12 & 10 & 33 & 0 & 1 \\ 
  C31\_C32 & Manuf. Furniture & 1 & 1 & 53 & 0 & 8 & 4 & 43 & 0 & 1 \\ 
  C33 & Repair-Installation & 17 & 9 & 29 & 0 & 8 & 2 & 45 & 0 & 1 \\ 
  D35 & Electricity-Gas & 31 & 3 & 21 & 0 & 10 & 5 & 40 & 0 & 1 \\ 
  E36 & Water & 19 & 3 & 33 & 0 & 4 & 5 & 46 & 0 & 1 \\ 
  E37-E39 & Sewage & 18 & 3 & 34 & 0 & 6 & 8 & 41 & 0 & 1 \\ 
  F & Construction & 5 & 3 & 47 & 0 & 14 & 9 & 32 & 0 & 1 \\ 
  G45 & Vehicle trade & 2 & 5 & 48 & 0 & 9 & 7 & 39 & 0 & 1 \\ 
  G46 & Wholesale & 19 & 3 & 33 & 0 & 4 & 25 & 26 & 0 & 1 \\ 
  G47 & Retail & 2 & 3 & 50 & 0 & 6 & 10 & 39 & 0 & 1 \\ 
  H49 & Land transport & 28 & 3 & 24 & 0 & 11 & 2 & 42 & 0 & 1 \\ 
  H50 & Water transport & 9 & 8 & 38 & 0 & 8 & 5 & 42 & 0 & 1 \\ 
  H51 & Air transport & 5 & 7 & 43 & 0 & 10 & 6 & 39 & 0 & 1 \\ 
  H52 & Warehousing & 12 & 9 & 34 & 0 & 9 & 7 & 39 & 0 & 1 \\ 
  H53 & Postal & 6 & 7 & 41 & 2 & 3 & 5 & 47 & 0 & 1 \\ 
  I & Accommodation-Food & 5 & 3 & 47 & 0 & 7 & 6 & 42 & 0 & 1 \\ 
  J58 & Publishing & 1 & 2 & 52 & 0 & 10 & 14 & 31 & 0 & 1 \\ 
  J59\_J60 & Video-Sound-Broadcasting & 2 & 2 & 51 & 0 & 9 & 5 & 37 & 7 & 1 \\ 
  J61 & Telecommunications & 26 & 11 & 18 & 0 & 7 & 5 & 42 & 2 & 1 \\ 
  J62\_J63 & IT & 16 & 13 & 26 & 0 & 7 & 6 & 42 & 0 & 1 \\ 
  K64 & Finance & 10 & 19 & 26 & 0 & 6 & 3 & 46 & 0 & 1 \\ 
  K65 & Insurance & 6 & 12 & 36 & 2 & 6 & 3 & 46 & 0 & 1 \\ 
  K66 & Auxil. Finance-Insurance & 5 & 7 & 41 & 4 & 6 & 4 & 45 & 0 & 1 \\ 
  L68 & Real estate & 1 & 3 & 51 & 0 & 7 & 5 & 43 & 0 & 1 \\ 
  M69\_M70 & Legal & 12 & 15 & 28 & 0 & 5 & 3 & 46 & 2 & 1 \\ 
  M71 & Architecture-Engineering & 6 & 10 & 39 & 0 & 4 & 2 & 49 & 0 & 1 \\ 
  M72 & R\&D & 1 & 2 & 52 & 0 & 4 & 3 & 48 & 0 & 1 \\ 
  M73 & Advertising & 1 & 7 & 47 & 0 & 5 & 2 & 48 & 0 & 1 \\ 
  M74\_M75 & Other Science & 1 & 8 & 46 & 0 & 4 & 1 & 50 & 0 & 1 \\ 
  N & Private Administration & 16 & 16 & 23 & 0 & 3 & 2 & 50 & 0 & 1 \\ 
  O84 & Public Administration & 6 & 3 & 45 & 2 & 5 & 2 & 48 & 0 & 1 \\ 
  P85 & Education & 1 & 4 & 50 & 0 & 6 & 8 & 41 & 0 & 1 \\ 
  Q & Health & 1 & 6 & 48 & 0 & 7 & 7 & 41 & 0 & 1 \\ 
  R\_S & Other Service & 1 & 1 & 50 & 5 & 4 & 0 & 51 & 0 & 1 \\ 
  T & Household activities & 0 & 0 & 54 & 2 & 0 & 0 & 55 & 0 & 0 \\ 
   \hline
\end{tabular}
}
\caption{Summary table of critical input ratings by IHS Markit analysts.
Columns below \textit{Input-based rankings} show how often an industry has been rated as critical (score=1), half-critical (score=0.5) or non-critical (score=0) input for other industries, or how often the input was rates as NA.
Columns under \textit{Industry-based rankings} give how often an input has been rated as with 1, 0.5, 0 or NA for any given industry.
Column $n$ indicates the number of analysts who have rated the inputs of any given industry. Industry \textit{T} uses no inputs and is therefore not rated.
} 
\label{tab:ihs_results}

\end{table}

\FloatBarrier

\FloatBarrier

\section{Sensitivity analysis}
\label{apx:senstivity}

In this appendix we perform sensitivity analysis of the economic model with respect to both supply and demand shocks (Appendices \ref{apx:shock_uncertainty} and \ref{apx:supply_demand}) and model parameters and assumptions (Appendices \ref{apx:prodfun} and \ref{apx:model_sensitivity}). For the latter, we follow a one-at-a-time sensitivity analysis approach \citep{borgonovo2016sensitivity}, in the sense that we start from the baseline scenario described in the main text and vary some assumptions while holding all other assumptions fixed to the baseline scenario (see Table \ref{tab:scenarios}). Further, in Appendix \ref{apx:sensitivity_correlation} we show how the various scenarios compare in terms of matching sectoral unemployment data from the U.S. states of Washington and Texas, see Section \ref{sec:econimpact}. Finally, in Appendix \ref{apx:io_models}, we compare our model results to those of traditional input-output models, namely the Leontief and Gosh models.

\begin{table}[ht]
\centering
\tiny
\begin{tabular}{lrllrrl}
  \hline
Scenario names & $\tau$ & production function & hiring-firing & $\gamma_H$ & $\gamma_F$ & cons. function \\ 
  \hline
baseline & $\tau=10$ & Leontief, important inputs non-critical & yes & 0.03 & 0.07 & muellbauer \\ 
  $\tau=1$ & $\tau=1$ & Leontief, important inputs non-critical & yes & 0.03 & 0.07 & muellbauer \\ 
  $\tau=5$ & $\tau=5$ & Leontief, important inputs non-critical & yes & 0.03 & 0.07 & muellbauer \\ 
  $\tau=15$ & $\tau=15$ & Leontief, important inputs non-critical & yes & 0.03 & 0.07 & muellbauer \\ 
  $\tau=20$ & $\tau=20$ & Leontief, important inputs non-critical & yes & 0.03 & 0.07 & muellbauer \\ 
  linear & $\tau=10$ & linear & yes & 0.03 & 0.07 & muellbauer \\ 
  leontief & $\tau=10$ & Leontief & yes & 0.03 & 0.07 & muellbauer \\ 
  important inputs critical & $\tau=10$ & Leontief, important inputs critical & yes & 0.03 & 0.07 & muellbauer \\ 
 important inputs half-critical & $\tau=10$ & Leontief, important inputs half-critical & yes & 0.03 & 0.07 & muellbauer \\ 
  no hiring-firing & $\tau=10$ & Leontief, important inputs non-critical & no & 0.03 & 0.07 & muellbauer \\ 
  fast labor adjustment & $\tau=10$ & Leontief, important inputs non-critical & yes & 0.50 & 1.00 & muellbauer \\ 
  slow labor adjustment & $\tau=10$ & Leontief, important inputs non-critical & yes & 0.01 & 0.02 & muellbauer \\ 
  fixed consumption & $\tau=10$ & Leontief, important inputs non-critical & yes & 0.03 & 0.07 & fixed \\ 
  keynesian consumption & $\tau=10$ & Leontief, important inputs non-critical & yes & 0.03 & 0.07 & keynesian \\ 
   \hline
\end{tabular}
\caption{Scenarios for model parameters and assumptions considered in this sensitivity analysis. See Sections Sections \ref{apx:prodfun} and \ref{apx:model_sensitivity} for a more detailed description of the various specifications.}
\label{tab:scenarios}
\end{table}

\subsection{First-order shocks uncertainty} \label{apx:shock_uncertainty}

Since there is substantial uncertainties in first-order shocks discussed in Section \ref{sec:pandemic_shock}, we test how sensitive model results are with respect to the shock initialisation considered in the main text. 
To do this, we first randomly perturb the supply and demand shocks for every industry. More specifically, we create perturbed supply and demand shock vectors by letting
\begin{equation} \label{eq:supplyshock_perturb}
    \bar{ \epsilon}^S_{i,0} = { \epsilon}^S_{i,0} (1+ \psi_i^S),
\end{equation}
and 
\begin{equation} \label{eq:demandshock_perturb}
    \bar{ \epsilon}_{i,0}^D = \epsilon_{i,0}^D (1+ \psi_i^D),
\end{equation}
where $\psi_i^S,\psi_i^D \sim N (0, \sigma)$.
We use different values for standard deviation, $\sigma \in \{0.01, 0.1, 0.2 \}$, representing a normal randomization of original values by 1-20\% standard deviations. 
We then initialise the model with the perturbed first-order shocks and run the lockdown simulations. 
We repeat this procedure 1,000 times and report median values, interquartile range (IQR) and the 95\% confidence bounds of aggregate output values. We did not investigate perturbing other final demand $f_{i,0}^d$.

The upper left panel of Fig. \ref{fig:shocked_shocks} presents the result of this analysis. Since results are qualitatively very similar for the explored standard deviation specifications, we only show the largest perturbation case with $\sigma=0.2$.
First note that the default model result (red line) follows very closely the median result (black line). Also, the IQR is only a narrow band around the reported default values. These results are reassuring as they indicate strong robustness of the model result against uncertainty in initial shock values for a large range of simulations. This picture changes when considering the 95\% confidence bounds instead. Here, the ribbon expands dramatically towards small values after around 110 time steps. This finding suggests that for a certain range of initial shocks our model would predict a substantial collapse of the economy. Since we do not observe similar nonlinearities for the IQR, this initial shock arrangement is not particularly likely given our estimates represent reasonable expected values of the ``true'' shocks. Also, the economic collapse happens only after almost four months of lockdown, a much longer time horizon as considered in the simulations for the main results. 
Nevertheless, the results emphasizes the importance of nonlinearities in the economic system by demonstrating how related initial economic shocks can be amplified in very different ways.

The upper right and lower panels of Fig. \ref{fig:shocked_shocks} show the same simulations but using exclusively perturbations on the supply and demand side, respectively. It is immediately evident that the large confidence bounds after four months of lockdown are driven by the supply side shock uncertainty. When perturbing only demand shocks and setting supply shocks to the default values (lower panel), there is very little variance in our model prediction.

\subsection{Economic impact of supply and demand shocks} \label{apx:supply_demand}

\begin{figure}[H]
    \centering
\includegraphics[width = 1\textwidth]{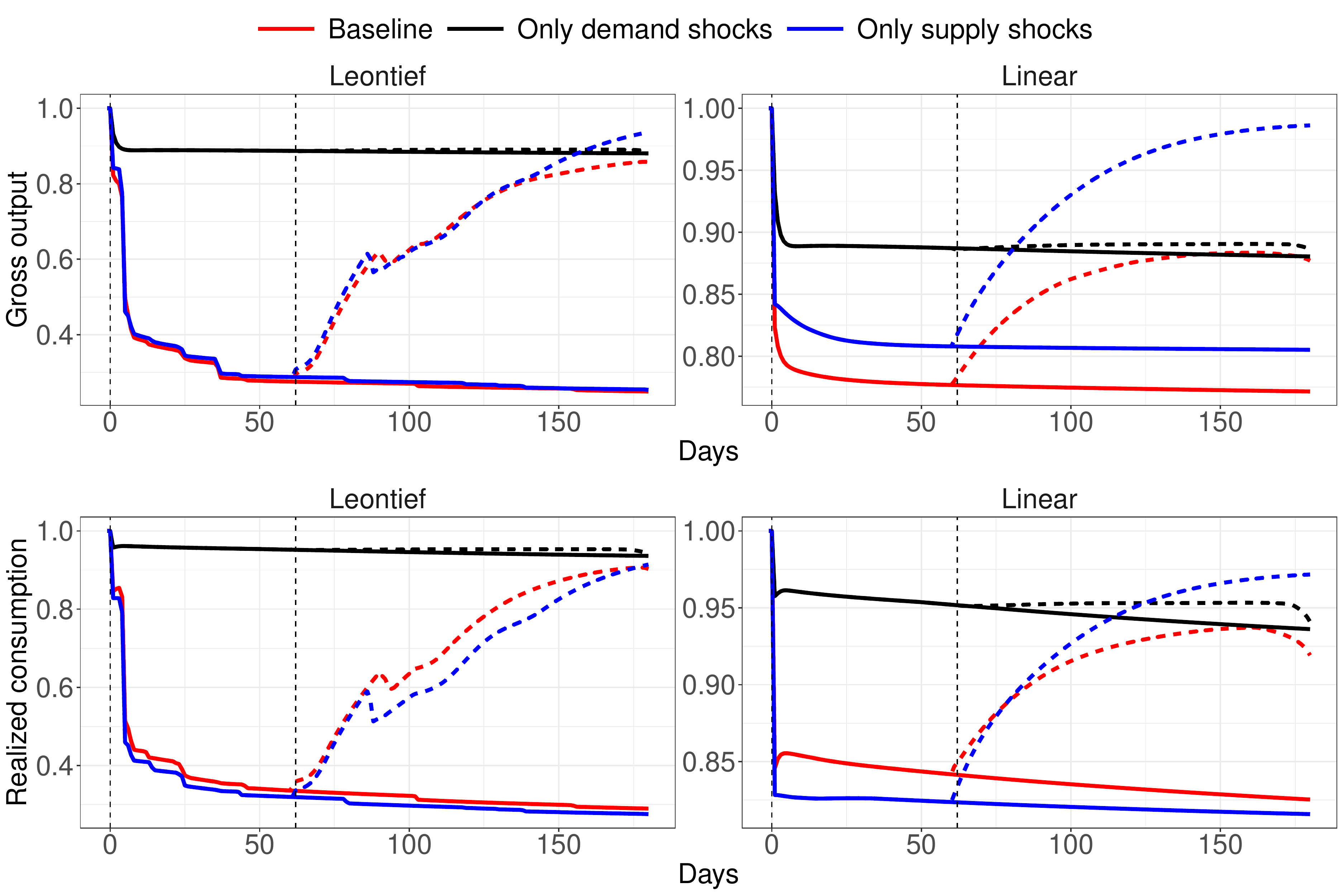} 
    \caption{
    {\bf Dynamic effect of supply shocks vs. demand shocks for different production functions.}
    As Fig. \ref{fig:shockimpact}, but using Leontief (left panels) and linear (right panels) production functions.
    Normalized values of gross output (upper panels) and realized consumption (lower panels) for different shock scenarios. 
    Baseline (red) denotes the model default setup where both supply and demand shocks are used. The blue/black line shows the case where only demand/supply shocks are switched on. The lockdown starts at $t=0$ and ends for all industries after two months at $t=60$.
    }
        \label{fig:shockimpact_leontief}
\end{figure}

We repeat the analysis in Section \ref{sec:prodnet} of running model simulations with only parts of the initial shocks being switched on for alternative production function specifications.
In the left and right panels of Fig. \ref{fig:shockimpact_leontief} we show simulation results for Leontief and linear production functions, respectively. We find for all production functions that supply shocks are substantially more severe than demand shocks, in particular for Leontief production.

For the Leontief production model, economic impacts on gross output are almost identical for the supply-shocks-only and baseline scenarios. There is a slightly less realized consumption when having only supply shocks present compared to both supply and demand shocks being switched on. 

In the case of linear production functions there is a clearer ordering of how severe demand, supply and both shocks together impact overall economic performance.
Here, impacts on gross output are smaller if only supply shocks are considered compared to the baseline case where both demand and supply shocks are switched on. This makes sense since there are no input bottlenecks in this case, making higher competition for given production levels less problematic. Nevertheless, realized final consumption is also smaller for the linear production model if only supply shocks are considered.

\subsection{Production function} \label{apx:prodfun}

\begin{figure}[H]
\includegraphics[trim = {0cm 0cm 0cm 0cm}, clip, width = 1\textwidth]{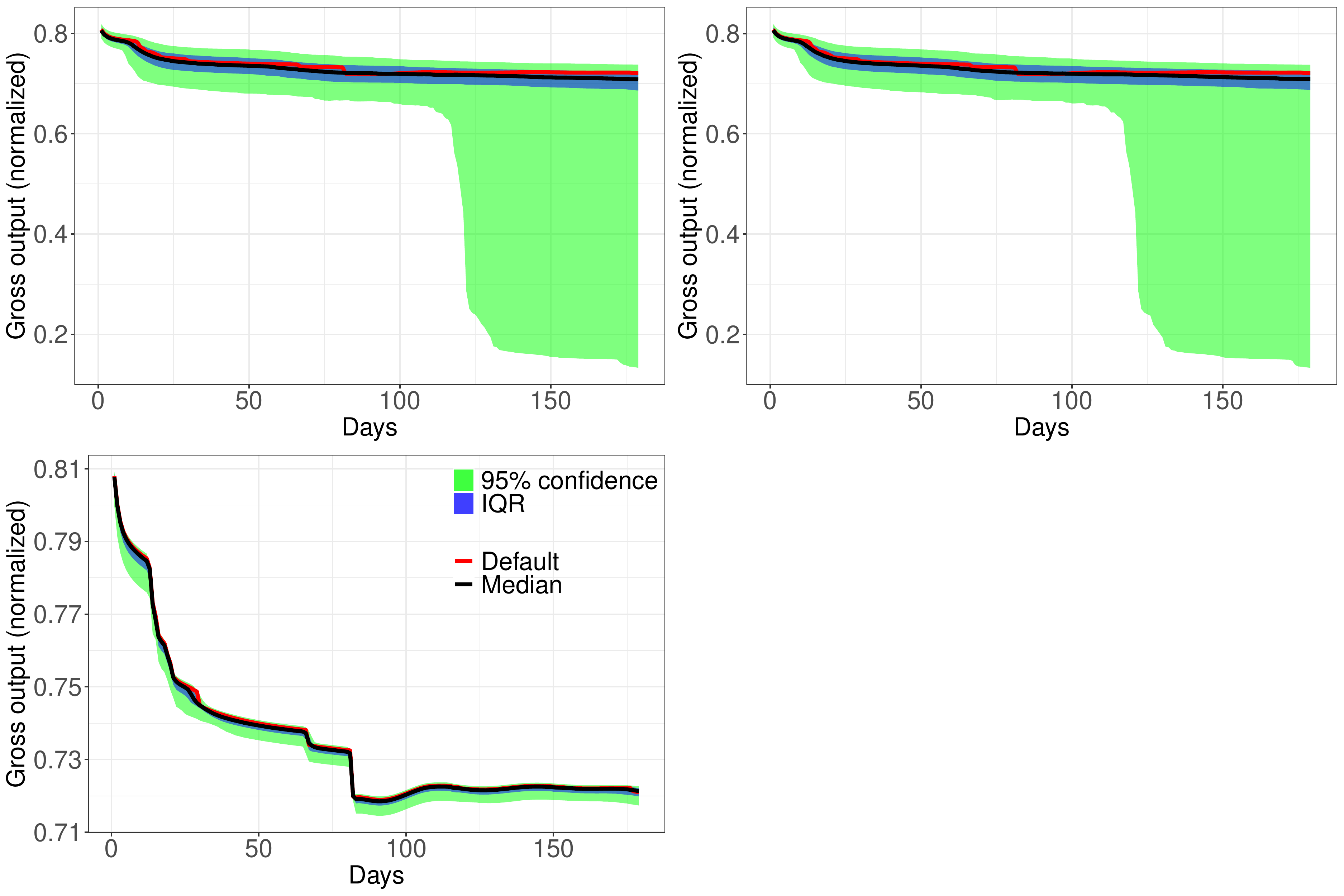}
    \caption{ 
    {\bf Impact of uncertainty in first-order shocks.}
    \emph{Upper left panel:} Perturbing initial supply and demand shocks. 
    \emph{Upper right panel:} Perturbing only supply shocks. 
    \emph{Lower panel:} Perturbing only demand shocks.
    The red line is the default model run reported in the main text, the black line the median of all 1,000 model runs with perturbed initial shock vectors. Green indicates the 95\% quantile and blue the interquartile range.
    }
        \label{fig:shocked_shocks}
\centering
\includegraphics[trim = {0cm 0cm 0cm 0cm}, clip,width=\textwidth]{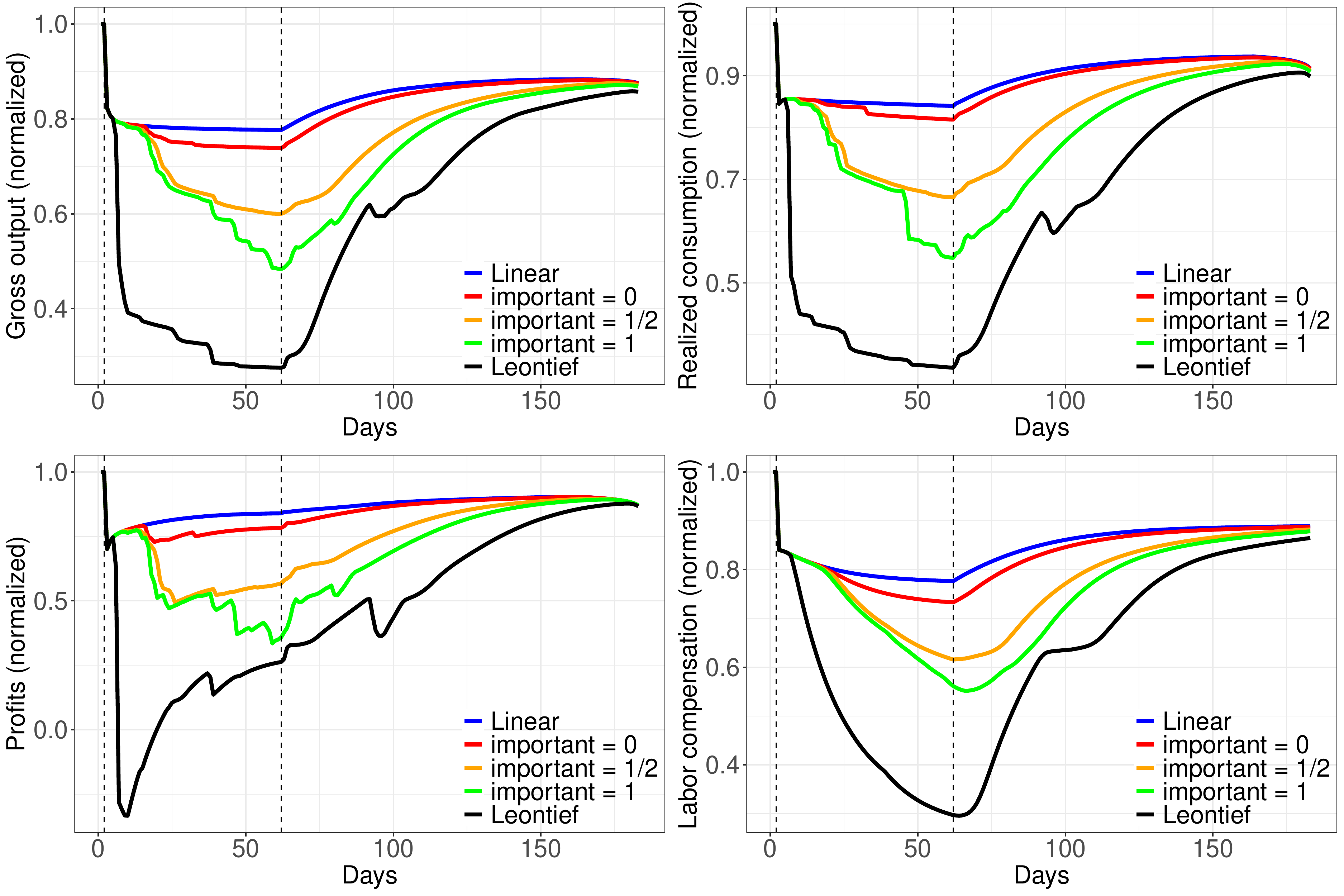}
    \caption{
    {\bf Recovery from the lockdown for four different production functions. }
    The same as Fig. \ref{fig:prodfun_base}, but here the lockdown ends at $t=62$.
    }
    \label{fig:prodfun_reopen}
\end{figure}

We re-run the same simulations as in Section \ref{sec:prodfun_simul}, but now open all industries after two months of lockdown to also compare recovery paths between different production function specifications.
Fig. \ref{fig:prodfun_reopen} shows the results of these simulations where the lockdown ends at $t=62$ (vertical dashed line). 
We find that after six months the five recovery paths converge for different production function specifications, although the transient looks very different for an extended period of time. Note that the economy does not fully recover after six months due to the slow rebouncing of pessimistic consumer expectations consumers and persistence of shocks in exports and investments (see Section \ref{sec:pandemic_shock}).

\subsection{Sensitivity analysis of model parameters} \label{apx:model_sensitivity}

To better understand how results are affected by particular model parameter choices, we conduct a series of sensitivity tests. We make `local' sensitivity tests, meaning that we take the default model setup and then vary a set of parameter to investigate how simulation results are affected.

We first present sensitivity tests on inventory adjustment parameter $\tau$ which plays an important role in intermediate demand; Eq. (\ref{eq:order_interm}). Note that a small $\tau$ represents quick adjustment behavior where firms aim to replenish run-down inventories essentially within a day. On the other hand, if $\tau$ is large, firms react slowly to changes in their input inventories, even when at risk of facing input bottlenecks.

\begin{figure}[H]
    \centering
\includegraphics[trim = {0cm 0cm 0cm 0cm}, clip, width = \textwidth]{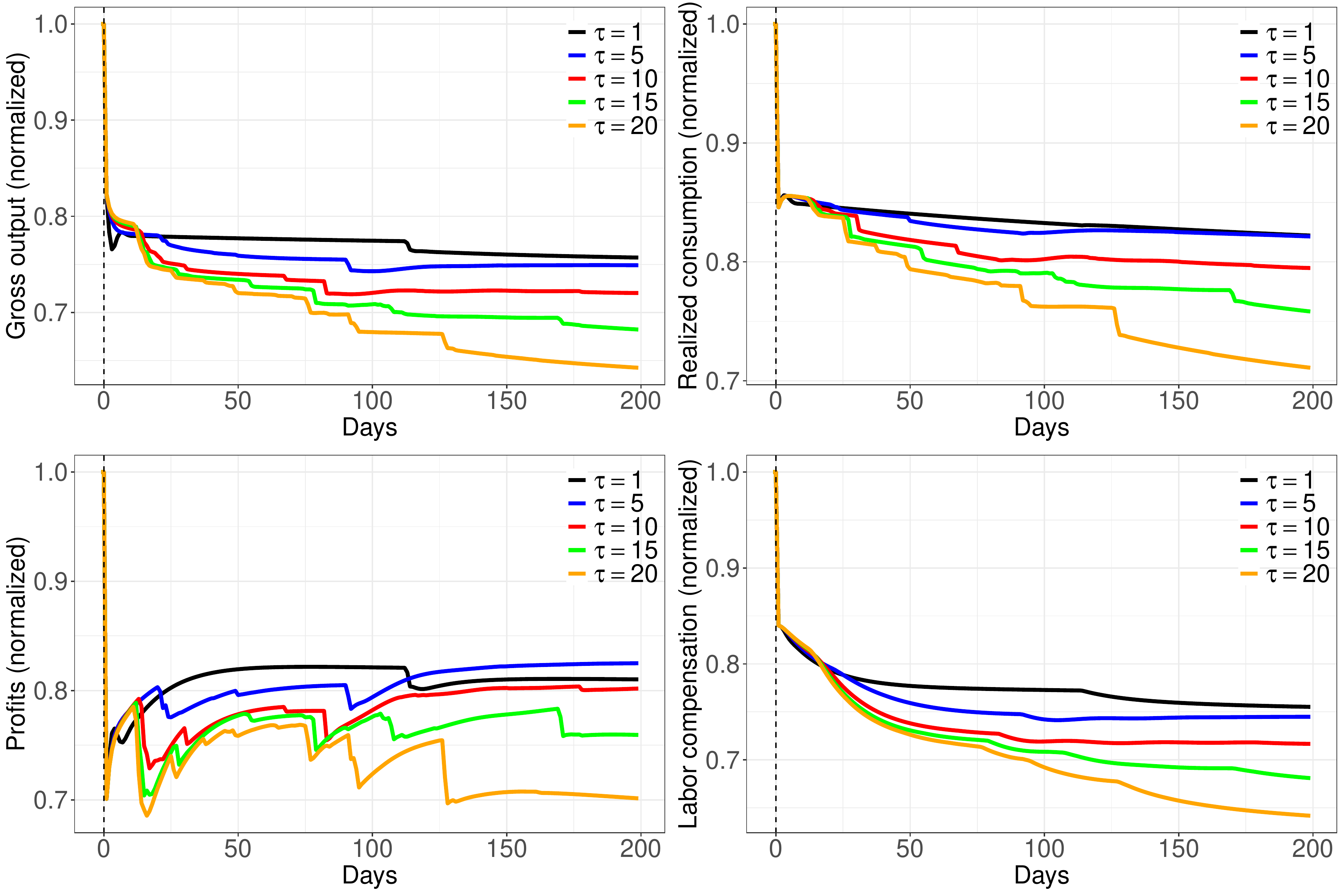}
    \caption{ Model results for different choices of the inventory adjustment speed parameter $\tau$. The baseline case is $\tau = 10.$ }
        \label{fig:tautest}
\end{figure}

We see in Fig. \ref{fig:tautest} how aggregate economic outcomes depend on parameter $\tau$. We find that small values of $\tau$, representing highly responsive firms, dampen adverse economic impacts, while negative impacts are larger if we assume higher sluggishness. 

We also make sensitivity tests with respect to different consumption functions. We test following specifications. First, we use the default consumption function inspired by \cite{muellbauer2020} which is discussed in detail in Section \ref{sec:demand}. As alternative we also consider a simpler consumption function where consumers demand simply a fixed portion of their current income (i.e. have a fixed marginal propensity to consume) which for brevity we call ``Keynesian'' consumption function.
As an even simpler specification we also consider a fixed consumption function where consumers demand a fixed portion of their initial income. For the two alternative consumption functions we choose marginal propensities to consume equal to one such that all of present or initial income is consumed.

\begin{figure}[H]
\centering
\includegraphics[trim = {0cm 0cm 0cm 0cm}, clip, width = \textwidth]{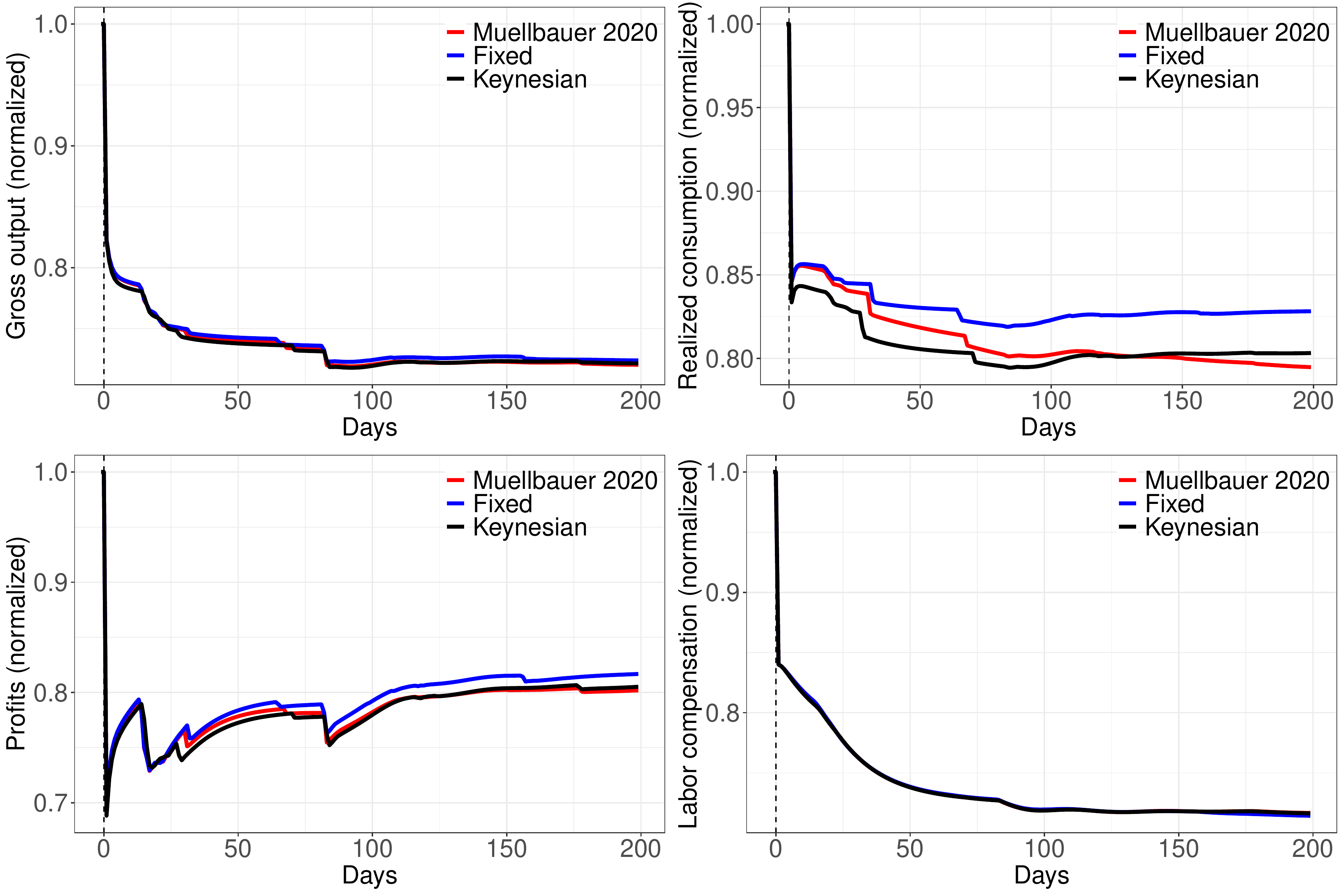}
    \caption{ Model results for different consumption functions. The default case is ``Muellbauer 2020''. }
        \label{fig:constest}
    \centering
\includegraphics[trim = {0cm 0cm 0cm 0cm}, clip, width = \textwidth]{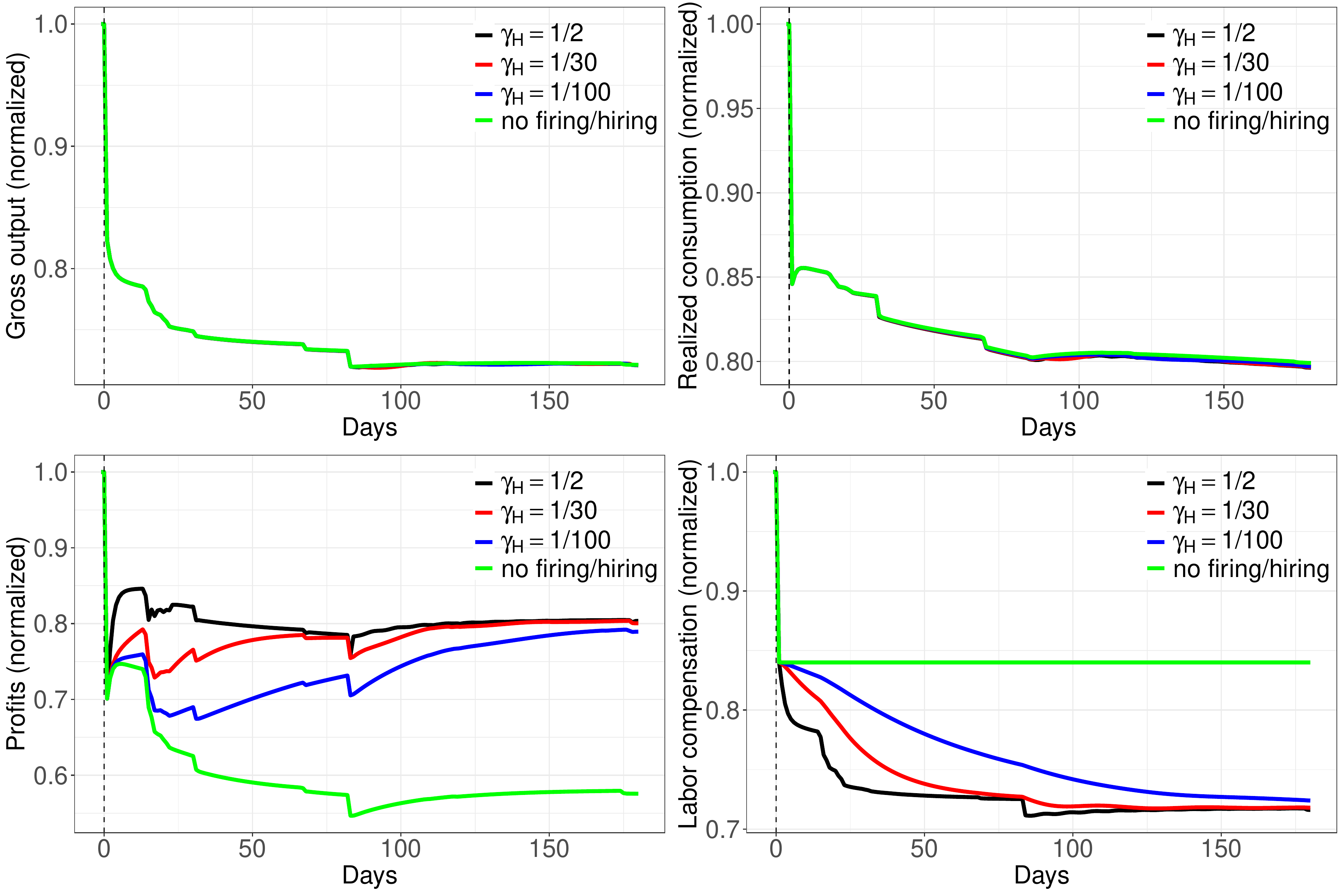}
    \caption{ The effect of hiring and firing speed, $\gamma_H$ and $\gamma_F$, on model results. The default case is $\gamma_H = 1/30$. In all simulations we used $\gamma_F = 2 \gamma_H$. }
        \label{fig:firetest}
\end{figure}

Model results for alternative consumption function specifications are shown in Fig. \ref{fig:constest}. 
There are only negligible differences between different production functions on gross output and labor compensation. Realized consumption is slightly higher for a fixed consumption function which is not surprising and somewhat artificially achieved since here consumers demand based on comparatively large initial income values.

We also investigate how model results depend on the speed of adjustment in labor inputs. In Section \ref{sec:demand} we introduced a parameter $\gamma_H$ which represents how quickly firms can hire employees in case they want to ramp up their productive capacities. Values of $\gamma_H$ close to one represent the case where hiring can happen very quickly, whereas values close to zero indicate that it is very hard for firms to hire new workers.
Similarly, we considered an equivalent parameter $\gamma_F$ for firing workers.

In Fig. \ref{fig:firetest} we show how model results are affected if different $\gamma_H$ values are used as well as if hiring and firing are completely ruled out. All these simulations use $\gamma_F = 2 \gamma_H$ to reflect the situation where firing of employees takes less them than hiring if allowed. We find almost no differences on gross output and realized consumption for all these cases. In line with intuition the exact specification of hiring and firing affects labor compensation and firms' profits. In case of no hiring and firing, labor compensation remains constant throughout the simulation, once the initial labor supply shock is applied. Labor compensation is smaller the easier it is for firms to fire (and hire) employees. This makes sense since firms which face production constraints other than capacity constraints will lay off employees, reducing overall labor income. The picture is reversed for profits. If there is no flexibility for firms in adjusting labor input, there is a larger negative impact on profits. The easier firms can lay off workers, the more they reduce costs on labor which they do not need to satisfy aggregate demand.

\subsection{Sensitivity of comparison to empirical data}
\label{apx:sensitivity_correlation}

\begin{figure}[H]
\centering
\includegraphics[trim = {0cm 0cm 0cm 0cm}, clip,width=0.7\textwidth]{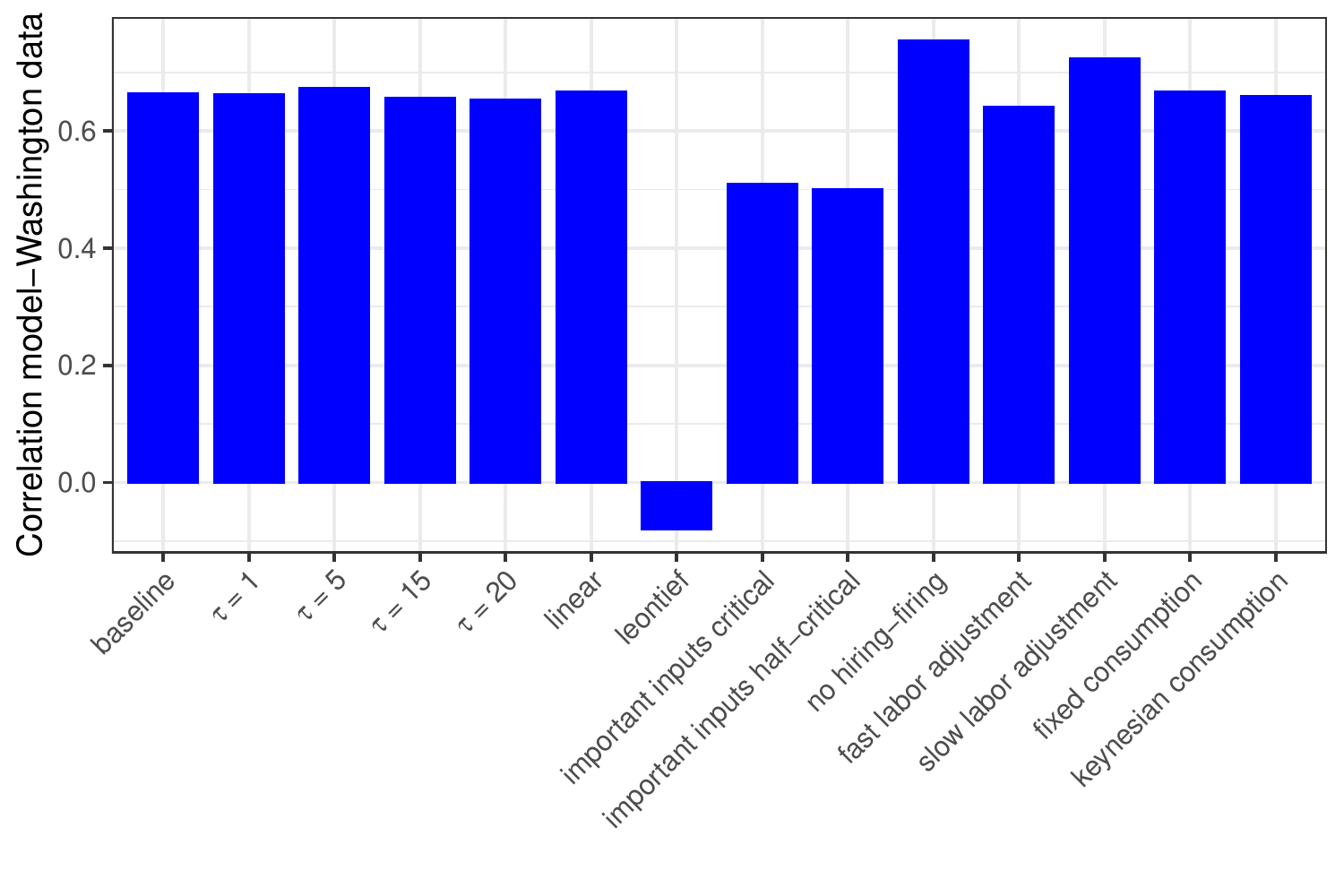}
    \caption{For various scenarios, we show the weighted correlation coefficient between model predictions for rises in unemployment in various sectors and empirical data coming from Washington State. See Table \ref{tab:scenarios} for a definition of the various scenarios. }
    \label{fig:robustness_correlations}
\end{figure}

In Section \ref{sec:econimpact}, we compared model predictions to data coming from the U.S. states of Washington and Texas, when running the model in the baseline scenario. In this section, we consider the other scenarios outlined in Table \ref{tab:scenarios} and described in the previous appendices. We do not find much difference in terms of the relative performance of each scenario when either comparing to Washington or Texas, or using the Pearson or weighted correlation coefficients. We thus report in Fig. \ref{fig:robustness_correlations} only results for Washington, using a weighted correlation coefficient to compare model predictions and empirical data. 

It is immediately apparent that the performance of the various scenarios is similar, except for the cases of the basic Leontief production function and of the Leontief production function with important inputs considered as critical or half-critical. In these cases, and especially in the Leontief case, performance is substantially lower, suggesting that our modeling choice of distinguishing between critical and non-critical inputs adds realism to our model. Correlation between model predictions and empirical data is somewhat lower in case no hiring or firing takes place (in the sense that workers are only furloughed due to the epidemic shock and not due to second-order effects), but this is a clearly unrealistic assumption. Given the combined uncertainties of comparing the model to the data, and the intrinsic uncertainty in these preliminary data, it would not be wise to select an unrealistic assumption based on a small increase in empirical performance.

Therefore, our choice of the baseline reflects a balance between ability to reproduce empirical patterns and prior belief in certain assumptions/parameter values. We use poor empirical performance to exclude the Leontief, ``important inputs critical'' and ``important inputs half-critical'' scenarios. We use instead our best judgement to exclude too fast or slow adjustments of inventories and labor force, full substitution of inputs in the linear production function, and too simple consumption functions such as the fixed or ``Keynesian'' ones. As shown in Appendices \ref{apx:prodfun} and \ref{apx:model_sensitivity}, in any case, model results tend to depend weakly on these specific assumptions, consistently with the little ability of data to distinguish between the respective scenarios.

\subsection{Comparison to traditional IO models}
\label{apx:io_models}

We also compare our model results to traditional input-output (IO) models. In particular, we compare the steady state of our model with two models, the demand-driven Leontief \citep{leontief1936} and the supply-driven Gosh model \citep{ghosh1958input}.
Since these simpler IO models do not include inventory effects, we set input inventories artificially high such that they do not restrict economic production.

In the Leontief model final demand is exogenous, and under the assumption of fixed production recipes, gross output per industry is endogenously determined by multiplying demand with the Leontief inverse \citep{miller2009input}. 
When considering only demand shocks, we can write the Leontief prediction as 
\begin{equation}
    x^L = (\mathbb{I} - A)^{-1} (c^{\text{shocked}} + f^\text{shocked} ).
\end{equation}
We also rerun our model with all supply shocks being switched off and only considering demand shocks.
We then compare the steady state results of our model with the Leontief prediction. 

\begin{figure}[H]
\centering
\includegraphics[trim = {0cm 0cm 0cm 0cm}, clip,width=\textwidth]{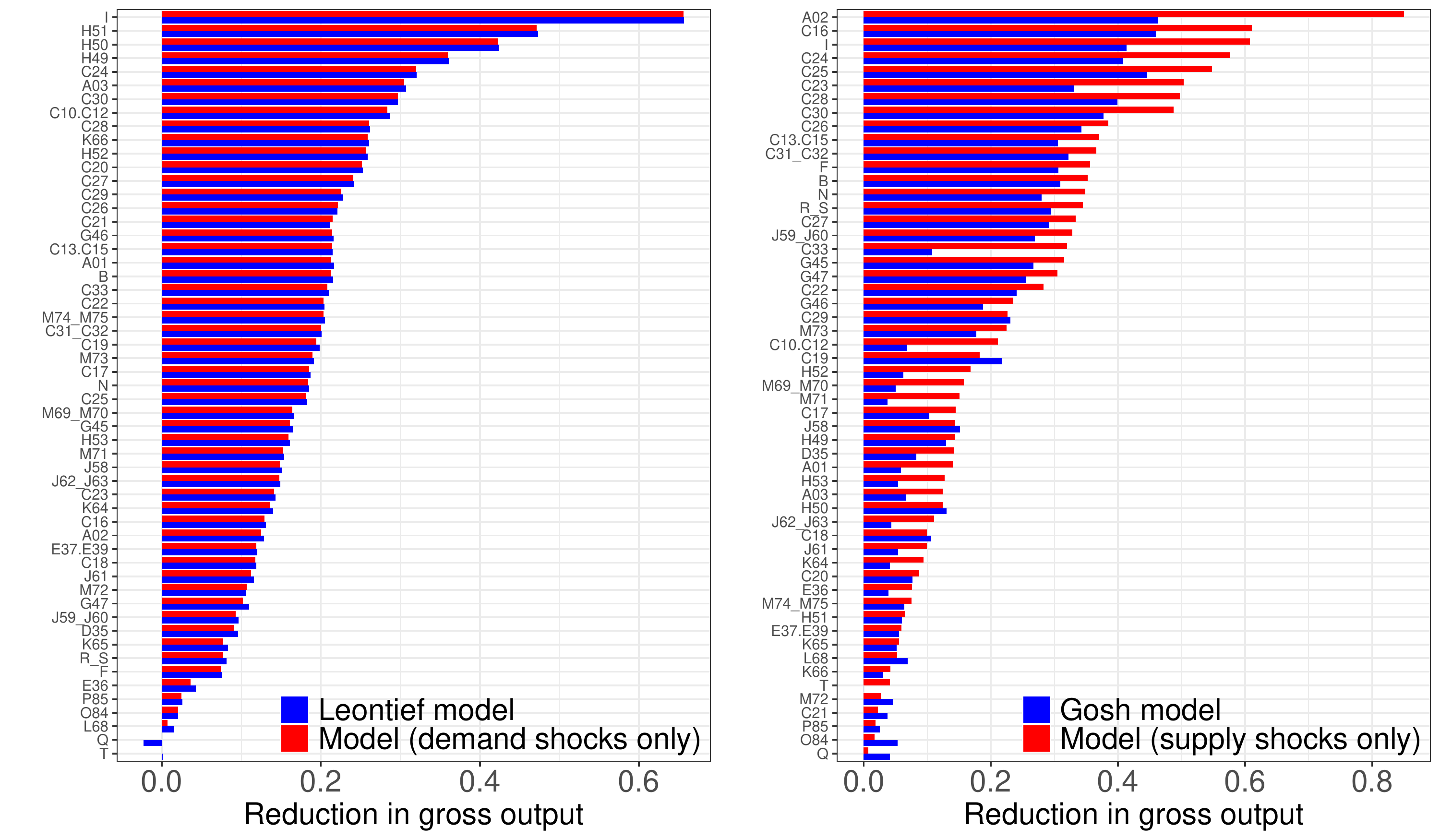}
    \caption{ Comparison between impacts on sectoral gross output obtained from traditional IO models with our model. The left panel compares the model steady state with the Leontief (only demand shocks are switched on), the right panel with the Gosh model (only supply shocks are switched on). 
    Note that our model was run with artificially high inventories, no hiring and firing, $\Delta s = 1$ and no income losses ($b=1$).
    }
    \label{fig:io_bars}
\end{figure}

Fig. \ref{fig:io_bars} (left panel) shows the reduction of sectoral gross output compared to the pre-shock state as barplots for our and the Leontief model. We find that our model very closely recovers the Leontief prediction in the steady state. Gross output per industry in the steady state of our model and the Leontief model have almost a correlation of one.
The differences between predicted sectoral reductions in gross output are almost zero in all cases. Only for Health (Q) they differ by 2.3\%, since the Leontief model would predict an increase as a result of positive demand shocks which cannot be satisfied in our model due to fixed maximum capacity constraints.

These results are very robust against using empirical inventories.
It should be noted that the Leontief model is static and we are comparing the steady state of our dynamic model. Thus, modeling the transient which is relevant for the short time-scales considered in the main text is not possible with the traditional Leontief model.

We do a similar comparison with the supply-driven Gosh model. There are no fixed production recipes in the Gosh model, but fixed ``allocation coefficients'' $B_{ij} = Z_{ij,0}/x_{i,0}$. Here, a change in gross output is due to a change in primary inputs, i.e. represented as value added. In the notation used here we can formulate the Gosh prediction as
\begin{equation}
    x^G = (\mathbb{I} - B^\top)^{-1} (l^\text{shocked} + e).
\end{equation}

We plot the Gosh predictions and the steady state results of our model with only supply shocks turned on in the right panel of Fig. \ref{fig:io_bars}. We find greater differences between the Gosh and our model for the supply shocks. This should not come as a surprise, since the Gosh model builds upon a very different production function.

Rankings of sectoral declines are still very correlated (Spearman correlation of 0.91). This is higher than the correlations between our model's steady state rankings of industries and the initial shock rankings (correlation of 0.87). Unsurprisingly, the Gosh model rankings are most similar as the ones obtained from using initial supply shocks only (correlation of 0.94).

These results are not very robust with respect to the specifications of the economic model considered here. Using empirical inventories in our model enlarges differences in model predictions tremendously.

We compared our model also to slightly more complex mixed endogenous/exogenous IO models \citep{dietzenbacher2015reflections, arto2015global} which simultaneously can take supply and demand shocks into account. Yet these models do not always guarantee positive solutions for variables such gross output and final consumption \citep[p.628]{miller2009input}. In particular when applying the large first-order supply and demand shocks of the pandemic to the UK economy, the mixed IO model does not yield feasible allocations.

\FloatBarrier

\section{Epidemic modelling} 
\label{apx:epidemic_modelling}

In this appendix we present our epidemic model where we divide contagion channels by activities. As we focus on the early stage of the epidemic, we do not explicitly model the number of recovered individuals $R$, although that plays a role to determine the total population $M$. We start denoting the number of susceptible and infected people in the pupils and students and in the non-working adults category by  $S^s$, $I^s$, $S^u$, $I^u$ respectively. Similarly, $S_i$ and $I_i$ denote the number of susceptible and infected workers of industry $i$.   It follows that the decrease in the overall susceptible population $S$ is given by 
\begin{equation}
    \frac{dS}{dt} = \frac{dS^s}{dt} + \frac{dS^u}{dt} + \sum_{i=1}^N \frac{dS_i}{dt}.
    \label{eq:SIR_allapx}
\end{equation}
In what follows, we compute the rate of infection of each population category by focusing on the different channels of contagion each person is exposed to. In these computations we assume homogeneous mixing of the population, meaning that the probability that a person  had contact with an individual that was infected is $\frac{I}{M}$, regardless on the channel they had contact in.

\paragraph{Normalizing contact-weighted shares by population}

As we discuss in Appendix \ref{apx:epidemiology_data} we have data on the share of intensity-weighted contacts in each activity of the overall population. For the derivation of the epidemiological model it is useful to renormalize these shares of weighted contacts (i.e. the $\beta$'s) by the population they come from. This is not necessary for $\beta_c(0)$ or $\beta_h(0)$, since consumption and other household interaction related contacts are spread evenly across the whole population. On the contrary, we do need to renormalize $\beta_s(0)$ by the student and pupil population $\eta_s$, so that
\begin{equation}
    \hat{\beta}_s(0) = \frac{\beta_s(0)}{\eta_s}.
\end{equation}
Another way to look at the equation above is to note that $\hat{\beta}_s(0)$ is the share of intensity-weighted contacts in school \textit{per unit} population, and to obtain the actual share of intensity-weighted contacts $\beta_s(0)$ one needs to multiply $\hat{\beta}_s(0)$ by the population share of students, $\eta_s$.   Similarly, we renormalize the work intensity-contacts across the workers of different industries as follows
\begin{equation}
    \hat{\beta}_w(0) = \frac{\beta_w(0)}{\sum_{l=1}^N  \eta_l b_{l,w}},
\end{equation}
where the normalization includes the $b_{i,w}$ factors i.e., the heterogenous distribution of intensity-weighted contacts across industries.

In the transport channel, we must distribute the contacts across the commuter population (i.e. workers and students). To account for a density effect (see below), we assume that the number of contacts scales with the square of the number of people in public transport, and use the normalization factor 
\begin{equation}
    \hat{\beta}_T(0) = \frac{\beta_T(0)}{\left(\eta^s + \sum_{i=1}^N \eta_i\right)^2},
\end{equation}
where $\eta^s + \sum_{i=1}^N \eta_i$ is the pre-lockdown share of the population that commutes.

\paragraph{Students and pupils}
To simplify notation we define $\mu^s$ as the fraction of the students and pupils population attending schools, which is given by 
$$\mu^s =  \left( \delta_s + (1 - \delta_s) \left( g \sum_{j=1}^N \delta_{i,w}\eta_i \right) \right).$$
We know that students and pupils are exposed to infection due to school attendance, transport, consumption, and other household interaction. We assume that pupils that go to school have the same amount of contacts in school as before lockdown\footnote{
We make this assumption considering that a) only certain schools are open so it is unclear to what extent the density in schools has decreased b) it is possible that pupils interact more with the few pupils left in school and thus the number of contacts can remain roughly constant.
}, while for transport we consider that the number of contacts decreases due to the reduced density of people in the bus/train. With these assumptions we obtain the following equation for the infection rate
\begin{eqnarray}
\frac{dS^s}{dt}  & = &  - \beta^*\Bigg[ \hat{\beta}_s(0) \mu^sS^s  \frac{I}{M}  +  \hat{\beta}_T(0) \mu^s S^s   \left( \mu^s \frac{I^s}{M} + \sum_{k=1}^N\delta_{k,w}\frac{I_k}{M} \right)  \nonumber \\
 &+&\beta_c(0) S^s \frac{I}{M} \sum_{k=1}^N \delta_k(t) b_{k,c}  + \beta_h(0) S^s \frac{I}{M}  \left( (1 - \delta_h)\kappa + \delta_h \right)   \Bigg],
\end{eqnarray}
where the first two terms correspond to the infections happening at school and transport and thus only apply to the fraction $\mu^s$ of the student population that goes to school. The third and fourth term correspond to infections happening while consuming or doing other household activities and therefore apply to the whole student population. $b_{k,c}$ is the consumption related contacts, while $\kappa$  is the share of social/family/friends contacts that are not avoidable by social distancing. Notice that since transport is shared with both students and workers, the transport term includes both $I^s$ and $I_k$. The fact that we are considering the fractions $\mu^s$ and $\delta_{k,w}$ of infected in the transport term reflects our assumption that density matters in particular in transports; note, for example, that we are not multiplying infected individuals by $\mu^s$ in the school term.   

We simplify the above equation using the mean field approximation $S_i \approx \eta_i S$, $I_i \approx \eta_i I$, $S^s \approx  \eta^s S$, and $I^s \approx \eta^s  I$ and obtain
\begin{eqnarray}
\frac{dS^s}{dt}  & = &  - \frac{\beta^*}{M} SI\Bigg[\hat{\beta}_s(0)  \mu^s\eta^s   + \hat{\beta}_T(0)    \left(    (\eta^s\mu^s)^2 + \eta^s\mu^s\sum_{k=1}^N\delta_{k,w}\eta_k \right)   \nonumber \\
 &+& \beta_c(0)\eta^s  \sum_{k=1}^N \delta_k(t) b_{k,c}  + \beta_h(0)\eta^s  \left( (1 - \delta_h)\kappa + \delta_h \right)   \Bigg].
  \label{eq:SIR_school}
\end{eqnarray}

\paragraph{Working population}
Workers are exposed to infection due to work, transport, consumption, and other household interaction. For a worker in industry $i$, the infection rate is
\begin{eqnarray}
\frac{dS_i}{dt}  & = &  -\beta^*  \Bigg[ \hat{\beta}_w(0) \delta_{i,w} S_ib_{i,w} \frac{I}{M}  + \hat{\beta}_T(0) S_i \delta_{i,w} \left( \mu^s \frac{I^s}{M} + \sum_{k=1}^N\delta_{k,w}\frac{I_k}{M} \right)  \nonumber \\
 &+& \beta_c(0)S_i \frac{I}{M} \sum_{k=1}^N \delta_k(t) b_{k,c}  + \beta_h(0) S_i \frac{I}{M}  \left( (1 - \delta_h)\kappa + \delta_h \right)   \Bigg],
\end{eqnarray}
where we have assumed that workers that go to work make the same amount of contacts at work as before lockdown, while for transport we consider that the number of contacts decreases due to the reduced density of people in public transport. We have made explicit that the work and transport infection channels only apply to the fraction $\delta_{i,w}$ of the working population in $i$ going to work and to the $\mu^s$ fraction of students going to school. As before, we use the mean field approximation $S_i \approx \eta_i S$, $I_i \approx \eta_i I$, $S^s \approx  \eta^s S$, and $I^s \approx \eta^s  I$ to simplify the equation to
\begin{eqnarray}
\frac{dS_i}{dt}  & = & - \frac{\beta^*}{M} SI\Bigg[  \hat{\beta}_w(0)\delta_{i,w} \eta_ib_{i,w}   + \hat{\beta}_T(0)\eta_i\delta_{i,w}  \left(  \mu^s \eta^s + \sum_{k=1}^N\delta_{k,w}\eta_k \right)  \nonumber \\
 &+& \beta_c(0)\eta_i \sum_{k=1}^N \delta_k(t) b_{k,c}  + \beta_h(0) \eta_i  \left( (1 - \delta_h)\kappa + \delta_h \right)   \Bigg].
\end{eqnarray}
We now sum across all $N$ industries to obtain
\begin{eqnarray}
\sum_{i=1}^N \frac{dS_i}{dt}  & = & - \frac{\beta^*}{M} SI\Bigg[ \hat{\beta}_w(0) \sum_{i=1}^N\delta_{i,w} \eta_ib_{i,w}  + \hat{\beta}_T(0) \left(\mu^s \eta^s \sum_{i=1}^N \delta_{i,w}\eta_i + \left(\sum_{i=1}^N \delta_{i,w}\eta_i\right)^2\right)  \nonumber \\
 &+& \beta_c(0)\sum_{i=1}^N \eta_i \sum_{k=1}^N \delta_k(t) b_{k,c}  +\beta_h(0) \sum_{i=1}^N \eta_i  \left( (1 - \delta_h)\kappa + \delta_h \right)   \Bigg].
  \label{eq:SIR_work}
\end{eqnarray}

\paragraph{Non-working adults}
By definition non-working adults are not exposed to the work or school infection channel. Furthermore, since we only consider work-commuting transport use, the non-working adults are not exposed to the transport infection channel either. It follows that the decrease in the susceptible population depends only on the consumption and other household interaction channel
\begin{equation}
    \frac{dS^u}{dt} = - \frac{\beta^*}{M} S I \Big[ \beta_c(0) \eta^u \sum_{k=1}^N \delta_{k,c}(t)  b_{k,c} + \beta_h(0) \eta^u \left( (1 - \delta_h)\kappa + \delta_h \right),  
    \Big]
    \label{eq:SIR_nonworking}
\end{equation}
where we have again used the approximation  $S^u \approx \eta^u S$.

\paragraph{Total population}
To get the infection rate of the overall population we substitute Eqs. \eqref{eq:SIR_school}--\eqref{eq:SIR_nonworking} in equation Eq. \eqref{eq:SIR_allapx}. It follows that 
\begin{eqnarray}
\frac{dS}{dt}  & = &  - \beta^* \frac{SI}{M}\Bigg[ \hat{\beta}_w(0)  \sum_{i=1}^N\delta_{i,w}\eta_ib_{i,w}+ \hat{\beta}_s(0)  \mu^s\eta^s \nonumber \\
 &+& \hat{\beta}_T(0) \left(\mu^s \eta^s +\sum_{i=1}^N \eta_i\delta_{i,w} \right)^2  \nonumber \\
 &+& \beta_c(0) \sum_{k=1}^N \delta_k(t) b_{k,c}  +  \beta_h(0)  \left( (1 - \delta_h)\kappa + \delta_h \right)   \Bigg],
\end{eqnarray}
where we have used the fact that $ \eta^s + \eta^u + \sum_{i=1}^N \eta_i = 1.$ The above terms correspond to the infection due to work, school, transport, consumption, and other household interaction respectively. We can write the above equation as
\begin{equation}
 \frac{dS}{dt}  = - \beta(t)\frac{SI}{M}
\end{equation}
where 
\begin{equation}
   \beta(t) =  \beta^* \Big( \beta_w(t)+\beta_s(t)+\beta_c(t)+\beta_T(t)+\beta_h(t) \Big),
\end{equation}
which is Eq. \eqref{eq:beta} of the main text. The $\beta$'s are given by
\begin{equation}
   \beta_w(t) = \hat{\beta}_w(0)  \sum_{i=1}^N\delta_{i,w}\eta_ib_{i,w}  = \beta_w(0) \sum_{i=1}^N\delta_{i,w}\frac{ \eta_ib_{i,w}}{\sum_{l=1}^N \eta_l b_{l,w}},
\end{equation}
\begin{equation}
   \beta_s(t) = \hat{\beta}_s(0)   \mu^s\eta^s = \beta_s(0)   \mu^s,
\end{equation}
\begin{equation}
   \beta_T(t) = \hat{\beta}_T(0) \left(\mu^s \eta^s +\sum_{i=1}^N \eta_i\delta_{i,w} \right)^2  = \beta_T(0) \left( \frac{  \mu^s \eta^s +\sum_{i=1}^N \eta_i\delta_{i,w} }{\eta^s + \sum_{l=1}^N \eta_l} \right)^2,
\end{equation}
\begin{equation}
   \beta_c(t) =   \beta_c(0) \sum_{k=1}^N \delta_k(t) b_{k,c},
\end{equation}
and
\begin{equation}
   \beta_h =   \beta_h(0)  \left( (1 - \delta_h)\kappa + \delta_h \right).
\end{equation}

\FloatBarrier

\section{Calibration of epidemic model}
\label{apx:epidemiology_data}
 
\subsection{Literature review}

In epidemiology, the main method to understand contact patterns is to use social contact surveys. A landmark study is the Polymod study \citep{mossong2008social}. Several other studies in the last decade have confirmed that, roughly speaking, people have about 10-20 non casual contacts per day, mostly at home and at work. By ``non-casual'' contact,  we mean contacts as defined by these studies, that is, either physical contact or non physical contact defined as ``a two-way conversation with three or more words in the physical presence of another person''. 

The Polymod study is very interesting for us because it reports \emph{where} contacts take place. Averaging across countries and pooling physical and non-physical contacts, 23\%, 21\%, 14\%, 3\%, and 16\% are made at home, at work, at school, while travelling, and during leisure activities, respectively. 

There are still significant uncertainties on the mode of transmission of SARS-CoV-2, and in particular whether it can diffuse through casual contact (whether simply `sharing air' is risky, because aerosolized particles, rather than just droplets, are risky). Fortunately, there have also been a couple of studies quantifying ``casual'' contacts, that is, contacts between anonymous individuals but that nevertheless involve enough proximity to result in a transmission. \citet{mikolajczyk2008collecting} report several studies where participants (students) were ``asked about aggregate numbers of contacts on six levels of proximity: intimate contacts, close contacts (same household), direct conversation ($>2$ min duration, max. 2 m distance), small group (with conversations, but less intensive than in direct conversations), larger group (seminary or lecture room) and occasional contacts (in the range of 2 m in local transportation, cinema, etc.).'' The number of conversational contacts (i.e. including intimate, close and direct conversation ) was sensibly below but in line with the Polymod study (6-13 contacts vs 10-20). Their Fig. 2 suggests that the number of contacts in small groups, large groups, and random contacts was roughly speaking 8, 30 and 40, with variations depending on study design. Roughly speaking, people have on average 10 close contacts per day but 80 casual (non-close) contacts.

A few studies have looked at social contact patterns to understand the diffusion of tuberculosis, which spreads very easily. Reading from their Fig. 3, the estimates of mean casual contacts per day obtained by \citet{mccreesh2019estimating} for South Africa are about 10 for transport (combining trains and large taxis), 6 for school and work, 5 for shops (spaza shop, shebeen and mall), 2 for home, and less than 2 for church and community hall. These estimates are for the mean number of contacts per day, but \citet{mccreesh2019estimating} also asked participants for the number of casual contact present during the visit to a location (Fig. S2), showing about 40 casual contacts in Malls and Trains. In many other categories relating to transport or shopping, the number of casual contacts is around 10-15.

In this paper, we use a study from Sweden \citep{stromgren2017place}. The study reports, for a variety of places, the likelihood that it is visited during an average day, the duration of the visit\footnote{
The duration of shop visits is highly consistent with the data reported by \citet{goldfarb2020retail}, who use mobile phone data for the US and report an average visit of 22 to 42 minutes across 11 categories of retail shops.
}, the number of people present, and the likelihood of physical contact, see Table \ref{table:stromgren}.

\subsection{Calibration}

\begin{table}[ht]
\small
\centering
\begin{tabular}{|l|p{32mm}|p{13mm}p{15mm}p{20mm}p{18mm}|p{12mm}|p{12mm}|}
  \hline
 &  & \emph{Visit} & \emph{Duration} & \emph{Crowd} & \emph{Physical} & \multicolumn{2}{c|}{\emph{Intensity}} \\ 
  \hline
Category & Stromgren et al's category & Visit likelihood & Duration (hours) & Median number people present& Likelihood physical contact & \multicolumn{2}{c|}{Weighted contacts} \\ 
  \hline
Work & Work & 21.20 & 7.60 & 20.00 & 55.80 & 0.29 & 0.29\\   
\hline
\multirow{2}{*}{School}  
 & Pre-school & 8.60 & 7.60 & 20.00 & 73.30 & 0.12 & \multirow{2}{*}{0.28} \\ 
 & School & 12.00 & 7.60 & 20.00 & 71.10 & 0.16 &\\ 
\hline
\multirow{4}{*}{Consume} &  Convenience store & 5.20 & 0.40 & 10.00 & 8.30 & 0.00 & \multirow{4}{*}{0.16}\\ 
&  Large store & 24.10 & 0.80 & 21.50 & 18.00 & 0.04 &\\ 
&  Restaurant & 9.40 & 1.40 & 30.00 & 30.80 & 0.04 &\\ 
 & Sports venue & 11.50 & 2.30 & 34.50 & 53.80 & 0.08 & \\ 
\hline
\multirow{1}{*}{Transport} &  Public transport & 16.30 & 1.00 & 40.00 & 8.30 & 0.06& 0.06\\ 
\hline
\multirow{4}{*}{Home} &  Home & 95.00 & 18.40 & 1.00 & 73.70 & 0.16 &  \multirow{4}{*}{0.21}\\ 
&  Car & 58.70 & 0.90 & 1.00 & 25.80 & 0.00 &\\ 
&  Public urban space & 6.60 & 1.80 & 20.00 & 28.30 & 0.02& \\ 
&  Friends and relatives & 21.00 & 5.10 & 3.00 & 80.10 & 0.03 &\\ 
   \hline
\end{tabular}
\caption{The columns \emph{Duration} and \emph{Crowd} for the rows School and Pre-school are inferred from the equivalent number in the row Work. ``Large store'' is short for ``Large and specialist store''. The source of raw data is  \citet{stromgren2017place}. Intensity-weighted contacts are our own calculations, see text. The last column shows the values calculated for Table \ref{table:sharesall},
Eq.s \eqref{eq:beta}-\eqref{eq:beta_0} in the main text.
}
\label{table:stromgren}
\end{table}
\normalsize

We used the data from Table \ref{table:stromgren} to create an intensity-weighted number of contacts. We define for each of the 12 places\footnote{
Note that we could have used the variable showing the likelihood of physical contact as proxy for the closeness of contact, as an additional factor in Eq. \eqref{eq:intensity}. We have done so in a robustness check and most results are similar, except for Sports Venue which becomes an even larger share of all consumption risks. We decided against using this additional variable in the current draft as we match this activity with the industry that contains cinemas, theatres, religious gatherings, etc. It is true that, like sports, these activities have a significant duration, but they are not as likely to involve physical contact. 
}.
\begin{equation}
\text{Intensity}_i= \frac{\text{Visit}_i \times \text{Duration}_i \times \text{Crowd}_i }{ \sum_{j=1}^{12}  \Big[ \text{Visit}_j \times \text{Duration}_j \times \text{Crowd}_j \Big]}
\label{eq:intensity}
\end{equation}
To compute the values in Table \ref{table:sharesall} for Eqs. \eqref{eq:beta}-\eqref{eq:beta_0}, we sum-up the relevant \emph{Intensity} variables.

\paragraph{Work.}
To calibrate $b_{iw}$, we create an index based on the \emph{physical proximity} and \emph{exposure to infection} index of each industry, which, as explained in Appendix \ref{apx:shocks}, we map from O'NET data. At the occupation level, physical proximity and exposure to infection range from 0 to 100 and are described as follows.
\begin{itemize}
    \item \emph{Exposure to disease and infection}. O*NET assigns a score to each occupation depending on the frequency with which workers in that occupation are exposed to disease and infection \textit{in normal times}. The scale runs from $0$, indicating that the worker is never exposed to 100, indicating that the worker is exposed every day. It is important to consider that this rating was done before the pandemic, and doesn't seem to properly take into account the properties of COVID-19
    \item \emph{Physical proximity}. O*NET also considers to what extent performing job tasks requires physical proximity. A score of $75$ implies being moderately close (at arm's length) and $100$ implies near touching.
\end{itemize}
To obtain a score at the industry level, we aggregate occupation-level scores using employment data from the BLS, which indicates the occupational composition of each industry and then map into the WIOD classification (see Appendix \ref{apx:shocks} for details). Our industry-specific infection risk is the average of physical proximity and exposure to infection. That is
\begin{equation}
    b_{iw} = \frac{1}{2}(\text{exposure to infection}_i + \text{physical proximity}_i).
\end{equation}

\begin{figure}
    \centering
\includegraphics[width = 0.5\textwidth]{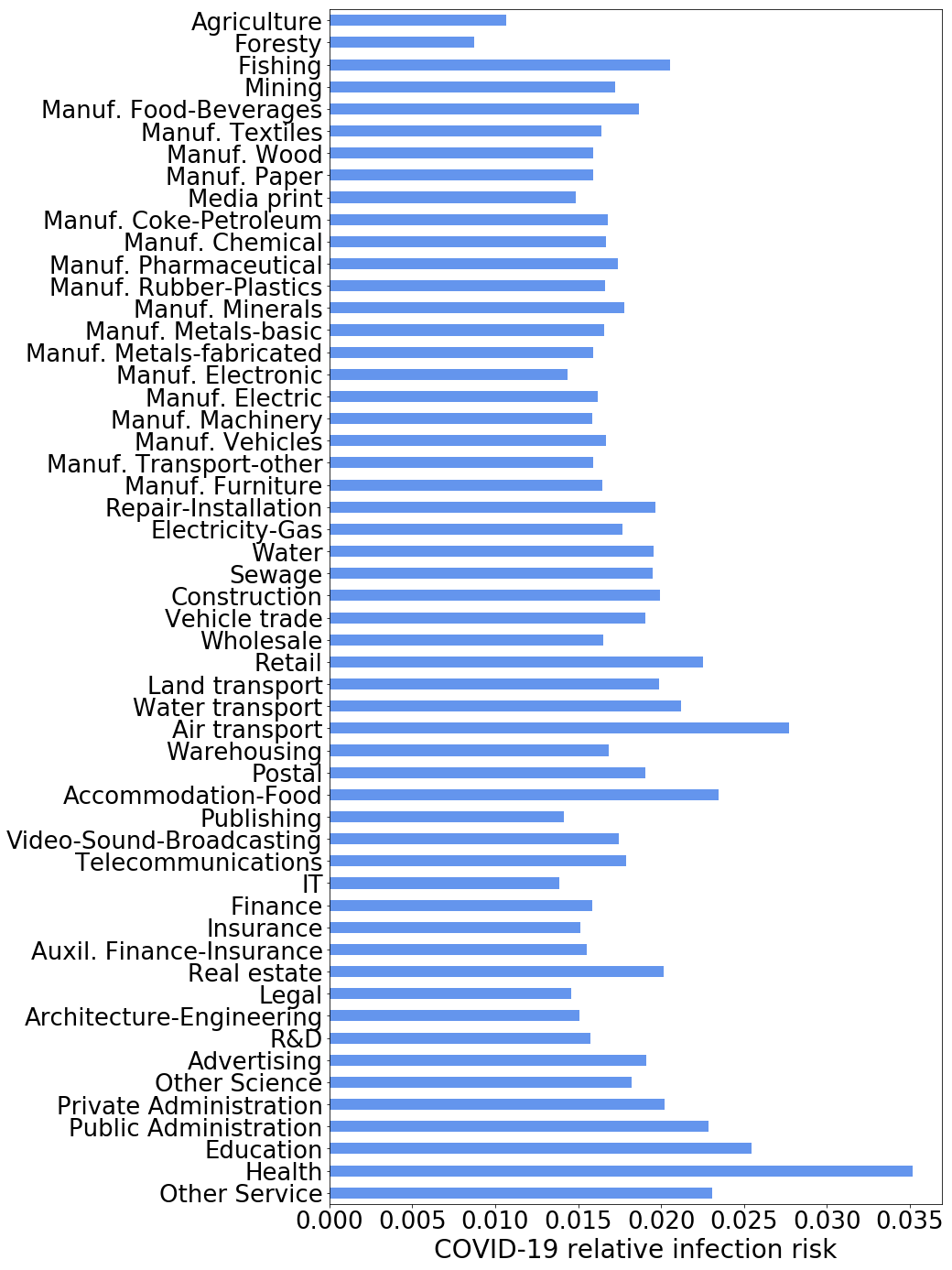}
    \caption{\textbf{COVID-19 relative infection risk of industries.} This index is constructed by taking the average of the exposure to infection and the physical proximity index of industries and then normalizing so that they sum up to one.}
        \label{fig:COVID_risk}
\end{figure}

\paragraph{Consumption.} We consider that, from Table \ref{table:stromgren}, there are three types of consumption activities: Shopping (Convenience stores and Large stores), Restaurants, and Sports Venue. We then map these into the WIOD but looking at the list of industries (Table \ref{tab:FO_shocks}) and assuming that all Shopping activity comes from the WIOD industry \emph{G47: Retail}; that all Restaurant activity comes from the Industry \emph{I:Accomodation-Food}; and all Sports Venue activity comes from the \emph{Other Services} activity.

\paragraph{Transport.} We consider the value from Table \ref{table:stromgren}. We note that \citet{stromgren2017place} observe an important divide between rural and urban places in terms of time spent in public transports. 

\paragraph{Home-related.} In the main text, we need to consider the impact of social distancing measures on $\beta_h$, the share of contacts that are unrelated to whether industries are open or not. We assume that during lockdown, the number of contacts at home stay the same, but the number of contacts with Friends and Relatives, within a family car, or in public urban spaces fall to zero. Since Home is responsible for 76\% (16/21) of the home related contacts, we take $beta_h(\text{lockdown})=\beta_h(t=0)*0.7$.

\paragraph{Population data.} To obtain the share of population in the special industries Schools and Out-of-the-labor-force, we use the ONS Current Population Survey \footnote{
\url{https://www.ons.gov.uk/peoplepopulationandcommunity/populationandmigration/populationestimates/datasets/populationestimatesforukenglandandwalesscotlandandnorthernireland}
}
According to these surveys $62\%$ of the population is employed and $23\%$ of the population is between 0-19 years old. Therefore, we assign $62\%$ of the population to the $i=1, ..., M-2$ working industries, $23\%$ to the school industry\footnote{
We assume that all people between 0-19 years old go to school.
}
and the rest to the retired industry (unemployed and all inactive are thus assigned into this industry). From the $62\%$ of the working population we assign them to the $i=1, ..., M-2$ working industries according to the shares of employment calculated from the WIOD employment data. Finally, 
again using the ONS CPS,
we compute that the share of 0-19 year old who are 14 or below is $ g = 17/23$.

\FloatBarrier

\newpage
\section{Notation}
\label{apx:notation}

\begin{table}[H]
\scriptsize
\begin{center}
    \begin{tabular}{ | p{2cm} | p{12cm} |}
    \hline
    Symbol  & Name  \\ 
    \hline
    $N$ & Number of industries  \\
    $t$ & Time index \\
    $t_\text{start\_lockdown}$ & Start date of lockdown\\
    $t_\text{end\_lockdown}$ & End date of lockdown \\
    $t_\text{end\_pandemic}$ & End date of pandemic \\
    $x_{i,t}$ & Gross output of industry $i$   \\
    $Z_{ij,t}$ & Intermediate consumption of good $i$ by industry $j$  \\
    $c_{i,t}$ & Household consumption of good $i$   \\
    $c_{i,t}^d$ & Demand of household consumption of good $i$   \\
    $f_{i,t}$ & Non-household final demand of good $i$  \\
    $f_{i,t}^d$ & Demand non-household final demand of good $i$  \\
    $l_{i,t}$ & Labor compensation to workers of industry $i$  \\
    $\pi_{i,t}$ & Profits of industry $i$   \\
    $e_{i,t}$ & ``All other'' (non intermediates or labor) expenses of industry $i$   \\
    $\tilde l_t$, $\tilde c_t$, $\tilde \pi_t$ , $\tilde x_t$ & Total labor compensation, consumption, profits and output    \\
    $d_{i,t}$ & Total demand for industry $i$   \\
    $O_{ij,t}$ & Orders (demand from industry $j$ to industry $i$)   \\
    $n_{j}$ & Number of days of targeted inventory for industry $j$   \\
    $A_{i,j}$ & Payments to $i$ per unit produced of $j$ (technical coefficients) \\
    $S_{ij,t}$ & Stock of material $i$ held in $j$'s inventory   \\
    $\tau$ & Speed of inventory adjustment   \\
    $\theta_{i,t}$ & share of goods from industry $i$ in consumption demand   \\
    $\bar \theta_{i,t}$ & share of goods from industry $i$ in consumption demand (unnormalized)  \\
    $\tilde{c}_{t}^d$ & Aggregate consumption demand   \\
    $\rho$ & speed of adjustment of aggregate consumption  \\
    $\epsilon_t$ & consumption exogenous shock   \\
    $\tilde{l}_{t}^p$ & expectations for permanent labor income   \\
    $m$ & share of labor income used to consume final domestic goods   \\
    $\xi_t$ & fraction of pre-pandemic labor income that households expect to retain in the long-run  \\
    $\xi_t^L$ & fraction of pre-pandemic labor income that households expect to retain in the long-run during the lockdown  \\
    $x_{i,t}^{\text{cap}}$ & industry production capacity based on available labor   \\
    $x_{i,t}^{\text{inp}}$ & industry production capacity based on available inputs   \\
    $\tilde{\epsilon}_{i,t}^S$ & exogenous supply shock to industry $i$   \\
    $\tilde{\epsilon}_{i}^D$ & relative changes in demand for goods of industry $i$ during lockdown   \\
    $\tilde{\epsilon}_{i,t}$ & relative changes in demand for goods of industry $i$   \\
    $\tilde \epsilon_t$ & aggregate consumption shock   \\
    $\Delta l_{i,t}$ &  desired change of labor supply of industry $i$\\
    $ l_{i,t}^\text{max}$ & maximum labor supply for industry $i$\\
    $\gamma_\text{H}$, $\gamma_\text{F}$ & Speed of upward/downward labor adjustment (hiring/firing)  \\
    $\Delta s$ & Change in saving rate   \\
    $\tilde l^*_t$ & Household income including social benefits   \\
    \hline
        \end{tabular}
\end{center}
\caption{Notation for the Economic model}
\label{tab:notation_econ}
\scriptsize
\end{table}

\begin{table}[H]
\scriptsize
\begin{center}
    \begin{tabular}{ | l | p{12cm} | }
    \hline
    Symbol  & Name  \\ 
    \hline
    $S^s, S^u, S_i$ & Number of Susceptible individuals in the student, adult non-working, and working population   \\
    $I^s, I^u, I_i$ & Number of Infected individuals in the student, adult non-working, and working population   \\
    $S = S^s+ S^u+ S_i$ & Number of Susceptible individuals   \\
    $I = I^s+ I^u+ I_i$ & Number of Infected individuals   \\
    $R$ & Number of Recovered individuals   \\
    $M =(S+I+R)$ & Number of individuals in the population   \\
    $\beta^*$ & Force of infection   \\
    $\gamma$ & Recovery rate   \\
    $R_0$ & Reproduction number   \\
    $\eta^s, \eta^u, \eta_i$ & Share of people in the student category, the adult non-working category and in  industry $i$  \\
    $\mu^s$ & Share of the student population that attends school  \\
    $\beta_w$ & Share of intensity-weighted contacts at work    \\
    $\beta_s$ & Share of intensity-weighted contacts in schools    \\
    $\beta_c$ & Share of intensity-weighted contacts in consumption  \\
    $\beta_T$ & Share of intensity-weighted contacts in transports  \\
    $\beta_h$ & Share of intensity-weighted contacts at home   \\
     \hline
    \end{tabular}
\end{center}
\caption{Notation for the Epidemic model}
\label{tab:notation_SIR}
\normalsize
\end{table}

\newpage

\FloatBarrier

\end{document}